\documentclass[aps,prb,twocolumn,showpacs,citeautoscript]{revtex4-1}

\usepackage{amsmath,amssymb}
\usepackage{bm}
\usepackage{graphicx}
\usepackage{epstopdf}
\usepackage{latexsym} 
\usepackage{subfigure}
\usepackage{color}
\usepackage{dsfont} 
\usepackage{wasysym}

\newcommand{\be}{\begin{equation}}
\newcommand{\ee}{\end{equation}}

\begin{document}
\title{Spin susceptibility anomaly in cluster Mott insulators on a partially-filled
anisotropic Kagome lattice: applications to LiZn$_2$Mo$_3$O$_8$}
\author{Gang Chen}
%\email{gchen@physics.utoronto.ca}
\affiliation{Department of Physics, University of Toronto, Toronto, Ontario, M5S1A7, Canada}
\author{Hae-Young Kee}
\affiliation{Department of Physics, University of Toronto, Toronto, Ontario, M5S1A7, Canada}
\affiliation{Canadian Institute for Advanced Research/Quantum Materials Program, Toronto, Ontario MSG 1Z8, Canada}
\author{Yong Baek Kim}
\affiliation{Department of Physics, University of Toronto, Toronto, Ontario, M5S1A7, Canada}
\affiliation{Canadian Institute for Advanced Research/Quantum Materials Program, Toronto, Ontario MSG 1Z8, Canada}
\affiliation{School of Physics, Korea Institute for Advanced Study, Seoul 130-722, Korea}
\date{\today}

\begin{abstract}
Motivated by recent experiments on the quantum-spin-liquid candidate material LiZn$_2$Mo$_3$O$_8$, we study a single-band extended Hubbard model on an anisotropic Kagome lattice with the 1/6 electron filling. Due to the partial filling of the lattice, the inter-site repulsive interaction is necessary to generate Mott insulators, where electrons are localized in clusters, rather than at lattice sites. We provide examples of such cluster Mott insulators and study the phase transitions between metallic states and cluster Mott insulators on an anisotropic Kagome lattice. It is shown that these cluster Mott insulators are generally U(1) quantum spin liquids with spinon Fermi surfaces. However, the nature of charge excitations in different cluster Mott insulators could be quite different and we show that there exists a novel cluster Mott insulator where charge fluctuations around the hexagonal cluster induce a plaquette charge order (PCO). The spinon excitation spectrum in this spin-liquid cluster Mott insulator is reconstructed due to the PCO so that only 1/3 of the total spinon excitations are magnetically active. The strong coupling limit of the same model is also analyzed via a Kugel-Khomskii-like model. Based on these results, we propose that the anomalous behavior of the finite-temperature spin-susceptibility in LiZn$_2$Mo$_3$O$_8$ may be explained by finite-temperature properties of the cluster Mott insulator with the PCO as well as fractionalized spinon excitations. Existing and possible future experiments on LiZn$_2$Mo$_3$O$_8$, and other Mo-based cluster magnets are discussed in light of these theoretical predictions.
\end{abstract}

\date{\today}

\pacs{75.10.Kt, 75.10.Jm}

\maketitle

\section{Introduction}
\label{sec1}

If there is no spontaneous symmetry breaking, the ground state of a Mott insulator with 
{\it odd} number of electrons per unit cell may be a quantum spin liquid (QSL)\cite{Hastings04}. 
The QSL is an exotic quantum phase of matter with a long-range quantum entanglement\cite{Wenbook},
which is characterized by fractionalized spin excitations 
and an emergent gauge structures at low energies\cite{Balents10}. 
It is now clear that some frustrated Mott insulating systems which are proximate to 
Mott transitions may provide physical realizations of QSL 
phases.\cite{Lee05,Motrunich05,Motrunich06,Podolsky09,PhysRevB.87.165120}
These U(1) QSLs arise from strong charge fluctuations in the weak Mott regime,
which can generate sizable long range spin exchanges or spin ring exchanges and 
suppress possible magnetic orderings.\cite{Motrunich05,Motrunich06}
Several QSL candidate materials, such as the 2D 
triangular lattice organic materials $\kappa$-(ET)$_2$Cu$_2$(CN)$_3$ 
and EtMe$_3$Sb[Pd(dmit)$_2$]$_2$,
and a 3D hyperkagome system Na$_4$Ir$_3$O$_8$\cite{Hiroi01,Shimizu03,Okamoto07}, 
are expected to be in this weak Mott regime.
These weak Mott-insulator U(1) QSLs are obtained as a deconfined phase of an
emergent U(1) lattice gauge theory\cite{Florens04,Lee05}, where 
the electron is fractionalized into spin-carrying spinons and charged bosons.
The charge excitations are gapped and
the low-energy physics of the QSLs is described by a spinon Fermi surface 
coupled to the emergent U(1) gauge field. 

In this work, motivated by the recent experiments on a new QSL candidate material
LiZn$_2$Mo$_3$O$_8$\cite{Sheckelton12,Sheckelton14,Mourigal14},
we consider a 1/6-filled extended Hubbard model with nearest-neighbor repulsions
and propose a U(1) QSL with spinon Fermi surfaces and a plaquette
charge order (PCO) as a possible ground state. 
The Mott insulators in partially filled systems arise due to
large nearest-neighbor repulsions and localization of the charge 
degrees of freedom in certain cluster units. Hence such Mott
insulators may be called ``cluster Mott insulators'' (CMIs). 

\begin{figure}[t]
{\includegraphics[width=8cm]{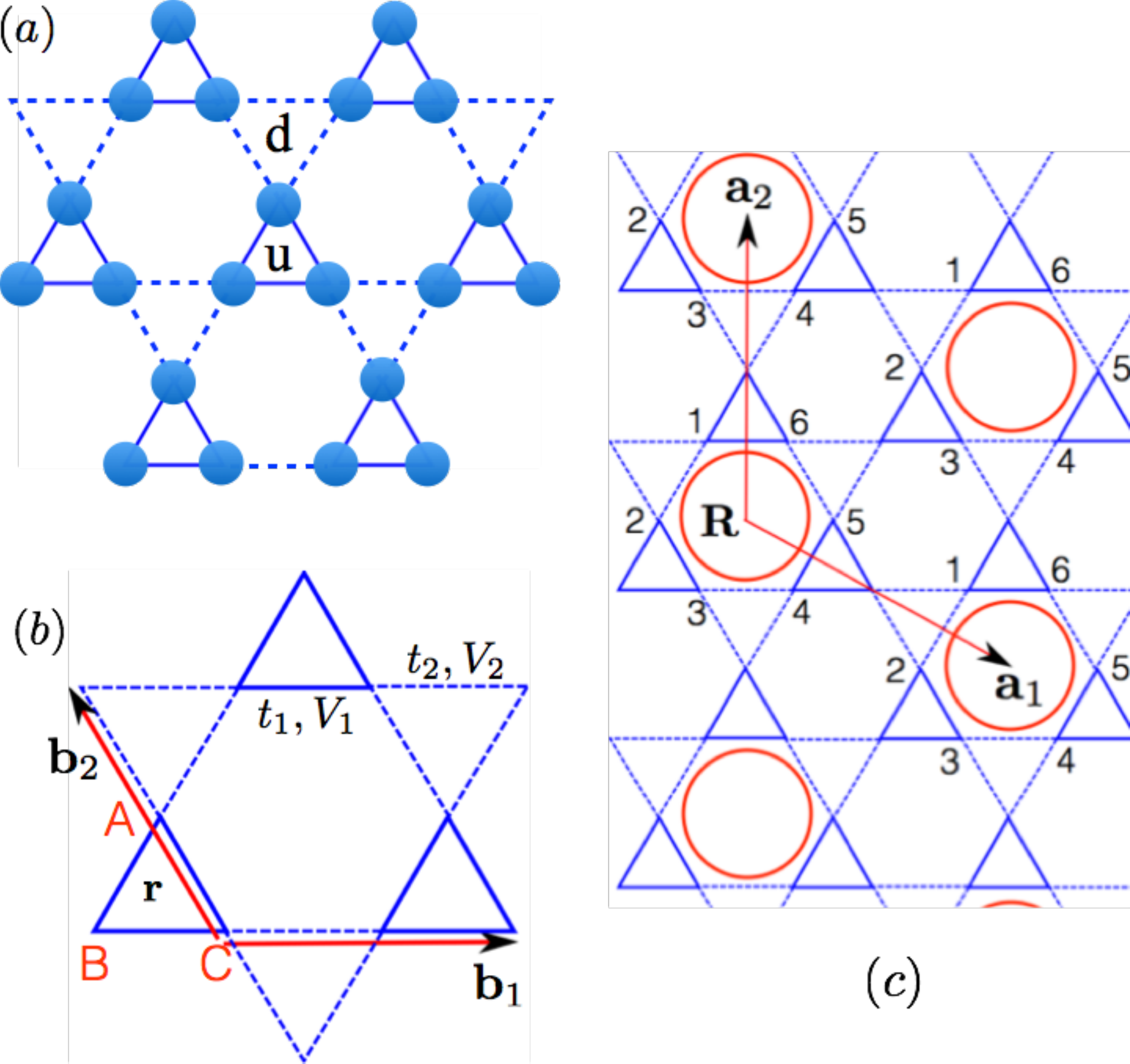}}
\caption{(Color online.) 
(a) Mo$_3$O$_{13}$ clusters are organized into a triangular lattice structure.
Oxygen atoms are not shown.  
The Mo sites form an anisotropic Kagome lattice. 
(b) 
${\bf b}_1, {\bf b}_2$ are two primitive lattice vectors that connect
neighboring unit cells.  ${\bf r}$ labels the Kagome unit cell and 
$\mu=\text{A,B,C}$ labels the 3 sublattices. 
(c) The PCO and the emergent triangular lattice (ETL)
(with lattice vectors ${\bf a}_1, {\bf a}_2$) in the type-II CMI. 
3 electrons hop resonantly in each hexagon that is marked by a
(red) circle. `${\bf R}$' labels the resonating hexagon or the unit cell
of the ETL and the `1,2,3,4,5,6' label the 6 vertices in the resonating hexagon. 
}
\label{fig1}
\end{figure}

In LiZn$_2$Mo$_3$O$_8$, as described in
Ref.~\onlinecite{Sheckelton12}, 
each Mo$_3$O$_{13}$ triangular cluster hosts one unpaired localized electron 
with $S={1}/{2}$ moment. These Mo$_3$O$_{13}$ clusters 
 are organized into 
a triangular lattice structure (see Fig.~\ref{fig1}).\cite{Sheckelton12,Sheckelton14,Mourigal14} 
No magnetic ordering is detected in neutron scattering, in
NMR and $\mu$SR measurement down to $\sim 0.1$K.\cite{Sheckelton12,Sheckelton14,Mourigal14}  
In particular, the spin susceptibility shows a very puzzling anomaly: 
below about 100K the spin susceptibility is governed by a different 
Curie-Weiss law with a much smaller 
Curie-Weiss temperature ($\Theta_{\text{CW}}^{\text L} = -14$K) 
from the high temperature one ($\Theta_{\text{CW}}^{\text H} = -220$K) and
 a much reduced Curie constant which is 1/3 of the high temperature one.  

%-----------------------------
%-----------------------------

In a recent theoretical work\cite{Flint13}, Flint and Lee 
considered the possibility of an emergent honeycomb lattice with 
weakly coupled dangling spins in the centers of the hexagons. 
In their description, the emergent honeycomb system may 
form a gapped QSL phase 
while the remaining weakly-coupled dangling spin moments 
comprise 1/3 of the total magnetic moments
and dominate the low-temperature magnetic properies
which then explains the ``1/3 anomaly'' in the spin susceptibility. 
Their theory invokes the lattice degrees of freedom to work in a way to 
generate the emergent honeycomb lattice for the spin system. 
Such a scenario might be plausible but needs to be confirmed by further experiments. 
In this paper, however, we explore an alternative explanation for the experiments 
that is based on electronic degrees of freedom and their interactions. 

Instead of working with the exchange model between the local spin moments,
we consider a single-band extended Hubbard model for the unpaired Mo electrons of
the Mo$_3$O$_{13}$ clusters. The Hubbard model is 
the parent model of the spin exchange interactions and may contain the crucial physics that 
is not described by the spin exchange model. 
Moreover, the Mo electrons are in $4d$ electron orbital states, and $4d$ electron systems
are often not in the strongly localized regime due to the substantial spatial extension of the 
orbital wavefunction. Therefore, we think it is more appropriate to model the system by a
Hubbard-like model. 
Our single-band extended Hubbard model is defined on the 
anisotropic Kagome lattice that is formed by the Mo sites (see Fig.~\ref{fig1}a)
and is given by
\begin{eqnarray}
H &=&  \sum_{ \langle ij \rangle \in \text{u}} 
[-t_1 (c^{\dagger}_{i\sigma} c^{\phantom\dagger}_{j\sigma} + h.c.)  + V_1 n_i n_j  ] 
+(\text{u}\leftrightarrow \text{d},1\leftrightarrow 2)
\nonumber \\
&+&  \sum_{i} \frac{U}{2} (n_i -\frac{1}{2})^2 ,
\label{eq1}
\end{eqnarray}
where the spin $S=1/2$ index $\sigma$ is implicitly summed, 
$c^{\dagger}_{i\sigma}$ ($c^{\phantom\dagger}_{i\sigma} $) creates 
(annihilates) an electron with 
spin $\sigma$ at lattice site $i$, and $t_1, V_1$ and $t_2, V_2$ are the 
nearest-neighbor electron hopping and interaction in the up-pointing triangles 
(denoted as `u') and the down-pointing triangles (denoted as `d') 
(see Fig.~\ref{fig1}b), respectively. $n_i = \sum_{\sigma} 
c^\dagger_{i\sigma}c^{\phantom\dagger}_{i\sigma}$ 
is the electron occupation number at site $i$. Since there exists only one electron
in each Kagome lattice unit cell, the electron filling for this Hubbard model is $1/6$. 
In Sec.~\ref{sec2},
we shall provide a motivation to consider this single-band Hubbard model 
using a quantum chemistry analysis. 

We include the on-site Hubbard-$U$ interaction as well as two inter-site 
repulsions $V_1$ and $V_2$ in the extended Hubbard model. 
Although the down-triangles are larger in size than the up-triangles 
in LiZn$_2$Mo$_3$O$_8$, because of the large spatial extension
of the $4d$ Mo electron orbitals we think it is necessary to 
include the inter-site repulsion $V_2$ for the down-triangles.
Moreover, for LiZn$_2$Mo$_3$O$_8$ we expect $t_1 > t_2$ and 
$U >V_1 >V_2$. Keeping the Hubbard-$U$ interaction as the 
largest energy scale in the model, we study the phase diagram in terms
of $t_1, t_2, V_1, V_2$.
% and explore possible proximate phases in 
%much broader parameter regimes in this paper. 

In the case of the fractional filling, the Mott localization is driven by the inter-site 
repulsions ($V_1, V_2$) rather than the on-site Hubbard interaction $U$ 
and the electrons are localized in the (elementary) triangles of the Kagome lattice 
instead of the lattice sites.
Because of the asymmetry between the up-triangles and down-triangles of the Kagome lattice,
the Mott localization in the up-triangles and down-triangles does not need to occur simultaneously. 
Therefore, {\it two types of CMIs are expected.}  

Refs.~\onlinecite{Sheckelton12,Flint13} assume that LiZn$_2$Mo$_3$O$_8$ 
is in the type-I CMI phase, where the electrons are only localized in the up-triangles while  
the electron number in the down-triangles remains strongly fluctuating. 
In our work,  
 we propose that the system may be more likely to be in the type-II CMI, 
where the inter-site repulsions ($V_1, V_2$) 
are strong enough to localize the electrons in both up-triangles and down-triangles
even though $t_1 > t_2$. 
The electron number in every triangle is then fixed to be one. 
Although a single electron cannot hop from one triangle to another, 
a (collective) ring hopping of 3 electrons on the perimeter 
of an elementary hexagon on the Kagome lattice is allowed 
and gives rise to a long-range
plaquette charge order (PCO) in the type-II CMI (see Fig.~\ref{fig1}c). 
The emergence of PCO in the type-II CMI is a {\it quantum effect} and cannot
be obtained from the classical treatment of the electron interaction. 

With the PCO, 1/3 of the elementary hexagons 
become resonating. 
%and host 3 electrons hopping collectively on the 
%perimeter of the hexagon, which resembles the Benzene molecule. 
As shown in Fig.~\ref{fig1}c, 
these resonating hexagons form an emergent triangular lattice (ETL). 
The PCO triples the original unit cell of the Kagome lattice, and
the localized electron number in the enlarged unit cell now
becomes 3, which is still {\it odd}. Therefore, the type-II CMI with
the PCO is not connected to a trivial band insulator and the QSL is 
still expected. In the resulting U(1) QSL, we obtain 9 mean-field spinon bands for 
the type-II CMI with the PCO, compared to the 3 spinon bands
in the U(1) QSL for the type-I CMI without the PCO. 
A direct band gap separates the lowest spinon band 
from other spinon bands in the presence of PCO.
%due to the PCO. 
The lowest spinon band is completely filled by 2/3 of the spinons,
leaving the remaining 1/3 of the spinons to partially fill the second and third 
lowest spinon bands. Because of the band gap, the only active degrees 
of freedom at low energies 
are the spinons in the partially filled spinon bands, and the fully-filled 
lowest spinon band is inert to external magnetic field at low temperatures 
as long as the PCO persists. Therefore, only 1/3 of the magnetic degrees of 
freedom are active at low temperatures. 
If one then considers the local moment formation starting 
from the band filling picture of the spinons (just like electrons occupying 
the same band structure) only the 1/3 of the spinons from the partially filled upper 
bands would participate in the local moment formation. This means 
the type-II CMI phase with the PCO would be continuously connected to 
the Curie-Weiss regime with the 1/3 Curie constant (compared to the case 
when all spinons can participate in the local moment formation) at high temperature.
This would explain the ``1/3 anomaly'' in the spin susceptibility data of 
LiZn$_2$Mo$_3$O$_8$.

Alternatively, we could consider the strong coupling regime where the 
PCO is very strong. Here the 3 electrons are strongly localized in each resonating 
hexagon and an effective local moment model appears.
The three electrons in one individual 
resonating hexagon are then locally entangled which leads to 
4-fold degenerate ground states.  This 4-fold degeneracy is 
characterized by one time-reversal-odd spin-1/2 and one time-reversal-even 
pseudospin-1/2 degrees of freedom.
It is then shown that the local quantum entanglement of the 
three electrons in each resonating hexagon only gives rise to 
one magnetically active spin-1/2 moment. 
The spins and pseudospins are weakly coupled and are 
described by a Kugel-Khomskii model\cite{Kugel82} on the emergent
triangular lattice (see Fig.~\ref{fig1}c). 
We show that this strong coupling result is also consistent with 
the ``1/3 anomaly''  in the spin susceptibility data of 
LiZn$_2$Mo$_3$O$_8$.

%-----------------------------
\begin{figure}[t]
{\includegraphics[width=7cm]{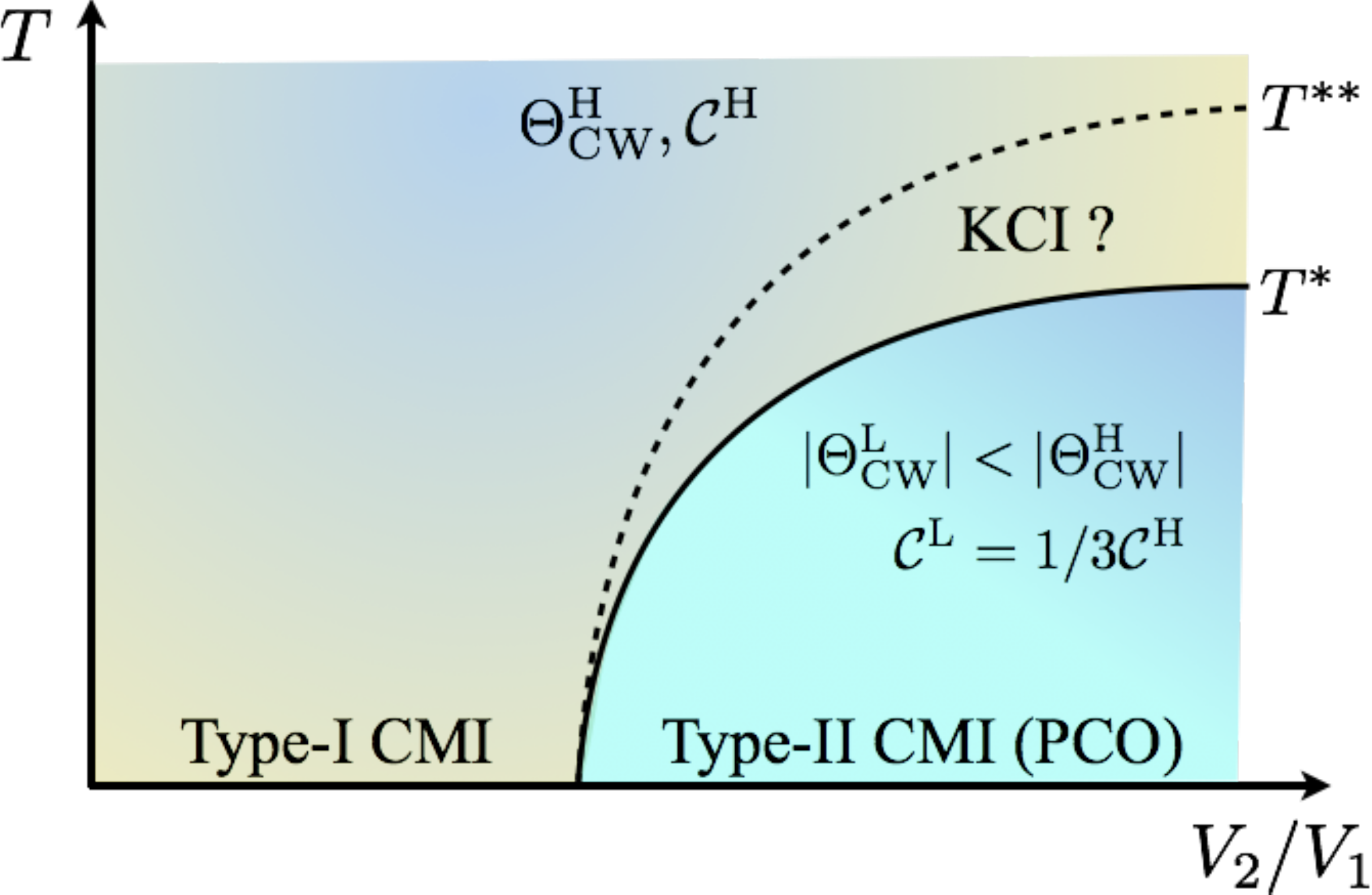}}
\caption{A schematic finite temperature phase diagram proposed for
LiZn$_2$Mo$_3$O$_8$. 
$T^{\ast}$ represents the finite-temperature transition for the PCO.
Between $T^{\ast}$ and a crossover temperature $T^{\ast \ast}$, the charge configurations are
highly degenerate and this phase is referred as the Kagome charge ice (KCI) in the text.
In LiZn$_2$Mo$_3$O$_8$, the intermediate KCI may be quite narrow. 
$\Theta_{\text{CW}}^{\text{H}}, {\mathcal C}^{\text H}$ 
($\Theta_{\text{CW}}^{\text{L}}, {\mathcal C}^{\text L}$) 
are the Curie-Weiss temperature and Curie constant at higher (lower) temperatures,
respectively. 
}
\label{fig2}
\end{figure}
%-----------------------------

The PCO in the type-II CMI breaks the discrete lattice symmetries of
the Kagome system. At the finite-temperature transition, 
the PCO is destroyed and the lattice symmetries are restored.
This transition should occur at a temperature $T^{\ast}$ that is on the
order of the electron ring-hopping energy scale. 
Moreover, this transition is found to be strongly first order in a clean system 
but would be smeared out in a disordered LiZn$_2$Mo$_3$O$_8$ sample.  
As shown in Fig.~\ref{fig2}, there exists another crossover temperature 
$T^{\ast\ast} \sim \mathcal{O}(V_2)$ above which the electron localization 
in the down-triangles is thermally violated. 
Between $T^{\ast}$ and $T^{\ast\ast}$,
the electron localization with one electron in each triangle is still obeyed 
but the PCO is destroyed. 
The electron occupation configuration in this intermediate regime
is extensively degenerate just like the spin configuration in 
a classical Kagome spin ice\cite{Wills02}, so we name this intermediate temperature phase
as ``Kagome charge ice'' (KCI). 
 Above the crossover temperature $T^{\ast\ast}$, the system 
 can be thought as the higher temperature regime of 
 the type-I CMI, where the charge localization occurs only in the up-triangles.
Besides the distinct finite-temperature charge behaviors, we also expect 
thermal crossovers in the spin susceptibility (see Fig.~\ref{fig2}). 
Above $T^{\ast}$, each electron would contribute
a local spin-1/2 moment, and hence, 
the anomalous low-temperature spin susceptibility 
changes into regular Curie-Weiss behavior whose 
Curie constant, $\mathcal{C}^{\text H}$,
is 3 times the low-temperature Curie constant,  
$\mathcal{C}^{\text L}$. 

The rest of the paper is structured as follows.
In Sec.~\ref{sec2}, we start from the molecular orbitals of the 
Mo$_3$O$_{13}$ clusters, introduce an appropriate atomic state representation, 
and provide a microscopic justification for the single-band Hubbard model in Eq.~\eqref{eq1}.
In Sec.~\ref{sec3A}, we formulate the charge sector of the type-II CMI 
as a compact U(1) gauge theory on the dual honeycomb lattice (DHL). 
Then in Sec.~\ref{sec3B}, we introduce a new slave-particle construction for the electron
to obtain the mean-field phase diagram that includes 
the type-I CMI, type-II CMI and a Fermi-liquid metal (FL-metal). 
In Sec.~\ref{sec3C}, we focus on the type-II CMI phase. 
We obtain the PCO in the charge sector by mapping the low-energy charge sector
Hamiltonian into a quantum dimer model on the DHL. 
We generalize the Levin-Wen string mean-field theory to study the reconstruction of 
the spinon band structure by the PCO. 
We explain the consequence of this reconstructed spinon band structure  
and discuss the low-temperature magnetic susceptibility. 
In Sec.~\ref{sec4}, we consider the strong coupling regime of the 
type-II CMI with the PCO and identify the structure of the local moments 
formed by the 3 electrons in an individual resonating hexagon.  
The interaction between these local moments is described by 
a Kugel-Khomskii model on the ETL. 
In Sec.~\ref{sec5A} and Sec.~\ref{sec5B}, we connect our theory to the experiments
on LiZn$_2$Mo$_3$O$_8$ and suggest possible future experiments.  
Finally in Sec.~\ref{sec5C}, we discuss other Mo based cluster magnets. 
Some details of the computations are included in the Appendices.

\section{Molecular orbitals and the Hubbard model}
\label{sec2}

\begin{figure}[t]
{\includegraphics[width=8cm]{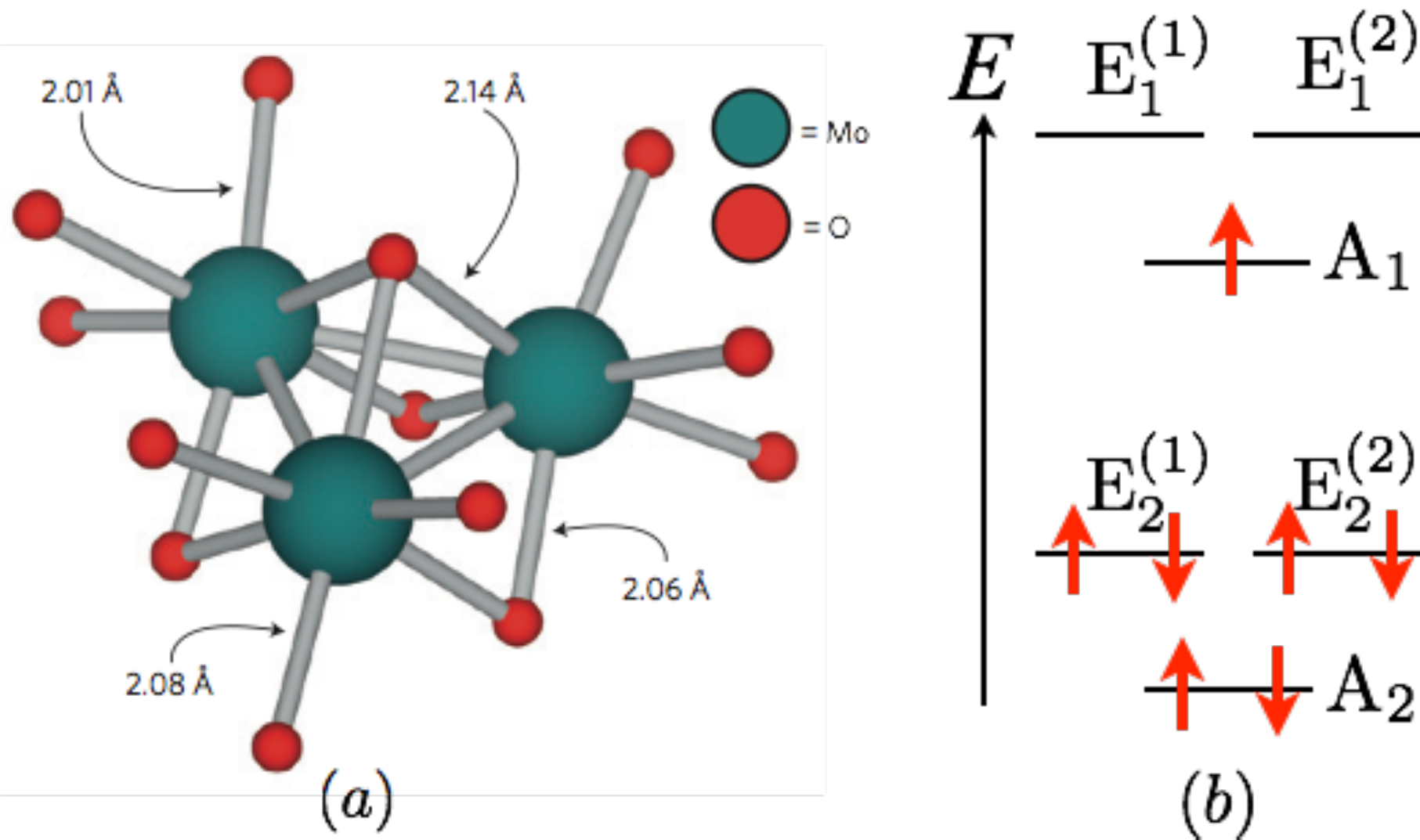}}
\caption{(Color online.) 
(a) The Mo$_3$O$_{13}$ cluster (adapted from Ref.~\onlinecite{Sheckelton12}).
(b) The schematic energy level diagram of the molecular orbitals for a single Mo$_3$O$_{13}$ cluster. 
The molecular orbitals are classified according to the irreducible representations of the 
C$_{3v}$ point group of the cluster\cite{Note1}.
The unfilled molecular orbitals at high energies are not shown. 
}
\label{figxx}
\end{figure}

As suggested by Refs.~\onlinecite{Cotton64,Sheckelton12}, the Mo electrons in an 
isolated Mo$_3$O$_{13}$ cluster form molecular orbitals
because of the strong Mo-Mo bonding. Among the 7 valence electrons in 
the cluster, 6 of them fill the lowest three molecular orbitals \{A$_2$, 
E$_2^{(1)}$, E$_2^{(2)}$\} in pairs, and the seventh electron remains 
unpaired in a totally symmetric A$_1$
molecular orbital with equal contributions from all three Mo atoms (see Fig.~\ref{figxx}). 

%-----------------------------
%-----------------------------

We first consider the molecular orbital states in the group
\{A$_1$, E$_1^{(1)}$, E$_1^{(2)}$\}. 
This group can be described by a linear combination of 
an atomic state $|\psi_1\rangle$ at each Mo site (which is in turn
a linear combination of five $4d$ atomic orbitals).
\begin{eqnarray}
|{\text A}_1 \rangle &=& \frac{1}{\sqrt{3}} 
\big[ | {\psi}_1 \rangle_{\text A} +  | {\psi}_1 \rangle_{\text B}
+  | {\psi}_1 \rangle_{\text C}\big],
\label{eqA2}\\
|{\text E}_1^{(1)} \rangle &=& \frac{1}{\sqrt{3}} \big[ | {\psi}_1 \rangle_{\text A} + 
e^{ i \frac{2\pi}{3}} | {\psi}_1 \rangle_{\text B}
+ e^{ - i \frac{2\pi}{3}} | {\psi}_1 \rangle_{\text C}\big],
\label{eqE2a}
\\
|{\text E}_1^{(2)}  \rangle &=& \frac{1}{\sqrt{3}} \big[ | {\psi}_1 \rangle_{\text A} +
e^{- i \frac{2\pi}{3}}   | {\psi}_1 \rangle_{\text B}
+ e^{ i \frac{2\pi}{3}} | {\psi}_1 \rangle_{\text C}\big],
\end{eqnarray}
where $\mu $($= \text{A,B,C}$) labels the three Mo sites in the cluster 
and the atomic state $| {\psi}_1 \rangle_{\mu}$ is the contribution
from the Mo atom at $\mu$. The atomic states $| {\psi}_1 \rangle_{\mu}$ 
at different Mo sites are related by the 3-fold rotation about the center of the cluster. 
Likewise, the fully-filled \{A$_2$, E$_2^{(1)}$, E$_2^{(2)}$\} and other unfilled 
molecular orbitals at higher energies are constructed from 
the atomic state $|\psi_2\rangle$ and other atomic states
$|\psi_j\rangle$ ($j=3,4,5$), respectively. Here,
the {\it atomic states} $\{ | {\psi}_j \rangle_{\mu} \}$ ($j=1,2,3,4,5$)
represent a {\it distinct} orthonormal basis from the five $4d$ {\it atomic orbitals} 
that are the eigenstates of the local Hamiltonian of the MoO$_6$ octahedron.  

We group the molecular orbitals based on the atomic state from which they are 
constructed. In this classification, for example, \{A$_1$, E$_1^{(1)}$, E$_1^{(2)}$\}
fall into one group while \{A$_2$, E$_2^{(1)}$, E$_2^{(2)}$\} fall into another group
as they are constructed from two different atomic states. 

In LiZn$_2$Mo$_3$O$_8$, the different molecular orbitals 
of the neighboring clusters Mo$_3$O$_{13}$ overlap and form molecular bands. 
To understand how the molecular orbitals overlap with each other, we 
consider the wavefunction overlap of different atomic states $|\psi_j\rangle$. 
Since the down-triangle has the same point group symmetry as the up-triangle in LiZn$_2$Mo$_3$O$_8$,
the wavefunction overlap of the atomic states in
the down-triangles should approximately resemble the one in the up-triangles.   
More precisely, the wavefunction of the atomic state (e.g. $|\psi_1\rangle$)
has similar lobe orientations both inward into and outward from the Mo$_3$O$_{13}$
cluster, with different spatial extensions due to the asymmetry 
between up-triangles and down-triangles.  
Consequently, the orbital overlap between the molecular orbitals from the same group
 is much larger than the one 
between the molecular orbitals  
from the different groups.  
Therefore, each molecular band cannot be formed by one single molecular orbital
but is always a strong mixture of the three molecular orbitals in the same group. 

We now single out the three molecular bands that are primarily formed by 
the group of \{A$_1$, E$_1^{(1)}$, E$_1^{(2)}$\} molecular orbitals. 
There are four energy scales associated with these three
molecular orbitals and bands:  \\
(1) the energy separation $\Delta E $ 
between the  \{A$_1$, E$_1^{(1)}$, E$_1^{(2)}$\}  group
 and other groups of orbitals (both filled and unfilled), \\
(2) the {\it total} bandwidth $W$ of the three molecular bands
formed by the \{A$_1$, E$_1^{(1)}$, E$_1^{(2)}$\} molecular orbitals, \\
(3) the intra-group interaction between two electrons on 
any one or two orbitals of the \{A$_1$, E$_1^{(1)}$, E$_1^{(2)}$\} group,\cite{note2} \\
(4) the inter-group interaction between the electron 
on an orbital of the \{A$_1$, E$_1^{(1)}$, E$_1^{(2)}$\} group
and the other electron on an orbital of a different group.
It is expected, from the previous wavefunction overlap argument, that the inter-group 
interaction is much weaker than the intra-group interaction 
and thus can be neglected at the first level of approximation\cite{Note3}. 

In this paper, we assume that the energy separation $\Delta E$
is larger than the total bandwidth $W$ and the intra-group interaction. 
In this regime, 
the large $\Delta E$ separates these three molecular bands 
from other molecular bands (both filled and unfilled) so that 
the fully filled \{A$_2$, E$_2^{(1)}$, E$_2^{(2)}$\} orbitals remain
fully-filled and the unfilled molecular orbitals remain unfilled even after they
form bands. 
Moreover, the large $\Delta E$ also prevents a band-filling reconstruction 
due to the interaction (in principle, the system can gain interaction energy by 
distributing the electrons evenly among different groups of orbitals). 
Therefore, we can ignore both the fully-filled and unfilled molecular bands
and just focus on the three partially filled bands. 
It also means one will have to consider three-band model 
with all of \{A$_1$, E$_1^{(1)}$, E$_1^{(2)}$\} orbitals
on the triangular lattice formed by the Mo$_3$O$_{13}$ clusters. 
In this case, alternatively one could simply consider  
atomic states as the starting point. Then the relevant model
would be a single-band Hubbard model based on the atomic state $|\psi_1\rangle$ 
at each Mo site of the anisotropic Kagome lattice.
We take the latter approach in this paper.
Finally, since only one atomic state $|\psi_1\rangle$ is involved at 
each Mo site, the orbital angular momentum of the electrons are trivially quenched
so that we can neglect the atomic spin-orbit coupling
at the leading order\cite{William14}. 

The corresponding single-band Hubbard model is given by Eq.~\eqref{eq1},
where we include the on-site and nearest-neighbor electron interactions. 
Now it is clear that the physical meaning of the electron operator $c^\dagger_{i\sigma}$ 
($c^{\phantom\dagger}_{i\sigma}$) in Eq.~\eqref{eq1} 
is to create (annihilate) an electron on the 
state $|\psi_1 \rangle_i$ with spin $\sigma$ at the Kagome lattice site $i$.

\section{Generic phase diagram}
\label{sec3}

As explained in Sec.~\ref{sec1}, the extended Hubbard model in Eq.~\eqref{eq1} 
can support two types of CMIs with distinct 
electron localization patterns.  Besides the insulating phases, 
the model includes a FL-metal when the interaction is weak. 
To describe different phases and study the Mott transitions in this model, we first
employ the standard slave-rotor representation for the 
electron operator,\cite{Florens04,Lee05} 
$c^\dagger_{i\sigma} = f^\dagger_{i\sigma} e^{i \theta_i}$,
where the bosonic rotor ($e^{i\theta_i}$) carries the electron charge and
the fermionic spinon ($f^\dagger_{i\sigma}$) carries the spin quantum number.  
To constrain the enlarged Hilbert space, we introduce an angular momentum variable $L_i^z$, 
$L_i^z = [\sum_{\sigma} f^\dagger_{i\sigma} f^{\phantom\dagger}_{i\sigma} ] - {1}/{2}$,
where $L_i^z$ is conjugate to the rotor variable with $[\theta_i,L_j^z] = i \delta_{ij}$. 
Moreover, since the on-site interaction $U$ is assumed to be the biggest energy scale,
in the large $U$ limit the double electron occupation is always suppressed. 
Hence, the angular variable $L_i^z$ primarily takes 
$ L_i^z = 1/2$ ($-1/2$) for a singly-occupied (empty) site. 

Via a decoupling of the electron hopping term into the spinon and rotor sectors, 
we obtain the following two coupled Hamiltonians for the spin and charge sectors, respectively,
\begin{eqnarray}
H_{\text{sp}} &=& 
-\sum_{\langle ij \rangle} t^{\text{eff}}_{ij}
 (f^{\dagger}_{i\sigma} f^{\phantom\dagger}_{j\sigma} + h.c.)
 - \sum_{i} h_i  
 f^\dagger_{i\sigma} f^{\phantom\dagger}_{i\sigma} 
 \label{eq5}
\\
H_{\text{ch}} &=& - \sum_{\langle ij \rangle } 
2{J}^{\text{eff}}_{ij} \cos (\theta_i -\theta_j) 
+ \sum_{\langle ij \rangle  } 
V_{ij} (L_i^z +\frac{1}{2})
\nonumber \\
&& \times (L_j^z +\frac{1}{2}) 
+ \sum_i [\frac{U}{2} (L_i^z)^2  + h_i (L_i^z +\frac{1}{2})], 
\label{eq6}
\end{eqnarray}
where 
$t_{ij}^{\text{eff}} = t_{ij} \langle e^{i \theta_i - i \theta_j} \rangle 
\equiv |t_{ij}^{\text{eff}}| e^{i a_{ij}},
{J}_{ij}^{\text{eff}} = t_{ij} \sum_{\sigma} 
\langle f^\dagger_{i\sigma} f^{\phantom\dagger}_{j\sigma} \rangle 
\equiv   |  { J}_{ij}^{\text{eff}}  | e^{- i a_{ij}}$
and $t_{ij} = t_1$ ($t_2$), $V_{ij} = V_1$ ($V_2$) 
for the bond $ij$ on the up-triangles (down-triangles).
$h_i$ is a Lagrange multiplier that imposes 
the Hilbert space constraint. Here, we have chosen the couplings to 
respect the symmetries of the Kagome lattice. 
The Hamiltonians $H_{\text{sp}}$ and $H_{\text{ch}}$ 
are invariant under an internal U(1) gauge transformation, 
$f^{\dagger}_{i\sigma} \rightarrow f^{\dagger}_{i\sigma} e^{- i \chi_i},
\theta_i \rightarrow \theta_i + \chi_i$, and $a_{ij} \rightarrow a_{ij} + \chi_i - \chi_j$. 
This internal U(1) gauge structure is then referred as the U(1)$_{\text{sp}}$ gauge field
in the following. 

Since the electron is not localized on a lattice site in the CMIs, 
it will be shown that the rotor variable $e^{i\theta_i}$ is insufficient to 
describe all the phases in a generic phase diagram, except for
certain special limits for the type-I CMI which we analyze 
in Appendix.~\ref{secapp1}. 
To remedy this issue, we will extend the slave-rotor representation
to a new parton construction for the electron operator in the following sections
and then generate the phase diagram.

\subsection{Charge sector of type-II CMI as a compact U(1) gauge theory}
\label{sec3A} 

To introduce a new parton construction, we 
  need to first understand the low-energy physics of the 
charge sector, especially in the type-II CMI. 
We will show the charge localization pattern in the type-II CMI 
leads to an emergent compact U(1) lattice
gauge theory description for the charge-sector quantum fluctuations. 
In the slave-rotor formalism, the charge-sector Hamiltonian is given by 
\begin{eqnarray}
H_{\text{ch}} & =&  
\sum_{\langle ij\rangle } 
{- 2J_{ ij}^{\text{eff}}}\cos (\theta_i -\theta_j) 
+ V_{ ij}( L_i^z + \frac{1}{2})  (L_j^z + \frac{1}{2}) 
\nonumber \\
&&+ \sum_i h_i (L_i^z + \frac{1}{2}),
\end{eqnarray}
where we have dropped the $U$ interaction term because 
$L_i = \pm 1/2$ in the large $U$ limit.  
This charge sector Hamiltonian can be thought as a Kagome lattice 
spin-1/2 XXZ model in the 
presence of an external magnetic field upon
identifying the rotor operators as the spin ladder operators,
$e^{\pm i \theta_i } = L^{\pm}_i $
where 
\begin{equation} L^{\pm}_i |  L_i^z = \mp \frac{1}{2} \rangle = | L_i^z
=  \pm \frac{1}{2} \rangle.
\end{equation}
Thus the corresponding effective spin-$L$ model reads
\begin{eqnarray}
H_{\text{ch}} & = &
 \sum_{ \langle ij \rangle   } \big[ 
{- J_{ ij}^{\text{eff}}} ( L^+_i L^-_j + h.c. ) + V_{ ij} L_i^z L_j^z
\big]
\nonumber \\
&& + B^{\text{eff}} \sum_i L_i^z, 
\end{eqnarray}
in which we have made a uniform mean-field approximation such that 
$h_i + 3(V_1+V_2) \equiv B^{\text{eff}}$. 
The 1/6 electron filling is mapped to the total ``magnetization'' condition
${N_s}^{-1} \sum_i L_i^z = - \frac{1}{6}$, 
where $N_s$ is the total number of Kagome lattice sites.

%-----------------------------
\begin{figure}[ht]
\subfigure[]{\includegraphics[width=3.6cm]{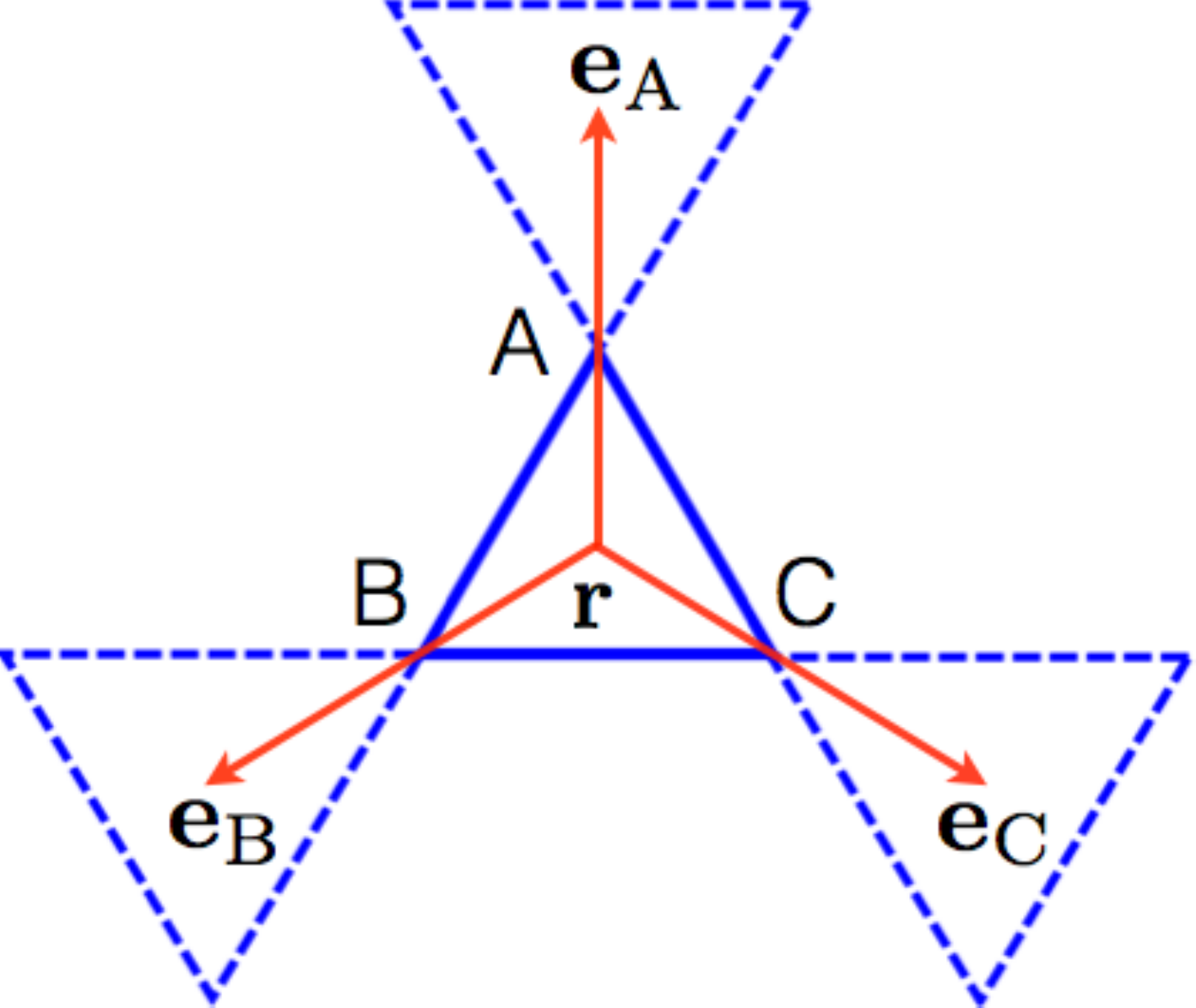}}
\hspace{0.5cm}
\subfigure[]{\includegraphics[width=4.2cm]{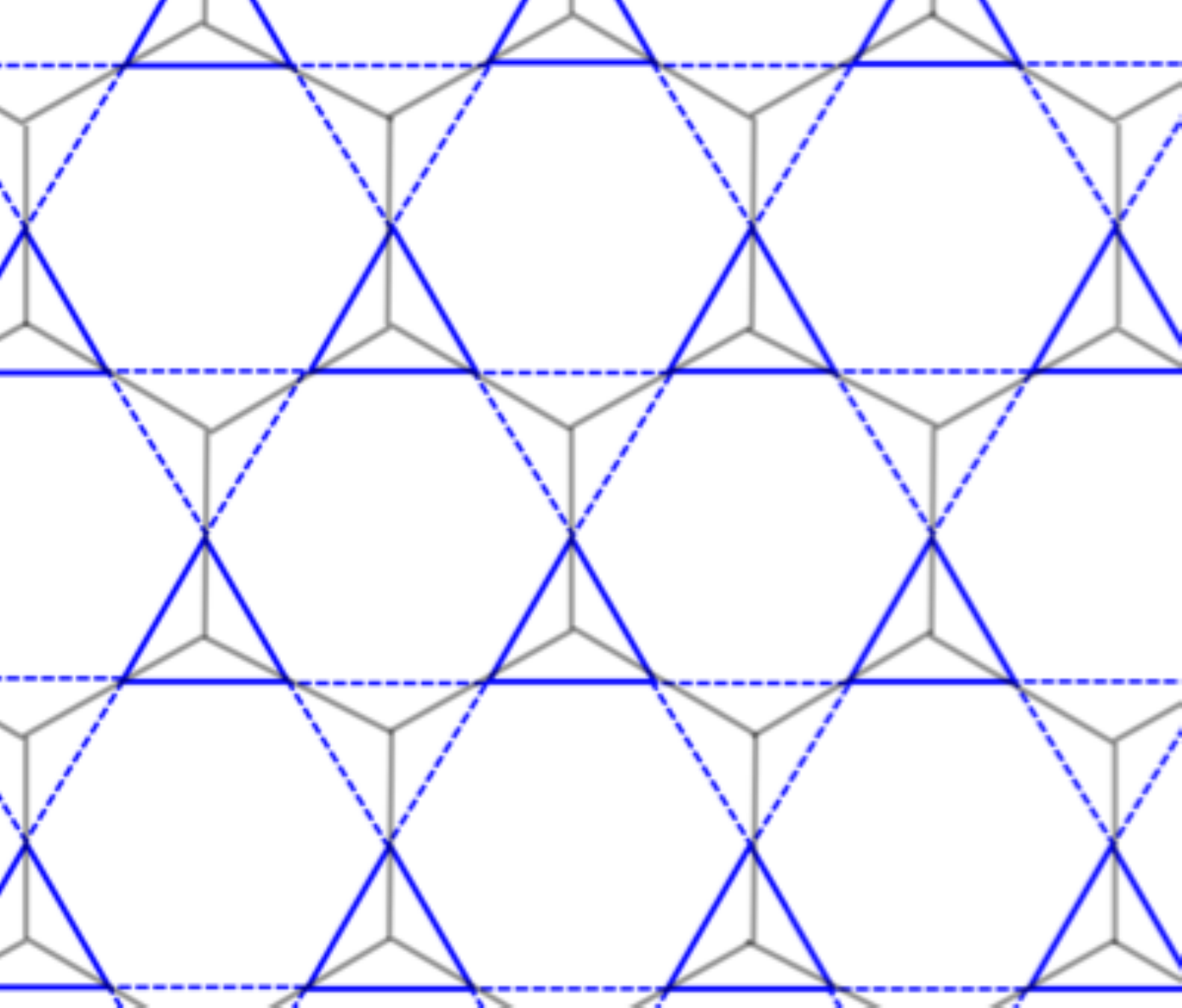}}
\caption{(Color online.) (a) { 
${\bf e}_{\text A}, {\bf e}_{\text B}$ and ${\bf e}_{\text C}$} are three vectors that 
connect the center of an up-triangle to the centers of the neighboring down-triangles. 
(b) The centers of the triangles on the Kagome lattice form a DHL. 
}
\label{fig4}
\end{figure}
%-----------------------------

\begin{figure}[ht]
{\includegraphics[width=7cm]{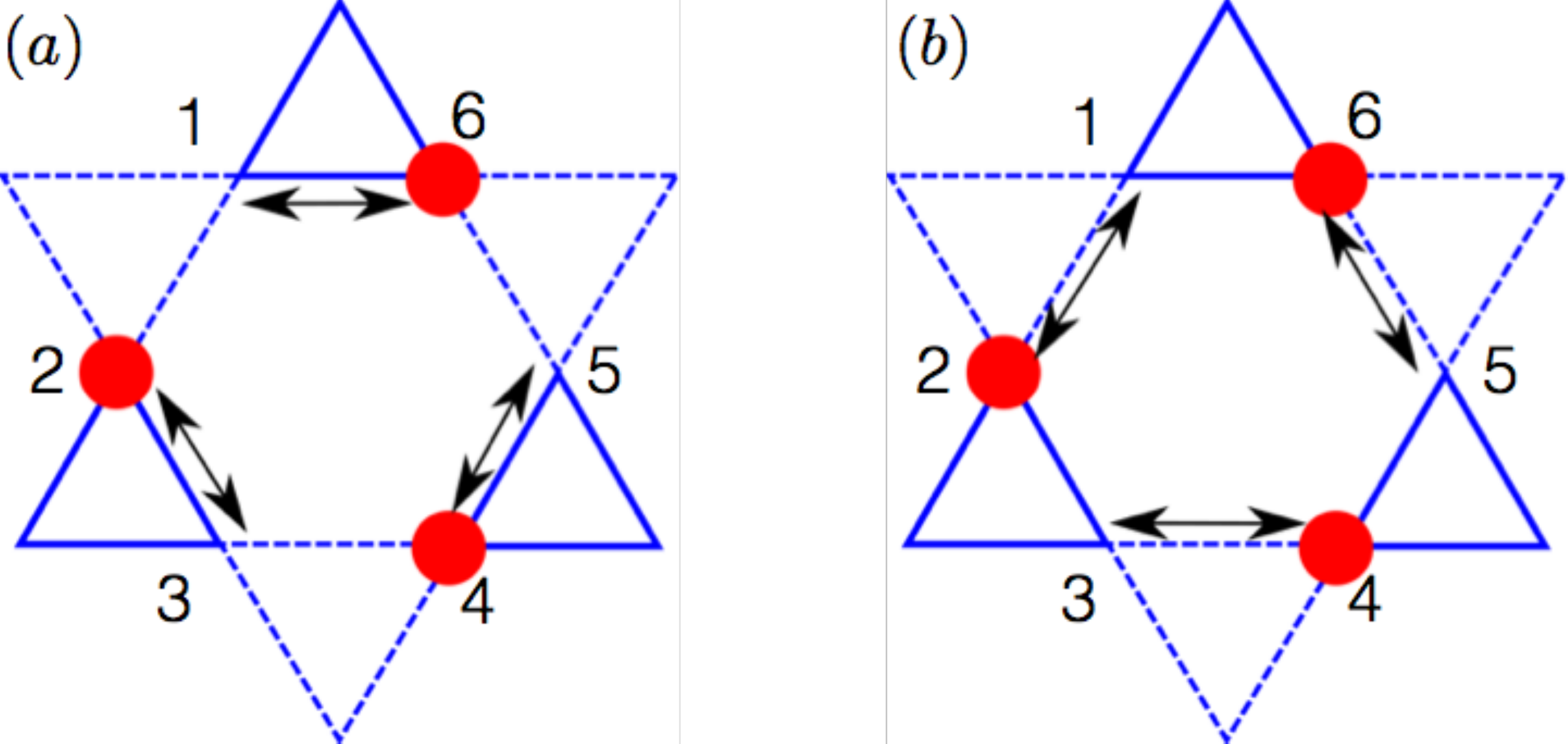}}
\caption{(Color online.) 
The two collective hopping processes that contribute to the 
ring electron hopping or the ring exchange in Eq.~\eqref{Hring}. 
The (red) solid ball represents the electron or the charge rotor. 
}
\label{fig5}
\end{figure}

The type-II CMI appears when the interactions $V_1, V_2$ are dominant 
over the hoppings $t_1, t_2$. In terms of the effective spin $L_i^z$, the 
electron charge localization condition 
in the type-II CMI is
\begin{eqnarray}
\sum_{  i \in \text{u}} L^z_i = - \frac{1}{2},\quad 
\sum_{  i \in \text{d}} L^z_i = - \frac{1}{2}. 
\label{eq29}
\end{eqnarray}
In the type-II CMI, the allowed effective spin configuration is ``2-down 1-up'' in every triangle. 
These allowed classical spin configuration are extensively degenerate. 
The presence of the transverse effective spin exchanges lifts the classical ground state 
degeneracy and the effective interaction can be obtained from  
a third-order degenerate perturbation theory. 
The resulting effective ring exchange Hamiltonian is given as 
\begin{equation}
H_{\text{ch,ring}} = - \sum_{\hexagon } J_{\text{ring}} 
 (L^+_1 L^-_2 L^+_3 L^-_4 L^+_5 L^-_6 + h.c. ),
\label{Hring}
\end{equation}
where ``$\hexagon$'' refers to the elementary hexagon of the 
Kagome lattice,  
$J_{\text{ring}} = 
\frac{6 (J_1^{\text{eff}})^3 }{ V_2^2 } + \frac{6 (J_2^{\text{eff}} )^3 }{ V_1^2}
$ and ``1,2,3,4,5,6'' are the 6 vertices on the perimeter of the elementary hexagon 
on the Kagome lattice (see Fig.~\ref{fig5}). 

%-----------------------------
%-----------------------------

We now map the effective Hamiltonian $H_{\text{ch,ring}}$ into 
a compact U(1) lattice gauge theory on the DHL.  
We introduce the lattice U(1) gauge fields ($E, A$) by 
defining\cite{Hermele04}
\begin{eqnarray}
L^z_{{\bf r},\mu} & \equiv & {  L^z_{{\bf r} + \frac{{\bf e}_{\mu}}{2}}  }= E_{ {\bf r}, {\bf r} + {\bf e}_{\mu} }, \\
\quad L^{\pm}_{{\bf r},\mu} &\equiv& L^{\pm}_{{\bf r} + \frac{{\bf e}_{\mu}}{2}}= e^{ \pm i A_{ {\bf r},{\bf r} +{\bf e}_{\mu} }}
\label{Gauge2}
\end{eqnarray}
where ${\bf r} \in \text{u}$, $E_{{\bf r}{\bf r}\rq{}} = - E_{ {\bf r}\rq{} {\bf r}}$, and $
A_{ {\bf r}{\bf r}\rq{} } = - A_{{\bf r}\rq{}{\bf r}}$. 
The centers (labelled as ${\bf r},{\bf r}\rq{}$) 
of the triangles form a dual honeycomb lattice (see Fig.~\ref{fig4}). 
The fields $E$ and $A$ are identified as the electric field and the vector gauge field of 
the compact U(1) lattice gauge theory and 
$ [  E_{ {\bf r}, {\bf r} + {\bf e}_{\mu} },  A_{ {\bf r}, {\bf r} + {\bf e}_{\mu} } ] = - i . 
$ With this identification, the local ``2-down 1-up'' 
charge localization condition in Eq.~\eqref{eq29} 
is interpreted as the ``Gauss\rq{} law\rq{}\rq{} 
for the emergent U(1) lattice gauge theory.
The effective ring exchange Hamiltonian $H_{\text{ch,ring}}$
reduces to a gauge ``magnetic'' field term on the DHL,
\begin{eqnarray}
H_{\text{ch,ring}} = -2J_{\text{ring}}
\sum_{\varhexagon} \cos ( \Delta \times A ),
\label{eq35}
\end{eqnarray}
where $\Delta \times A$ is a lattice curl defined on 
the `$\varhexagon$' that refers to the elementary hexagon 
on the honeycomb lattice.
As this internal gauge structure emerges at low energies
in the charge sector, in the following we will refer this gauge 
field as the U(1)$_{\text{ch}}$ gauge field.  

\subsection{Slave-particle construction and mean-field theory}
\label{sec3B} 

Since the gauge theory in the charge sector 
is a compact U(1) gauge theory defined on a 2D lattice, 
it would be confining due to the well-known non-perturbative instanton effect
if all the elementary excitations (except for ``photon") is gapped.  
However, in our case, the spinon excitations are gapless and possess spinon Fermi surfaces.
While these spinons do not directly couple to U(1)$_{\rm ch}$ gauge field, they would
interact with charge excitations in terms of U(1)$_{\rm sp}$ gauge field and then can indirectly
couple to U(1)$_{\rm ch}$ gauge field via the charge excitations.  
Thus, a deconfined phase of the U(1)$_{\text{ch}}$ gauge field may still be allowed if spinon Fermi
surface fluctuations can suppress instanton events, which would then support fractionalized charge excitations 
inside the type-II CMI.  Resolving this issue requires non-perturbative computations and is left for a future work. 
In the following, we introduce a new parton formulation  
and obtain the mean-field phase diagram for the extended
Hubbard model. In this subsection, we shall first ignore the instanton effect (and the related
charge sector symmetry breaking) which will become important 
in the type-II CMI when we consider quantum fluctuations beyond the mean-field theory. 

\subsubsection{Generalized parton construction}
\label{sec3B1} 

Before introducing the new parton formalism, we would like to explain the connection and 
the difference between the current problem and the fractional charge liquid (FCL) Mott insulating phase
 in our previous work for a
3D pyrochlore lattice Hubbard model with a 1/4 or 1/8 electron filling.\cite{Chen14} 
Similar to the type-II CMI in the Kagome lattice case discussed here, 
the low-energy physics of the charge sector in the FCL 
is described by a compact lattice U(1) (or U(1)$_{\text{ch}}$) 
gauge theory on a 3D diamond lattice. 
Because it is defined in 3D, the U(1)$_{\text{ch}}$ gauge field for the 
pyrochlore lattice case can easily be deconfined in the Mott insulating phase,
which supports the charge quantum number fractionalization
in the FCL. Therefore, in the absence of instanton effect, we can use the same
construction here and represent the electron creation operator as 
\begin{equation}
c^{\dagger}_{ {\bf r}\mu\sigma} 
{ \equiv c^{\dagger}_{ {\bf r}+\frac{{\bf e}_\mu}{2},\sigma}}
= f^\dagger_{ {\bf r}\mu\sigma }
\bar{\Phi}^{\dagger}_{\bf r} \bar{\Phi}^{\phantom\dagger}_{{\bf r} + {\bf e}_{\mu}}
l^+_{ {\bf r},{\bf r}+{\bf e}_{\mu}},
\label{SlaveP}
\end{equation}
where ${\bf r}\in \text{u}$, $f^\dagger_{ {\bf r}\mu\sigma} \equiv f^{\dagger}_{ {\bf r}+{{\bf e}_\mu}/{2},\sigma}$ is
the same fermionic spinon creation operator in the slave-rotor representation, 
$\bar{\Phi}_{\bf r}^{\dagger}$ ($\bar{\Phi}_{\bf r}^{\phantom\dagger}$)
is the creation (annihilation) operator for the bosonic charge excitation 
in the triangle that is centered at ${\bf r}$, and 
$l^+_{ {\bf r}, {\bf r} + {\bf e}_{\mu} }$ is an open string operator 
of the U(1)$_{\text{ch}}$ gauge field that connects the two charge 
excitations in the neighboring triangles at ${\bf r}$ and 
${\bf r} + \hat{\bf e}_{\mu}$. In the following, we use 
the string or the U(1)$_{\text{ch}}$ field interchangeably. 
This parton representation for the electron operator is connected 
to the slave-rotor representation by identifying\cite{Sungbin12,Savary12}
$
 L^+_{{\bf r},\mu} =  
\bar{\Phi}^{\dagger}_{\bf r} \bar{\Phi}^{\phantom\dagger}_{{\bf r} + {\bf e}_{\mu}}
 l^+_{ {\bf r},{\bf r}+{\bf e}_{\mu}},
l^{\pm}_{{\bf r},{\bf r}+{\bf e}_{\mu}} = 
 | l^{\pm}_{{\bf r},{\bf r}+{\bf e}_{\mu}}  | 
e^{\pm i A_{{\bf r},{\bf r}+{\bf e}_{\mu}} },$
for ${\bf r} \in \text{u}$. 
The original Hilbert space constraint in the slave-rotor representation is also needed here. 
To match with the underlying lattice U(1)$_{\text{ch}}$ 
gauge theory description, we define the following 
operator, \cite{Sungbin12,Savary12}
\begin{eqnarray}
Q_{\bf r} &=&\frac{\eta_{\bf r}}{2}+  \eta_{\bf r}  \sum_{\mu} 
L^z_{  {\bf r},{\bf r} + \eta_{\bf r} {\bf e}_{\mu}  }
\equiv 
\frac{\eta_{\bf r}}{2} + 
\eta_{\bf r}  \sum_{\mu} 
l_{  {\bf r},{\bf r} + \eta_{\bf r} {\bf e}_{\mu}  }
\end{eqnarray}
that measures the local U(1)$_{\text{ch}}$ (electric) gauge charge. 
Here, $\eta_{\bf r} = +1$ ($-1$) for ${\bf r} \in \text{u}$ (${\bf r} \in \text{d}$)
and $  l_{  {\bf r},{\bf r} + \eta_{\bf r} {\bf e}_{\mu}  } =  L^z_{  {\bf r},{\bf r} 
+ \eta_{\bf r} {\bf e}_{\mu}}$. 
We further supplement this definition with a Hilbert space constraint\cite{Sungbin12,Savary12} 
\begin{eqnarray}
&&[\bar{\Phi}^{\phantom\dagger}_{\bf r}, Q_{\bf r} ] = \bar{\Phi}^{\phantom\dagger}_{\bf r},
\quad [\bar{\Phi}^{\dagger}_{\bf r},  Q_{\bf r} ] =    - \bar{\Phi}^{\dagger}_{\bf r},
\end{eqnarray}
such that the new representation of the electron operator is restricted to the 
physical Hilbert space. For the ground state, we have $Q_{\bf r} = 0$ 
in every triangle.  An effective spin-flip operator or the local charge excitation 
$L^{+}_{{\bf r},\mu}$ (with ${\bf r} \in \text{u}$) 
creates two neighboring defect triangles (at ${\bf r}$ and ${\bf r}+{\bf e}_{\mu}$) 
that violate the ``2-down 1-up'' charge localization condition 
of the type-II CMI. 
Thus, these two defect triangles carry the U(1)$_{\text{ch}}$ gauge charges
$Q_{\bf r} = +1$ and $Q_{{\bf r}+{\bf e}_{\mu}} = -1$,
which are assigned to the operators $\bar{\Phi}_{\bf r}^{\dagger}$ and 
$\bar{\Phi}_{{\bf r}+{\bf e}_{\mu}}$, respectively.  
Under an internal U(1)$_{\text{ch}}$ gauge transformation, 
$\bar{\Phi}^{\dagger}_{\bf r}  \rightarrow \bar{\Phi}^{\dagger}_{\bf r}\, e^{i \chi_{\bf r}}, 
\bar{\Phi}_{{\bf r} + {\bf e}_{\mu}}  \rightarrow \bar{\Phi}_{{\bf r} + {\bf e}_{\mu}} 
 e^{- i \chi_{{\bf r} + {\bf e}_{\mu}}}$, and $
 A_{{\bf r},{\bf r} + {\bf e}_{\mu}}  \rightarrow 
  A_{{\bf r},{\bf r} + {\bf e}_{\mu}} 
  e^{-i \chi_{\bf r} + i \chi_{{\bf r} + {\bf e}_{\mu}}}$.

\subsubsection{Type-I \& Type-II CMIs and mean-field phase diagram}
\label{sec3B2}

Using the new parton construction, 
the microscopic Hubbard model becomes
\begin{eqnarray}
H &=&
 -t_1 
\sum_{{\bf r} \in \text{u}}
 \sum_{\mu\neq \nu}
l^+_{{\bf r}, {\bf r} + {\bf e}_{\mu} }  l^-_{{\bf r}, {\bf r} + {\bf e}_{\nu}} 
f^{\dagger}_{ {\bf r}\mu\sigma } f^{\phantom\dagger}_{ {\bf r}\nu\sigma} 
\bar{\Phi}^{\dagger}_{ {\bf r} + {\bf e}_{\mu} } 
\bar{\Phi}^{\phantom\dagger}_{ {\bf r} + {\bf e}_{\nu} }
\nonumber \\
&&
-t_2 
\sum_{{\bf r} \in \text{d}} 
\sum_{\mu\neq \nu}
l^+_{ {\bf r} - {\bf e}_{\mu},{\bf r} }  l^-_{ {\bf r} - {\bf e}_{\nu}, {\bf r}} 
f^{\dagger}_{ {\bf r}\mu\sigma } f^{\phantom\dagger}_{ {\bf r}\nu\sigma} 
\bar{\Phi}^{\dagger}_{ {\bf r} - {\bf e}_{\mu} } 
\bar{\Phi}^{\phantom\dagger}_{ {\bf r} - {\bf e}_{\nu} }
\nonumber \\
&& + \frac{V_1}{2} \sum_{{\bf r} \in \text{u} } Q_{\bf r}^2 
+ \frac{V_2}{2} \sum_{{\bf r} \in \text{d} } Q_{\bf r}^2,
\end{eqnarray}
which is supplemented with the Hilbert space constraint. 
In a mean-field theory treatment, we 
decouple the electron kinetic terms and obtain four mean-field Hamiltonians
for each sector, 
\begin{eqnarray}
H_{\text{ch}}^{\text u} &=& -\bar{J}_1 
\sum_{{\bf r}\in \text{d} }
\sum_{ \mu \neq \nu}
\bar{\Phi}^{\dagger}_{{\bf r} - {\bf e}_{\mu}} 
\bar{\Phi}^{\phantom\dagger}_{{\bf r}-{\bf e}_{\nu} } 
+ \frac{V_1}{2} \sum_{{\bf r} \in \text{u}} Q_{\bf r}^2,
\\
H_{\text{ch}}^{\text d} &=& -\bar{J}_2 
\sum_{{\bf r}\in \text{u} }
\sum_{ \mu \neq \nu}
\bar{\Phi}^{\dagger}_{{\bf r} + {\bf e}_{\mu}} 
\bar{\Phi}^{\phantom\dagger}_{{\bf r}+{\bf e}_{\nu}} 
+ \frac{V_2}{2} \sum_{{\bf r} \in \text{d}} Q_{\bf r}^2,
\\
H_{\text{sp}} &=& - 
\sum_{\mu \neq \nu} [
\bar{t}_1 \sum_{{\bf r}\in \text{u} }
 f^{\dagger}_{{\bf r}\mu\sigma }
 f^{\phantom\dagger}_{{\bf r}\nu\sigma } +\bar{t}_2 
 \sum_{{\bf r}\in \text{d} } 
 f^{\dagger}_{{\bf r}\mu\sigma }
 f^{\phantom\dagger}_{{\bf r}\nu\sigma } ],
 \\
 H_{A} &=& - \sum_{ \mu \neq \nu} \big[\bar{K}_1 \sum_{{\bf r} \in \text{u} } 
 l^+_{{\bf r}, {\bf r} + {\bf e}_{\mu} }  l^-_{{\bf r}, {\bf r} + {\bf e}_{\nu}} 
 \nonumber \\
&&\quad\quad\,\, + \bar{K}_2  \sum_{{\bf r} \in \text{d} } 
 l^+_{ {\bf r} - {\bf e}_{\mu},{\bf r} }  l^-_{ {\bf r} - {\bf e}_{\nu}, {\bf r}} \big],
\end{eqnarray}
where 
\begin{eqnarray}
\bar{J}_1 &=& t_2 
\langle l^+_{ {\bf r} - {\bf e}_{\mu},{\bf r} } \rangle
\langle l^-_{ {\bf r} - {\bf e}_{\nu}, {\bf r} }  \rangle
\sum_{\sigma}
\langle f^{\dagger}_{ {\bf r}\mu\sigma } f^{\phantom\dagger}_{ {\bf r}\nu\sigma } \rangle,
\quad {\bf r} \in \text{d} ,
\nonumber
\\
\bar{J}_2 &=& 
t_1
\langle l^+_{ {\bf r}, {\bf r}+{\bf e}_{\mu} } \rangle 
\langle l^-_{  {\bf r}, {\bf r}+{\bf e}_{\nu} } \rangle
\sum_{\sigma}
\langle f^{\dagger}_{ {\bf r}\mu\sigma } f^{\phantom\dagger}_{ {\bf r}\nu\sigma } \rangle,
 \quad {\bf r} \in \text{u} ,\nonumber
\\
\bar{t}_1 &=& 
t_1 \langle l^+_{ {\bf r}, {\bf r}+{\bf e}_{\mu} } \rangle  
 \langle l^-_{ {\bf r}, {\bf r}+{\bf e}_{\nu} } \rangle  
\langle  \bar{\Phi}^{\dagger}_{{\bf r} + {\bf e}_{\mu}} 
\bar\Phi^{\phantom\dagger}_{{\bf r}+{\bf e}_{\nu}} 
\rangle,
\quad {\bf r} \in \text{u},\nonumber
\\
\bar{t}_2 &=&
t_2
\langle l^+_{ {\bf r} - {\bf e}_{\mu},{\bf r} } \rangle
\langle l^-_{ {\bf r} - {\bf e}_{\nu}, {\bf r} }  \rangle
\langle  \bar\Phi^{\dagger}_{{\bf r} - {\bf e}_{\mu}} 
\bar\Phi^{\phantom\dagger}_{{\bf r} - {\bf e}_{\nu}} 
\rangle,
\quad {\bf r} \in \text{d}, \nonumber
\\
\bar{K}_1 &=& t_1
\sum_{\sigma}
\langle f^{\dagger}_{ {\bf r}\mu\sigma } 
f^{\phantom\dagger}_{ {\bf r}\nu\sigma } \rangle
\langle  \bar\Phi^{\dagger}_{{\bf r} + {\bf e}_{\mu}} 
\bar\Phi^{\phantom\dagger}_{{\bf r}+{\bf e}_{\nu}} 
\rangle,\quad\quad {\bf r} \in \text{u}, \nonumber
\\ 
\bar{K}_2 &=&
 t_2
\sum_{\sigma}
\langle f^{\dagger}_{ {\bf r}\mu\sigma } 
f^{\phantom\dagger}_{ {\bf r}\nu\sigma } \rangle
\langle  \bar\Phi^{\dagger}_{{\bf r} - {\bf e}_{\mu}} 
\bar\Phi^{\phantom\dagger}_{{\bf r}-{\bf e}_{\nu}} 
\rangle,\quad\quad {\bf r} \in \text{d}. \nonumber
\end{eqnarray}
Here we explain a few things related to the above mean-field equations.  
First, in the above choices of mean-field couplings, we have assumed that these
couplings respect all the symmetries of the original Hubbard model on the Kagome lattice. 
Second, the Lagrange multipliers, which are used to fix the Hilbert space constraints, are 
expected to vanish from the same argument as noted in Appendix.~\ref{secapp1}, 
so we do not explicitly write them out in the mean-field Hamiltonians. 
Third, the electron hoppings on the bonds of down-triangles (up-triangles) mediate the 
tunnelling of the charge bosons in the up-triangles (down-triangles). Therefore,
the charge bosons on the up-triangles and down-triangles do not mix, and hence, we have 
two separate charge boson mean-field Hamiltonians $H^{\text u}_{\text{ch}}$
and $H^{\text d}_{\text{ch}}$ that are defined in the up-triangle and 
down-triangle subsystems, respectively.

%-----------------------------
\begin{figure}[t]
\subfigure[\, $t_1 = t_2$]{\includegraphics[width=6cm]{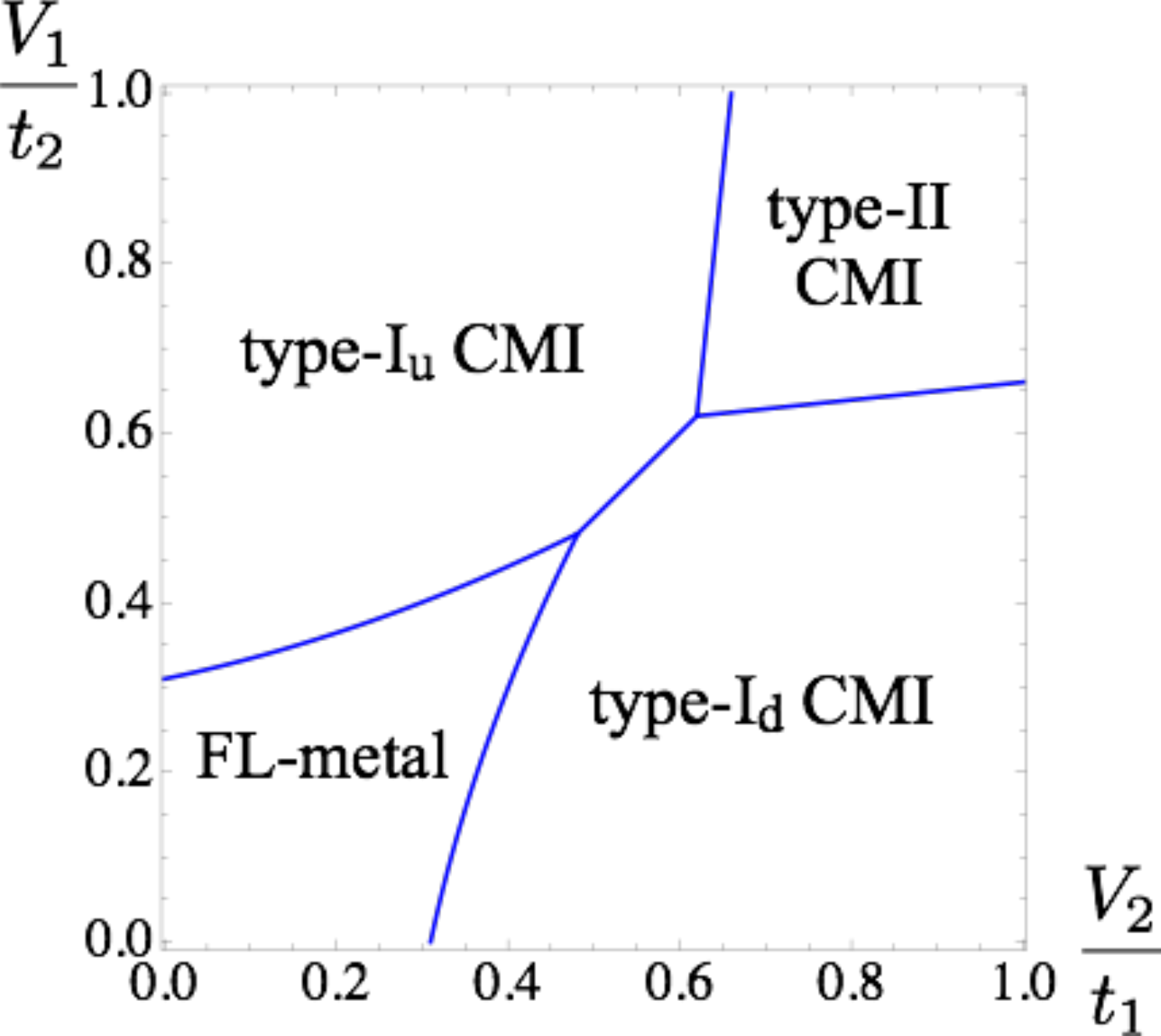}}
\subfigure[\, $t_1 =2 t_2$ ]{\includegraphics[width=6cm]{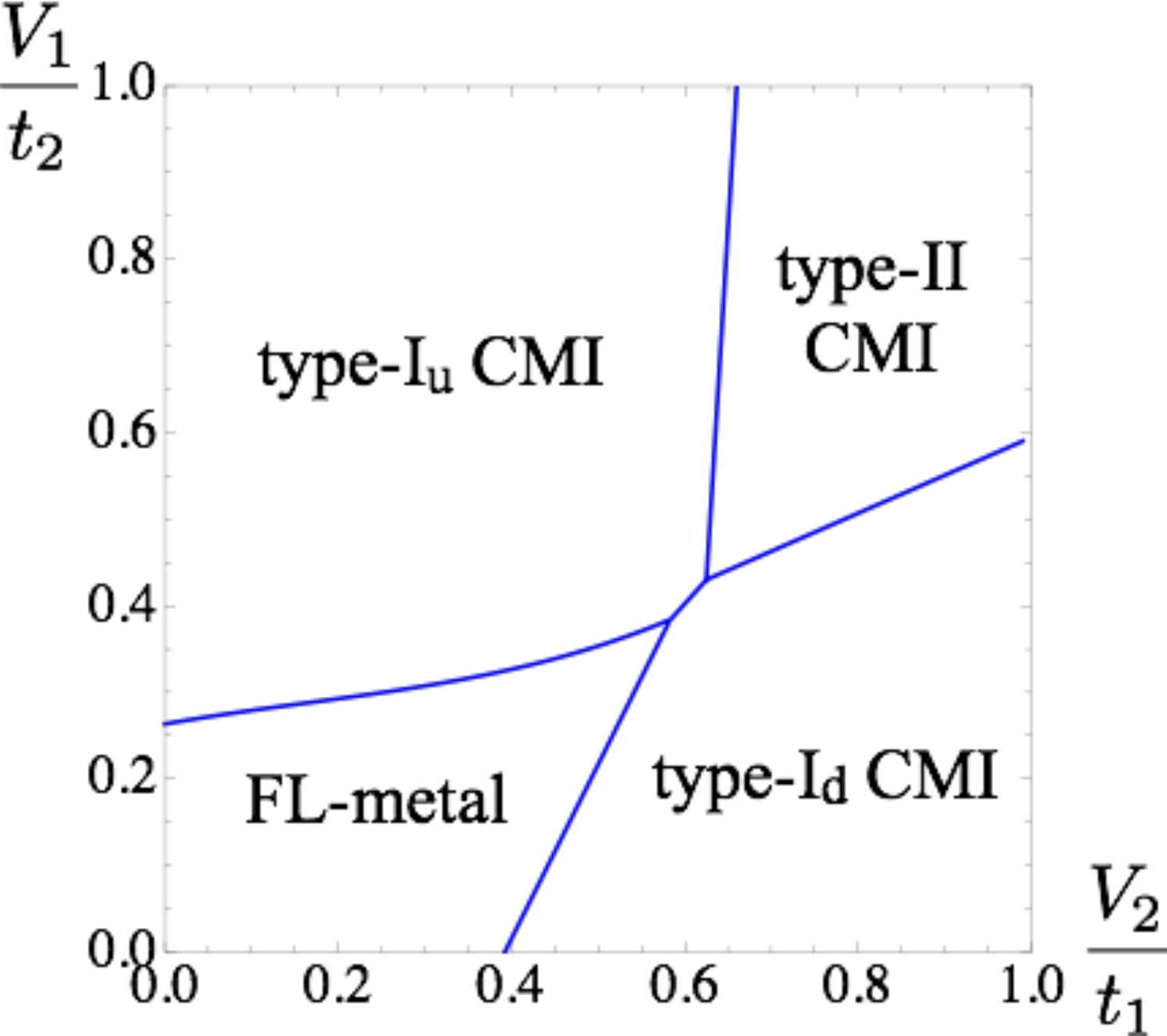}}
\caption{(Color online.) The phase diagram of the extended Hubbard model for different choices of 
parameters. }
\label{fig7}
\end{figure}
%-----------------------------
\begin{table}[t]
\centering
\begin{tabular}{ll}
\hline\hline 
Type-II CMI \quad \quad & $\langle \Phi_{\bf r} \rangle = 0$ for ${\bf r} \in \text{u,\;d}$.
\vspace{1mm}
\\
Type-I$_{\text u}$ CMI \quad \quad   & $\langle \Phi_{\bf r} \rangle = 0$ for ${\bf r} \in \text{u}$,  $\langle \Phi_{\bf r} \rangle \neq 0$ for ${\bf r} \in \text{d}$.
\vspace{1mm}
\\
Type-I$_{\text d}$ CMI \quad \quad   & $\langle \Phi_{\bf r} \rangle \neq 0$ for ${\bf r} \in \text{u}$,  $\langle \Phi_{\bf r} \rangle = 0$ for ${\bf r} \in \text{d}$.
\vspace{1mm}
\\
FL metal &  $\langle \Phi_{\bf r} \rangle \neq 0$ for ${\bf r} \in \text{u}$,  $\langle \Phi_{\bf r} \rangle \neq 0$ for ${\bf r} \in \text{d}$.
\vspace{1mm}
\\
\hline\hline
\end{tabular}
\caption{The description of the charge sector of the four different phases 
in the slave-particle formalism.}
\label{Tab1}
\end{table}

The string (or U(1)$_{\text{ch}}$ gauge-field) sector mean-field Hamiltonian is a simple 
effective spin-1/2 ferromagnetic XY model on the Kagome lattice, which is solved classically. 
The low-energy charge sector ring hopping Hamiltonian in Eq.~\eqref{eq35} 
favors a zero net U(1)$_{\text{ch}}$ gauge flux, which we choose 
for the U(1)$_{\text{ch}}$ gauge flux seen by the charge bosons in the mean-field theory. 
Therefore, the mean-field U(1)$_{\text{ch}}$ gauge flux is obtained 
based on a non-mean-field perturbative argument.  
This treatment is used  
to obtain the phase of the quantum spin ice in the 3D pyrochlore lattice
in Ref.~\onlinecite{Sungbin12}.  
Of course, this treatment misses the phases in other flux sectors which
may potentially be stabilized by other larger couplings. 
To resolve this issue, one needs much more
exhaustive analysis of the energetics of different flux sectors 
which is beyond the scope of this work. For the phases in which 
we are interested, we expect such a treatment is sufficient. 
In the following, we fix the gauge for the mean-field value of 
the string operator by requiring
$ \langle l^\pm_{ {\bf r}{\bf r}' } \rangle  = \Delta,
$ where $\Delta$ is a real parameter. In the semi-classical approximation of the string
mean-field Hamiltonian $H_{A}$, we can set $\Delta =1/2$. 

We solve the mean-field Hamiltonians self-consistently
(see Appendix.~\ref{secapp2B}). 
The mean-field phase diagrams for different choices of the couplings
are depicted in Fig.~\ref{fig7}. The four different phases correspond to 
different behaviors of the charge bosons that we list in Tab.~\ref{Tab1}. 
When the charge bosons from both up-triangles and down-triangle subsystems are condensed,
the FL-metal phase is realized. 
When they are both gapped and uncondensed, we have the type-II CMI. 
When the charge bosons from one triangle subsystem are condensed and the other
is uncondensed, we have the type-I CMI. 
Here we introduce a subindex `u' or `d' to the type-I CMI to indicate
which triangles the electrons are localized in. 
All the transitions are continuous in the mean-field theory, 
except the transition between the type-I$_{\text u}$ and 
the type-I$_{\text{d}}$ CMIs which is strongly first order. 
Beyond the mean-field theory, the transition between 
the FL-metal and the type-I CMIs will probably remain
continuous and quantum XY type\cite{Senthil08July1,Senthil08July2} 
while the transition into the type-II CMI may depend on the 
detailed charge structure inside the type-II CMI. 

Now we explain the phase boundaries.
We begin with the phase boundary between the type-I$_{\text u}$ CMI and 
the FL-metal. As we increase $V_2/t_1$, the effective electron hopping 
on the up-triangle bonds gets suppressed
which effectively enhances the kinetic energy gain through the down-triangle bonds. 
As a result, a larger $V_1/t_2$ is required to drive
a Mott transition. A similar argument applies to the phase boundary 
between the type-I$_{\text u}$ and the type-II CMIs. 
A larger $V_2/t_1$ is needed to compete with the kinetic energy gain 
on the upper triangle bonds for a larger $V_1/t_2$ in the type-I$_{\text u}$ 
CMI and to drive a transition to the type-II CMI. 
For $t_1 > t_2$, electrons are more likely to be localized in the up-triangles 
to gain the intra-cluster kinetic energy. Thus, a smaller $V_1/t_2$ is needed to 
drive a Mott transition and a larger $V_2$ is needed to drive the system 
from the type-I$_{\text u}$ to the type-II CMIs. 
 
In the type-I CMI, the spin sector forms a U(1) QSL with a spinon Fermi surface,
which is the same as the U(1) QSL in Appendix.~\ref{secapp1} with $V_2 =0$. 
The U(1)$_{\text{sp}}$ gauge field also behaves similarly.
To be concrete, we first discuss the type-I$_{\text u}$ CMI. 
The U(1)$_{\text{sp}}$ gauge field acquires a mass when the condensation of the charge bosons
in down-triangles occurs. The two fractionally-charged charge bosons $\Phi$ 
then are combined back into the original unit-charged charge rotor $e^{i \theta}$. 
While the charge fractionalization does not exist in the type-I CMI, 
the spin-charge separation still survives. 
The condensation of the charge bosons from the down-triangles leads to the local 
``metallic'' clusters in the up-triangles such that the localized electron can move 
more or less freely within each up-triangle. 
Because of the local ``metallic'' clusters, only the U(1)$_{\text{sp}}$ gauge field living
on the down-triangle bonds that connect the up-triangles 
remains active and continues to fluctuate at the low energies. 
The low-energy physics is described by the 
spinon Fermi surface coupled with a fluctuating U(1)$_{\text{sp}}$ gauge field,
 leading to a U(1) QSL in the triangular lattice formed by the up-triangles. 

In the type-II CMI, the charges remain strongly fluctuating
which is described by the compact U(1)$_{\text{ch}}$ gauge theory in Sec.~\ref{sec3A}. 
The mean-field theory, however, does not capture this charge quantum fluctuation 
inside the type-II CMI and thus cannot not give a reliable prediction 
for the behaviors of charge excitations and spinons. We will discuss this issue in Sec.~\ref{sec3C}. 
Nevertheless, the mean-field theory does  
obtain qualitatively correct phase boundaries.
 
\subsection{Type-II CMI with the plaquette charge order}
\label{sec3C}

In this subsection, we focus on the type-II CMI and discuss the PCO due to the leading
quantum fluctuation in the charge sector. 
Then we explain how the spin physics is influenced by the charge
sector in the type-II CMI. 

\subsubsection{The plaquette charge order via quantum dimer model}
\label{sec3C1}

\begin{figure}[t]
{\includegraphics[width=8.5cm]{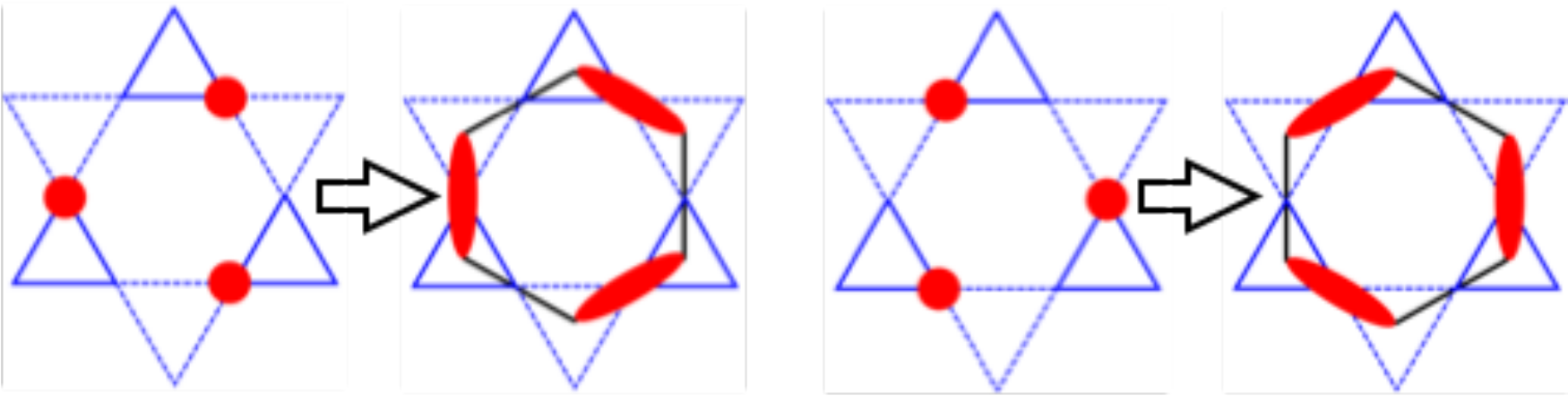}}
\caption{(Color online.) The electron charge configurations and the corresponding dimer coverings
on the DHL. }
\label{fig8}
\end{figure}

\begin{figure}[t]
{\includegraphics[width=6.5cm]{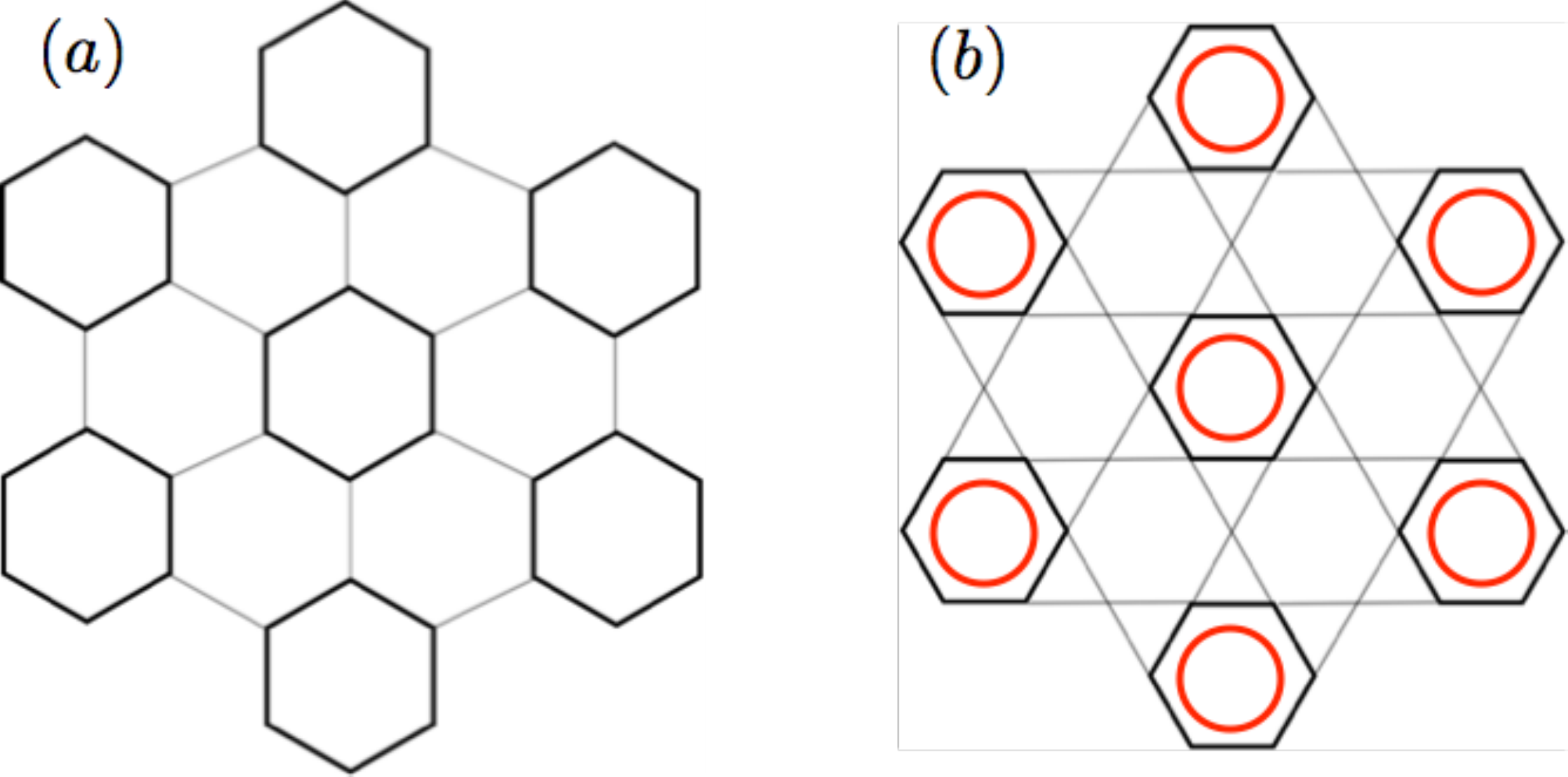}}
\caption{
($a$) The plaquette dimer ordering pattern on the DHL. The dark hexagons
have a higher probability to be occupied by the dimer. 
($b$) The corresponding PCO on the Kagome lattice. 
We mark the resonating hexagons with both dark bonds and the red circles.   
}
\label{fig9}
\end{figure}

As we point out in Sec.~\ref{sec3A}, the extensive degeneracy of the 
classical charge ground state of the type-II CMI is lifted by
the ring hopping Hamiltonian $H_{\text{ch,ring}}$ 
in Eq.~\eqref{Hring}, which arises from
the collective tunnelling of the 3 electrons on the perimeter of the
elementary hexagons on the Kagome lattice. 
This ring hopping Hamiltonian $H_{\text{ch,ring}}$ 
is the {\it dominant interaction} in the charge sector of the type-II CMI
and should be treated first. 
We now map $H_{\text{ch,ring}}$ into a quantum dimer model on the DHL. 
As depicted in Fig.~\ref{fig8}, a dimer is placed on the corresponding 
link of the DHL if the center of the link (or the Kagome lattice site) is occupied 
by an electron charge. The rotor operator $L_i^{\pm}$ simply adds or removes 
the dimer that is centered at the Kagome lattice site $i$. 
So $H_{\text{ch,ring}}$ is mapped into the quantum dimer model
with only a resonant term,
\begin{equation}
H_{\text{ch,ring}} = - \sum_{ \varhexagon }
J_{\text{ring}} 
( | \varhexagon_{\text A}\rangle \langle \varhexagon_{\text B} | + 
| \varhexagon_{\text B}\rangle \langle \varhexagon_{\text A} | )
\label{eq64}
\end{equation}
where $| \varhexagon_{\text A} \rangle$ and $| \varhexagon_{\text B} \rangle$
refer to the two dimer covering configurations in the elementary hexagon $\varhexagon$
of the DHL as shown in Fig.~\ref{fig8}. 
In Ref.~\onlinecite{Moessner01}, Moessner, Sondhi and Chandra have studied 
the phase diagram of the quantum dimer model on the honeycomb lattice quite extensively. 
In the case with only the resonant term, they found a 
translational symmetry breaking phase with a plaquette dimer order, 
in which the system preferentially gains dimer resonating (or kinetic) energy 
through the resonating hexagons on the DHL (see Fig.~\ref{fig9}). 
The plaquette dimer order of the quantum dimer model is then mapped back to 
the plaquette charge order (PCO) on the Kagome lattice (see Fig.~\ref{fig9}). 
This is a quantum mechanical effect and cannot be obtained from treating the 
inter-site electron interactions in a classical fashion. 

With the PCO, the electrons are preferentially hopping around the 
perimeters of the resonating hexagons on the Kagome lattice. 
These resonating hexagons are periodically arranged, forming an emergent
triangular lattice (ETL). Due to the translational symmetry breaking, 
this ETL has a larger unit cell than the original Kagome lattice.   

%-----------------------------
%-----------------------------
%
We note this PCO has already been obtained in the same Hubbard model 
in an isotropic Kagome lattice with 1/6, 1/3 and 2/3 electron filling 
in certain parameter regimes in previous works.\cite{Pollmann08,Fiete11,Pollmann14,Ferhat14} 
The result was obtained either through perturbatively mapping to the 
quantum dimer model or by a Hatree-Fock mean-field calculation. 
In particular, Ref.~\onlinecite{Isakov08} applied the quantum Monte Carlo technique 
to simulate a hardcore boson Hubbard model on an isotropic Kagome lattice and
discovered a direct weakly-first-order Mott transition from the superfluid phase 
to the type-II CMI with the PCO for 1/3 and 2/3 boson fillings.

%-----------------------------
%-----------------------------

\subsubsection{Effective low energy theory of the type-II CMI}
\label{sec3C2}

We now turn to the spin sector physics of the type-II CMI. 
It has been shown by exact diagonalization in Ref.~\onlinecite{Pollmann08}
that the 1/6-filled extended Hubbard model in
an isotropic Kagome lattice supports a ferromagnetic state for the 
type-II CMI in the limit $V_1 = V_2 \gg t_1 = t_2$ and 
$U\rightarrow \infty$. 
The underlying physical mechanism for such a ferromagnetic state is 
similar to the Nagaoka's ferromagnetism. 
This ferromagnetic state, however, is very unstable to the introduction
of the antiferromagnetic (AFM)
spin interaction between the electron spins.   
In the actual material, AFM spin (exchange) interaction is always
present, and moreover, the experiments did not find any evidence of ferromagnetic ordering. 
So we do not consider the possibility of the ferromagnetic ordering
in the extreme limit that is considered in Ref.~\onlinecite{Pollmann08}. 

\begin{figure}[ht]
{\includegraphics[width=8.5cm]{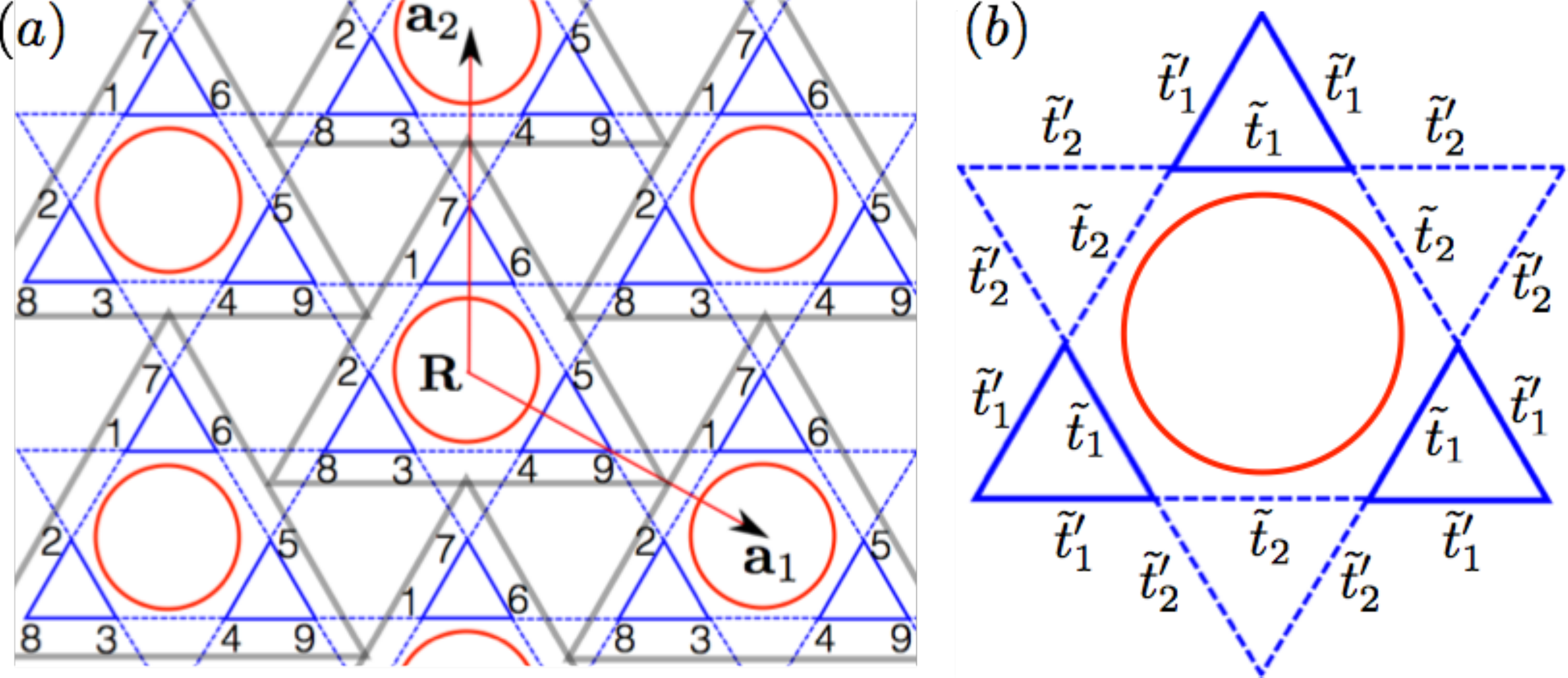}}
\caption{(a)
The Kagome lattice is partitioned into unit cells 
on the ETL.  
The unit cell (marked by a gray triangle) contains 9 sublattices
that are labelled by ``1,2,3,4,5,6,7,8,9''. 
(b) The spinon hoppings on the bonds that surround 
around a resonating hexagon. }
\label{fig10}
\end{figure}

To understand how the PCO in the charge sector influences the spin sector.
we first consider the low-energy effective ring hopping model 
in the type-II CMI that is expressed in terms of the original electron operators,
\begin{eqnarray}
H_{\text{ring}} &=& 
- \sum_{\hexagon} \sum_{\alpha \beta \gamma }
\big[
{\mathbb K}_1  (c^{\dagger}_{1\alpha} c^{\phantom\dagger}_{6\alpha} 
                                 c^{\dagger}_{5\beta} c^{\phantom\dagger}_{4\beta}  
  c^{\dagger}_{3\gamma} c^{\phantom\dagger}_{2\gamma}    + h.c.) 
\nonumber \\
&+& 
{\mathbb K}_2  (c^{\dagger}_{1\alpha} c^{\phantom\dagger}_{2\alpha} 
                                 c^{\dagger}_{3\beta} c^{\phantom\dagger}_{4\beta}  
                                  c^{\dagger}_{5\gamma} c^{\phantom\dagger}_{6\gamma}  
+ h.c.) 
\big],
\label{eq65}
\end{eqnarray}
where $ {\mathbb K}_1 = 6t_1^3/V_2^2$ and ${\mathbb K}_2 = 6t_2^3/V_1^2 $
are readily obtained from the same third-order degenerate perturbation calculation as 
in Sec.~\ref{sec3A}.  Here, $\alpha,\beta,\gamma = \uparrow,\downarrow$,
and
``1,2,3,4,5,6'' are the 6 vertices in the elementary hexagon of the Kagome lattice
and should not be confused with the sublattice labellings in Fig.~\ref{fig10}
for the ETL. Using the slave-rotor representation for the electron operator 
$c^\dagger_{i\alpha} =f_{i\alpha}^\dagger e^{i \theta_i} 
\equiv  f_{i\alpha}^\dagger L_i^+ $, the 
%electron 
ring hopping
model $H_{\text{ring}}$ can be decoupled as 
\begin{widetext}
\begin{eqnarray}
\tilde{H}_{\text{ring}}=
-  \sum_{\hexagon} \mathbb{K}_1
[ L_1^+ L_2^- L_3^+ L_4^- L_5^+ L_6^- \times
M_{165432} 
  + h.c.]
+ \mathbb{K}_2 
 [ L_1^+ L_2^- L_3^+ L_4^- L_5^+ L_6^- \times
 M_{123456}  + h.c.] \ ,
 \label{eq66a}
\end{eqnarray}
\end{widetext}
where we have singled out the charge sector and treated the 
spinon sector in a mean-field fashion and 
$ M_{ijklmn}  =
\sum_{\alpha\beta\gamma}
\langle f^{\dagger}_{i\alpha} f^{\phantom\dagger}_{j\alpha} 
  f^{\dagger}_{k\beta} f^{\phantom\dagger}_{l\beta}  
  f^{\dagger}_{m\gamma} f^{\phantom\dagger}_{n\gamma}  \rangle,$
where the lattice sites $i,j,k,l,m,n$ are arranged either clockwise 
or anti-clockwise.
Here $M_{ijklmn}$ is evaluated in the spinon mean-field ground state, which 
we explain below. For a time-reversal invariant system, we expect
$M_{ijklmn} = M_{ijklmn}^{\ast}$. 
If we further assume the translational invariance for the spinon sector, 
the resulting charge sector model is equivalent to $H_{\text{ch,ring}}$
in Eq.~\eqref{Hring} and also to the quantum dimer model in Eq.~\eqref{eq64}
except for the different couplings. 
Since the PCO breaks the translational symmetry of the Kagome lattice,
the spinon sector should be influenced by this symmetry breaking in the charge sector. 
In the following, we want to understand
how the spinon band structure is affected by the underlying PCO 
in the charge sector and how the modified spinon band structure 
feeds back into the charge sector. To this end, we take the enlarged unit cell of the
ETL and introduce the following spinon mean-field  
(hopping) Hamiltonian (see Fig.~\ref{fig10}),
\begin{widetext}
\begin{eqnarray}
 \tilde{H}_{\text{sp}}  
  &=&
 \sum_{{\bf R}}
  -\tilde{t}_1 \big[ 
  f^\dagger_{1\sigma } ({\bf R}) f^{\phantom\dagger}_{6\sigma } ( {\bf R})
+     f^\dagger_{ 2\sigma }( {\bf R}) f^{\phantom\dagger}_{3\sigma } ({\bf R})
+    f^\dagger_{ 4\sigma } ({\bf R}) f^{\phantom\dagger}_{5\sigma } ({\bf R})+ h.c.
     \big]
\nonumber 
\\
&&\quad\,\,\,
     -\tilde{t}_2 \big[ f^\dagger_{1\sigma } ({\bf R}) f^{\phantom\dagger}_{2\sigma } ( {\bf R})
     +  f^\dagger_{3\sigma } ({\bf R}) f^{\phantom\dagger}_{4\sigma } ( {\bf R})
     +  f^\dagger_{5\sigma } ({\bf R}) f^{\phantom\dagger}_{6\sigma } ( {\bf R})
     + h.c.\big] 
     \nonumber \\
&&\quad\,\,\,     
      - \tilde{t}_1' \big[
         f^\dagger_{1\sigma } ({\bf R}) f^{\phantom\dagger}_{7\sigma } ( {\bf R})
   +       f^\dagger_{6\sigma } ({\bf R}) f^{\phantom\dagger}_{7\sigma } ( {\bf R})
   +        f^\dagger_{2\sigma } ({\bf R}) f^{\phantom\dagger}_{8\sigma } ( {\bf R})
   +       f^\dagger_{3\sigma } ({\bf R}) f^{\phantom\dagger}_{8\sigma } ( {\bf R})
   \nonumber \\
   &&\quad\quad\quad\quad 
    +        f^\dagger_{9\sigma } ({\bf R}) f^{\phantom\dagger}_{4\sigma } ( {\bf R})
   +      f^\dagger_{9\sigma } ({\bf R}) f^{\phantom\dagger}_{5\sigma } ( {\bf R})
   + h.c.
      \big] 
      \nonumber \\
  &&\quad\,\,\,
      -\tilde{t}_2' \big[
            f^\dagger_{9\sigma } ({\bf R} ) f^{\phantom\dagger}_{1\sigma } ( {\bf R} + {\bf a}_1)
+      f^\dagger_{9\sigma } ({\bf R}) f^{\phantom\dagger}_{2\sigma } ( {\bf R}+ {\bf a}_1)
 +  f^\dagger_{7\sigma } ({\bf R}) f^{\phantom\dagger}_{3\sigma } ( {\bf R} + {\bf a}_2)
      +   f^\dagger_{7\sigma } ({\bf R}) f^{\phantom\dagger}_{4\sigma } ( {\bf R} + {\bf a}_2) 
\nonumber \\
   && \quad\quad \quad \quad
         +   f^\dagger_{8\sigma } ({\bf R}) f^{\phantom\dagger}_{5\sigma } ( {\bf R} -{\bf a}_1-{\bf a}_2)       
         +   f^\dagger_{8\sigma } ({\bf R}) f^{\phantom\dagger}_{6\sigma } ( {\bf R} -{\bf a}_1-{\bf a}_2)       
+ h.c.      \big], 
\label{eq68}
\end{eqnarray}
\end{widetext}
where ${\bf R}$ labels the unit cell of the ETL and
``1,2,3,4,5,6,7,8,9'' label the 9 sublattices of the ETL. 
Moreover, in Eq.~\eqref{eq68}, the spinon hoppings are given by 
\begin{eqnarray}
\tilde{t}_1 &=& t_1 \langle L^+_1 ({\bf R}) L^-_6 ({\bf R})   \rangle 
\\
\tilde{t}_2 &=& t_2 \langle L^+_1 ({\bf R}) L^-_2 ({\bf R})   \rangle 
\\
\tilde{t}_1' &=& t_1  \langle L^+_1 ({\bf R}) L^-_7({\bf R})   \rangle 
\\
\tilde{t}_2' &=& t_2  \langle L^+_9 ({\bf R}) L^-_1({\bf R} + {\bf a}_1)   \rangle . 
\end{eqnarray}
For the type-I CMI, we should have $\tilde{t}_1 = \tilde{t}_1' $ 
and $ \tilde{t}_2 = \tilde{t}_2' $
due to the 3-fold rotational symmetry around the center
 of each triangle on the Kagome lattice. 
For the type-II CMI, however, we expect $\tilde{t}_1 > \tilde{t}_1'$
and $\tilde{t}_2 > \tilde{t}_2'$ due to the presence of the PCO. 
The PCO enhances the hoppings of the charge rotors and spinons
in the resonating hexagons and weakens the ones in the non-resonating hexagons. 
The enhanced spinon hopping in the resonating hexagons 
further strengthen the couplings of the $\tilde{H}_{\text{ring}}$
in the resonating hexagons through $M_{ijklmn}$.
Thus the PCO would become more stable if the coupling between
spinon and charge excitations is switched on. 

%-----------------------------
%-----------------------------

\subsubsection{Levin-Wen's string mean-field theory and the spinon band structure}
\label{sec3C3} 

We consider the combination of the spinon hopping model $\tilde{H}_{\text{sp}}$ 
with the ring model $\tilde{H}_{\text{ring}}$. 
We extend Levin-Wen's string mean-field theory\cite{Levin04} for the 
quantum dimer (or string) model to solve the coupled 
charge and spinon problem. In Levin and Wen's original work, 
this mean-field theory is designed for the 
 quantum dimer model. 
As discussed previously in Sec.~\ref{sec4A}, 
our ring hopping model $\tilde{H}_{\text{ring}}$ is
a quantum dimer model if the charge occupation is mapped to the 
dimer covering on the DHL. The new ingredients here are the presence of 
the spinon degrees of freedom, and the coupling and the mutual feedback 
between the spinons and dimers (or charge fluctuations). 

We describe the variational wavefunction that is used to optimize the 
 Hamiltonian $\tilde{H}_{\text{ring}}$ in Eq.~\eqref{eq66a}
for the dimer sector. 
Following Levin and Wen, the wavefunction is parametrized by a set of 
variational parameters $\{ z_{{\bf r} {\bf r}'} \}$ where $z_{{\bf r}{\bf r}'}$
is defined on each bond ${ {\bf r}{\bf r}'}$ of the DHL. 
These variational parameters $z_{{\bf r} {\bf r}'}$ are also termed as string fugacity
by Levin and Wen.\cite{Levin04} 
Notice that the bonds on the DHL are also parametrized by the 
Kagome lattice sites that are the midpoints of the bonds,
For convenience, we label the set of variational parameters as $\{z_i \}$ where
$i$ is the Kagome lattice site.  For each set of $ \{ z_i \}$, 
the dimer wavefunction is given by 
\begin{equation}
\Psi ( \{ z_i \} ) = \prod_i \frac{|0 \rangle_i  + z_i | 1 \rangle_i}{ (1+ |z_i |^2)^{\frac{1}{2}}},
\label{eqfct}
\end{equation}
where $|0 \rangle_i$ and $| 1 \rangle_i$ define the 
absence and presence of the electron
charge at the Kagome lattice site $i$ or the dimer 
on the corresponding link on the DHL, respectively. 
Moreover, we have the following relations by definition,
\begin{eqnarray}
&& n_i | 0 \rangle_i = 0,
\quad\quad\,\,   n_i | 1 \rangle_i = | 1 \rangle_i ,
\\
&& L_i^+ |0 \rangle_i = |1 \rangle_i ,
\quad  L_i^- |1 \rangle_i = |0 \rangle_i . 
\end{eqnarray}

We employ the symmetry of the PCO to reduce the number of free 
variational parameters in the dimer wavefunction $\Psi ( \{ z_i \} ) $. 
Using the symmetries of the ETL, we find that
only two variational parameters are needed
\begin{eqnarray}
&& z_{1} ({\bf R}) = z_{2} ({\bf R})  = z_{3}({\bf R})  = z_{4}({\bf R})
\nonumber  
\\ &&\quad\quad\quad   = z_{5} ({\bf R})  = z_{6} ({\bf R})  \equiv z,
\\
&& z_{7}({\bf R}) = z_{ 8}({\bf R}) = z_{ 9}({\bf R}) \equiv \tilde{z}, 
\end{eqnarray}
where we have used the sublattice labelling in Fig.~\ref{fig10}. We
have reduced the set of variational parameters in the wavefunction 
to $z$ and $\tilde{z}$. 
But $z$ and $\tilde{z}$ are not independent from each other. 
This is because of the charge 
localization constraint of the type-II CMI. In terms of the 
dimer language, this constraint is that every DHL site is
connected by only one dimer. To satisfy this constraint, 
we only need to require
$\langle n_1( {\bf R}) \rangle + \langle n_6( {\bf R}) \rangle + 
\langle n_7( {\bf R}) \rangle =1,$
where the expectation value is taken for the variational 
wavefunction $\Psi ( \{ z_i \})$. This relation connects $\tilde{z}$ to $z$. 

For the DHL quantum dimer model $H_{\text{ch,ring}}$ in Eq.~\eqref{eq64}, 
the mean-field (or variational) phase is obtained by evaluating the Hamiltonian
$H_{\text{ch,ring}}$ with respect to $\Psi(\{z_i\})$ and optimizing
the energy by varying $z$. 
This static variational approach, however, cannot directly produce the 
plaquette ordered phase of the quantum dimer model.  
What it gives is a translationally invariant mean-field ground state. 
To obtain the right result, Levin and Wen applied a dynamical 
variational approach. Namely, for each static variational ground state,
one checks the stability of this phase by considering the 
quantum fluctuation of this phase. In the model that they were considering, 
they found some modes in a translationally invariant state can become unstable and 
drive a dimer crystal ordering. We expect similar physics should happen to 
our quantum dimer model $H_{\text{ch,ring}}$. 

Unfortunately, the dynamical variational approach by Levin and Wen 
is not a self-consistent mean-field approach and cannot be extended 
to the combined spinon and charge problem that we are interested in here. 
Since we know our quantum dimer model $H_{\text{ch,ring}}$ gives 
the ground state with the PCO
and we have argued that coupling the charge (dimer) with the spinons
makes the PCO even more stable, 
so we now introduce the PCO into the system by explicitly breaking the lattice symmetry.
That is, we modify the ring hoppings ${\mathbb K}_1$ 
and ${\mathbb K}_2$ in $\tilde{H}_{\text{ring}}$ of Eq.~\eqref{eq66a}.  
For the resonating hexagons, we change 
\begin{equation}
\mathbb{K}_1 \rightarrow \mathbb{K}_1 (1+ \delta),
\quad  \mathbb{K}_2 \rightarrow \mathbb{K}_2 (1 + \delta) ,
\end{equation}
and for the non-resonating hexagons, we use
\begin{eqnarray}
&& \mathbb{K}_1 \rightarrow \mathbb{K}_1 (1 - \delta),
\quad \mathbb{K}_2 \rightarrow \mathbb{K}_2 (1 - \delta) ,
\end{eqnarray}
where $\delta$ (with $\delta >0$) is a phenomenological parameter that breaks
an appropriate lattice symmetry for the PCO. 
This modification of the ring hoppings captures the energy modulation in the system
when the PCO is present. This phenomenological way of introducing the PCO is similar in spirit to Henley's 
approach\cite{Henley87} to the order by disorder, where a phenomenological interaction is introduced 
into the energy or the free energy by hand to model the ground state selection. 

We now solve the combined Hamiltonian of 
$\tilde{H}_{\text{sp}}$ and $\tilde{H}_{\text{ring}}$
with the modified ring hoppings self-consistently. 
This mean-field approach is expected to
underestimate the PCO for any fixed $\delta$ and thus 
underestimates the reconstruction of the spinon 
band structure due to the PCO. 
Nevertheless, to understand the {\it generic} features of the spinon band structure 
in the presence of the PCO, we can simply vary the phenomenological parameter $\delta$
and study the spinon band structure from this self-consistent mean-field theory. 

\begin{figure}[ht]
\subfigure[\, $\delta = 0$]{\includegraphics[width=4.cm]{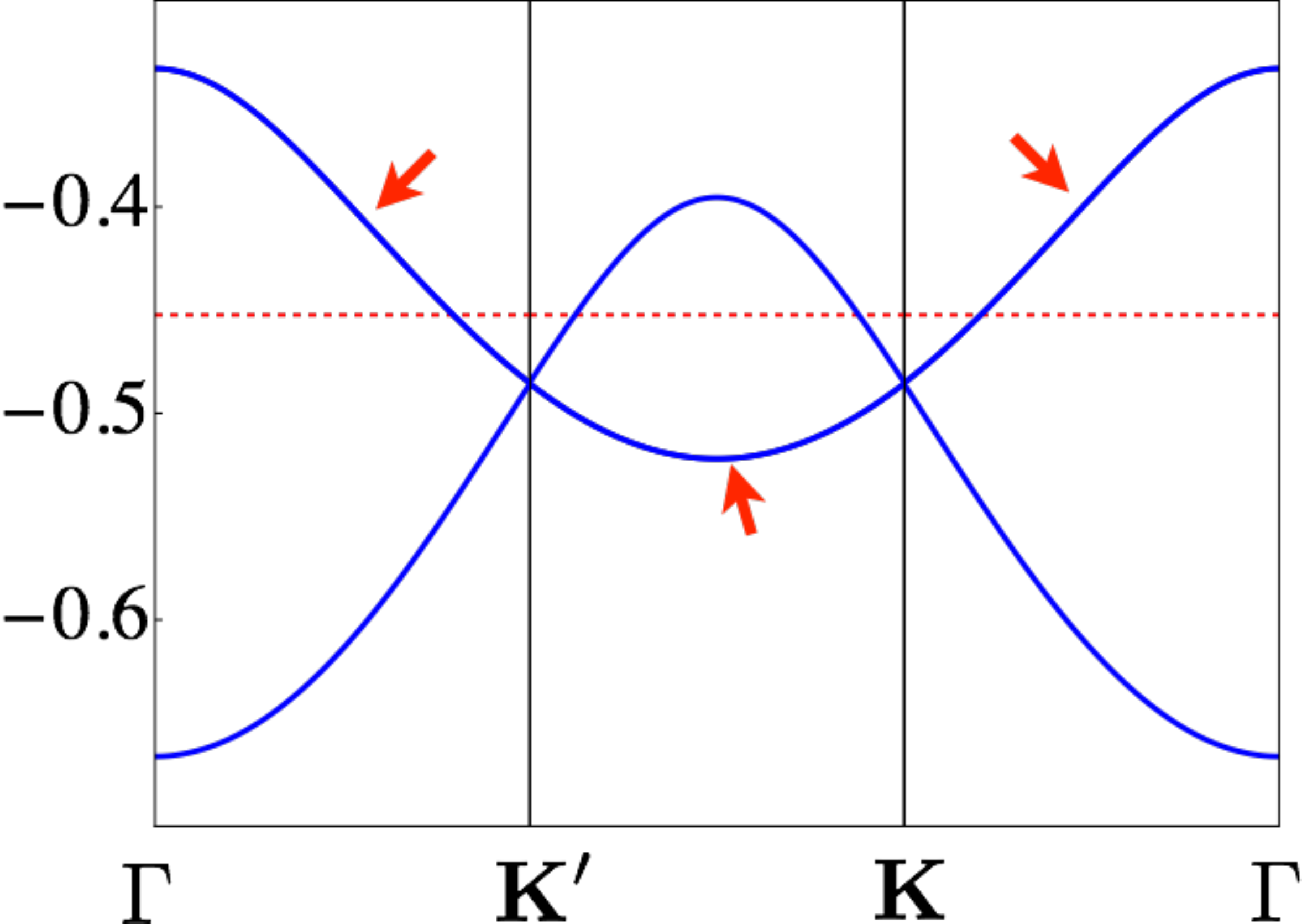}}
\hspace{0.4cm}
\subfigure[\, $\delta =0.3$ ]{\includegraphics[width=4.cm]{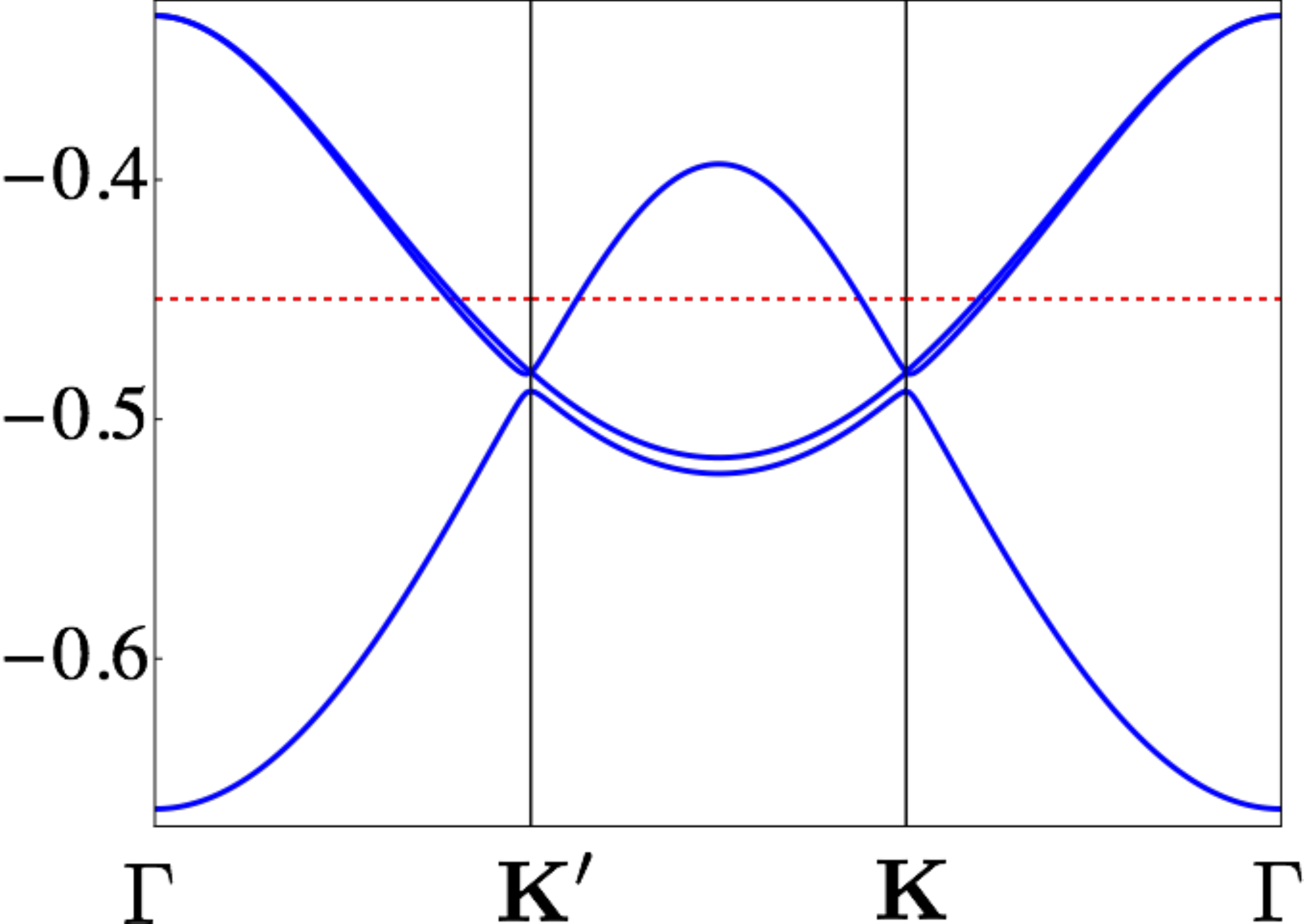}}
\subfigure[\, $\delta =0.7$]{\includegraphics[width=4.cm]{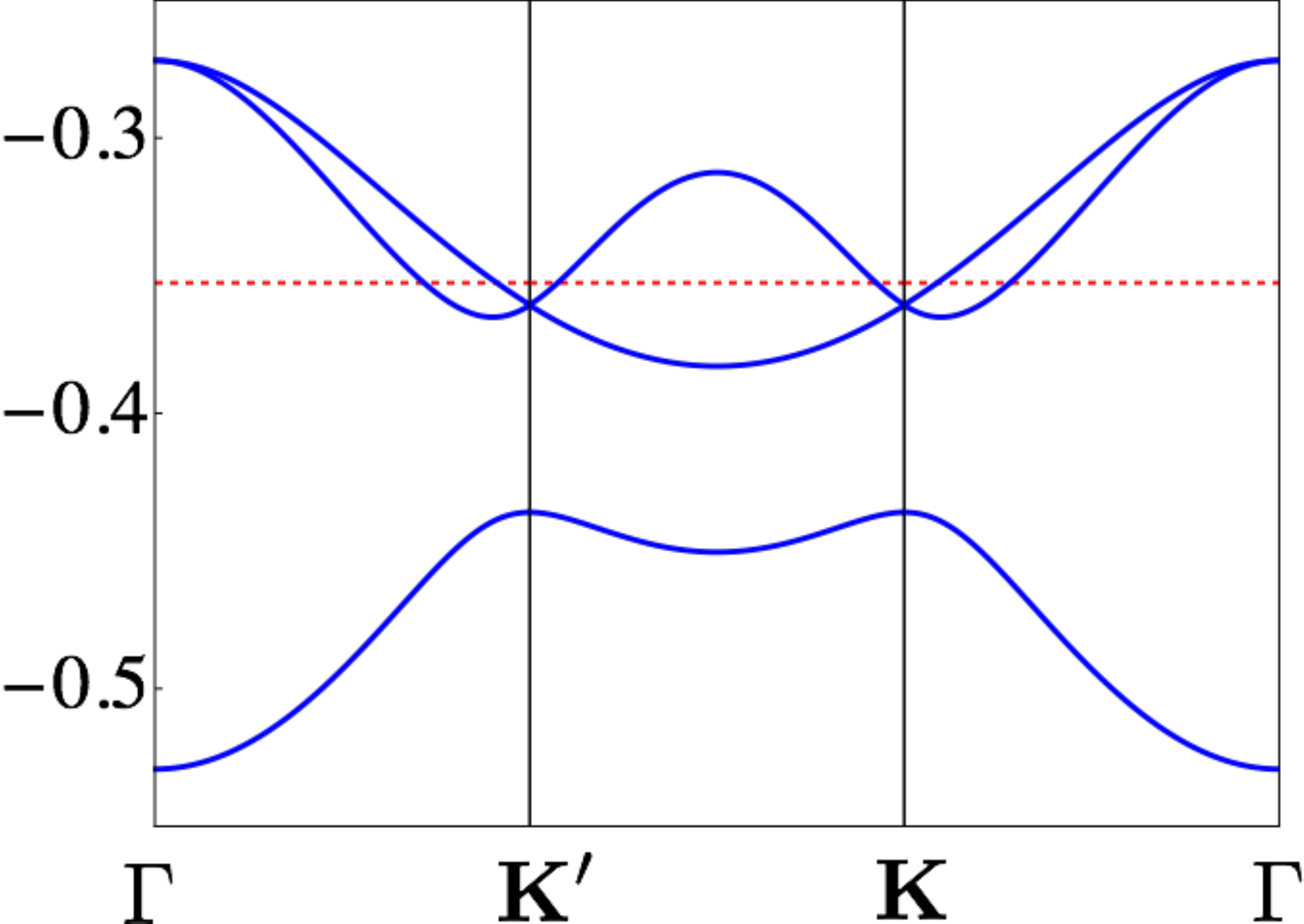}}
\hspace{0.9cm}
\subfigure[\, Brioulline zones]{\includegraphics[width=3.4cm]{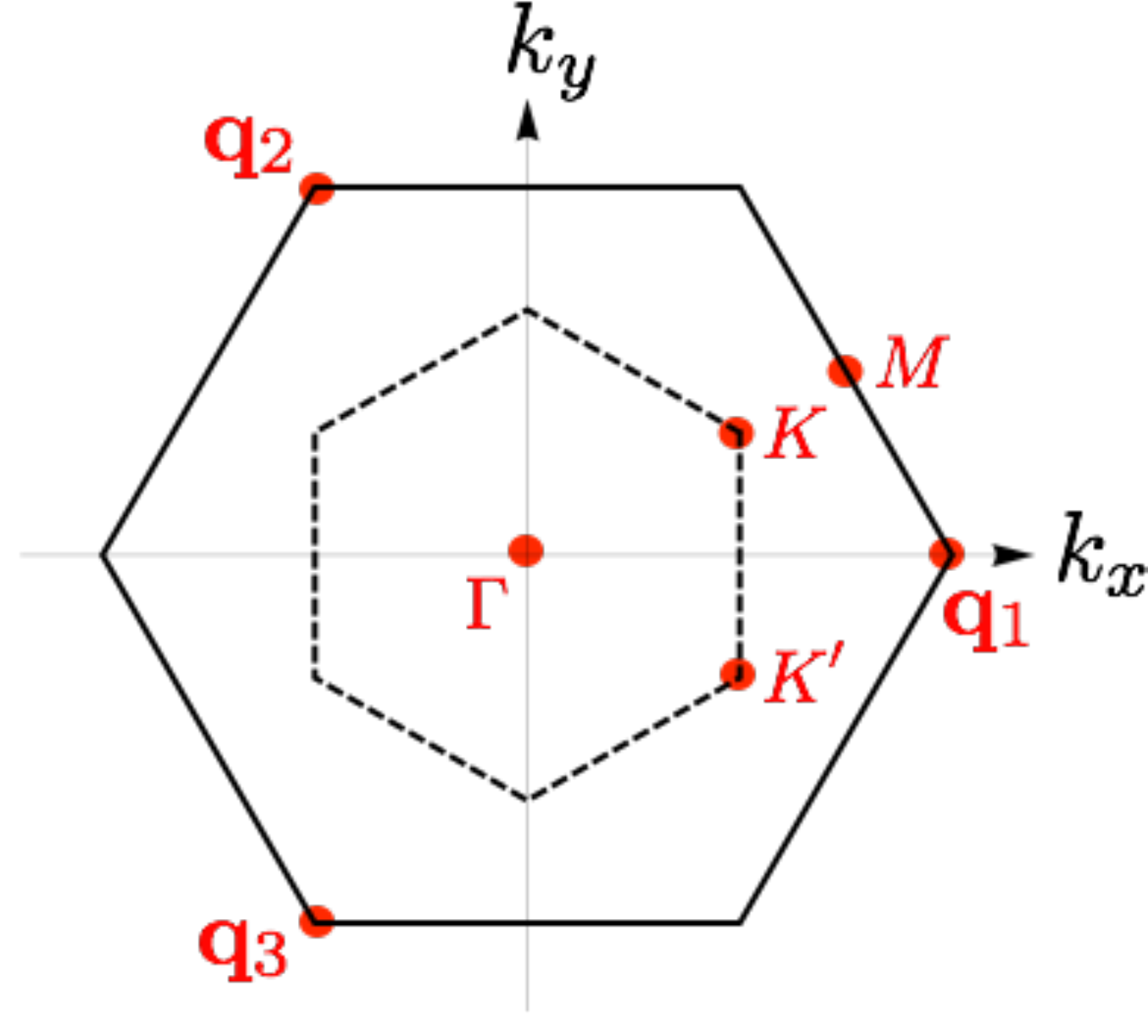}}
\caption{(Color online.) (a-c) The evolution of the spinon band structure as $\delta$ is varied.  
In the figure, we choose $\mathbb K_1 =4, \mathbb K_2 =1$ and $t_1=1, t_2 =0.5$. 
The (red) dashed line is the Fermi energy of the spinons. The energy unit is set to $t_1$.
The (red) arrow indicates a 2-fold degeneracy. 
The details of the band structures are discussed in the text. 
(d) The solid and dashed hexagons define the Brioullin
zones of the Kagome lattice (BZ1) and the ETL (BZ2). 
Setting $|{\bf b}_1| = |{\bf b}_2| =1$, we have 
${\bf K} = (\frac{2\pi}{3}, \frac{2\pi}{3\sqrt{3} }), 
{\bf K}' =(\frac{2\pi}{3}, -\frac{2\pi}{3\sqrt{3} })$
and ${\bf q}_1 = (\frac{4\pi}{3},0), {\bf q}_2 = ( - \frac{2\pi}{3}, \frac{2\pi}{\sqrt{3}}),
{\bf q}_3 =   ( - \frac{2\pi}{3},-\frac{2\pi}{\sqrt{3}})$.
}
\label{fig12}
\end{figure}

The evolution of the mean-field spinon band structure is depicted in Fig.~\ref{fig12}.
When $\delta = 0$, there is no PCO and the symmetry 
of the Kagome lattice is preserved.
The spinon band structure is similar to the one in a type-I CMI 
and contains 3 bands in the 1st Brioullin zone of the Kagome lattice 
(BZ1 in Fig.~\ref{fig12}d).
These 3 spinon bands are folded into 9 bands when we take a reduced-zone scheme 
and plot the spinon bands in the 1st Brioullin zone of the ETL (BZ2 in Fig.~\ref{fig12}d). 
As shown in Fig.~\ref{fig12}a, the mean-field spinon state is obtained by filling 
the lowest 3 bands as the rest 6 bands are much higher in energy and do not overlap 
with the lowest 3 bands. 
Moreover, the lowest band touches the second band 
on the zone boundary of the BZ2.

%-----------------------------

A finite $\delta$ creates the PCO and breaks the translational symmetry of the 
Kagome lattice. We have $\tilde{t}_1 > \tilde{t}_1'$
and $\tilde{t}_2 > \tilde{t}_2'$ as previously expected.  
The band touching of the lower 2 bands at the zone boundary of the BZ2
 is lifted by the level repulsion (see Fig.~\ref{fig12}b).   
A {\it direct band gap} is created between the lowest 2 bands. 
We emphasize this feature is generic and is not specific to the 
ring hoppings and electron hoppings that are chosen in Fig.~\ref{fig12}. 
As the parameter $\delta$ is further increased and the PCO becomes even stronger,
the direct band gap between the lowest 2 bands gets larger and larger
and eventually the lowest band is fully separated from the other bands by 
a full band gap (see Fig.~\ref{fig12}c). 

Even though the PCO enlarges the unit cell from 3 sites of the Kagome lattice to 
9 sites of the ETL, the spinon Fermi surface always exists.
This is because the number of electrons or spinons per unit cell remains odd in both cases. 
For the U(1) QSL in the type-I CMI
(in the type-II CMI with the PCO), the number of electrons per unit cell is 1 (3). 
Because of the band gap in the type-II CMI with the PCO, 
the lowest spinon band is completely filled by the spinons
which comprise 2/3 of the total spinon number, and 
the remaining 1/3 of spinons partially fill the upper 2 bands and 
give rise to the spinon Fermi surfaces.

\subsubsection{Spin susceptibility}
\label{sec3C4}
 
We now explain the low-temperature spin susceptibility 
for the type-II CMI with the PCO. 
Since the U(1) QSL ground state has spinon Fermi surfaces, 
we expect a finite (Pauli-like) spin susceptibility in the zero temperature limit.
At finite temperatures, one should recover the Curie-Weiss law for the 
spin susceptibility. What are the Curie constant and the Curie-Weiss temperature 
that characterize this Curie-Weiss law? 
The Curie constant is proportional to the number of the active local moments or 
unpaired localized electrons. Let us now consider the the local moment formation regime.
As long as the PCO is not destroyed by thermal fluctuations,
the lowest spinon band is fully filled and is inert to the external magnetic field,
and thus, only the 1/3 of the spinons from the partially filled upper bands contribute to the 
local moments, which comprise 1/3 of the total number of electrons in the system. 
Therefore, the low temperature Curie constant is 
\begin{equation}
{\mathcal C}^{\text L} = \frac{g^2 \mu_B^2 s(s+1)}{3k_{\text B}} \frac{N_{\Delta}}{3} 
\end{equation}
where $g\approx 2$ is the Land\'{e} factor\cite{Sheckelton12,Sheckelton14}, $s={1}/{2}$, and 
$N_\Delta$ is the total number of up-triangles in the system.   
From the electron filling fraction, we know $N_{\Delta} = N_e$.  
As we explain in Sec.~\ref{sec1}, because only 1/3 of the total spins are responsible for 
the low-temperature magnetic properties, the Curie constant is 
only 1/3 of the one at very high temperatures where all the 
electron spins are supposed to be active. 

As for the Curie-Weiss temperature, it is hard to make a quantitative prediction. 
But it is noted that the Curie-Weiss temperature is roughly set by the 
bandwidth of the active spinon bands in the QSL phase.
At low temperature PCO phase, the active spinon bands are the partially 
filled upper bands on the ETL rather than the lowest Kagome lattice spinon band 
for a type-I CMI. As one can see from Fig.~\ref{fig12}c, the bandwidth of the 
active spinon bands is significantly reduced when the PCO is present
compared to Fig.~\ref{fig12}a when the PCO is absent. 
As a result, we expect a much reduced Curie-Weiss temperature
for the type-II CMI with the PCO compared to the proximate type-I CMI.  
The above prediction may be consistent with the experimental results
in LiZn$_2$Mo$_3$O$_8$.\cite{Sheckelton12} 

\section{Local moments and Kugel-Khomskii model in the strong coupling limit}  
\label{sec4}

In Sec.~\ref{sec3}, we have established the existence of the PCO inside the 
type-II CMI and the reconstruction of the spinon band
structure caused by the PCO.
Using this, we provide a possible explanation of the low-temperature 
1/3 spin susceptibility in LiZn$_2$Mo$_3$O$_8$. 
In this section, we consider an alternative strong coupling regime
in which the PCO is strong such that the 3 resonating electrons are 
almost fully localized in the resonating hexagon and form the local moments 
which then interact with each other. 

\subsection{Structure of the local moments}
\label{sec4A}

To elucidate the nature of the local moments in each resonating hexagon,  
it is sufficient to isolate a single resonating hexagon and understand
the local quantum entanglement among the 3 resonating electrons. 
In this subsection, we consider two local interactions on the hexagon. 
The first interaction is already given in Eq.~\eqref{eq65} which is
the electron ring hopping model. The second interaction is the 
AFM exchange interaction between the electron 
spins. Since the electrons are always separated from each other by one lattice site, 
the AFM exchange is between the next nearest neighbors in the hexagon,
\begin{equation}
H_{\text{ex}}^0 =  J\sum_{\langle\langle ij\rangle\rangle}  
n_i n_j  ({\bf S}_i \cdot {\bf S}_j - \frac{1}{4} ),
\label{eq74}
\end{equation} 
where $i,j$ are the lattice sites that refer to the 6 vertices of the 
resonating hexagon (see Fig.~\ref{fig1}), $n_i$ is the electron occupation 
number at the site $i$ and ${\bf S}_i$ is the spin-1/2 operator of 
the electron spin at the site $i$. Because the electron position is not fixed in the hexagon, 
 the AFM interaction is active only when both relevant sites are occupied 
by the electrons and we need to include $n_i$ into the exchange interaction. 
The full Hamiltonian for an individual resonating hexagon plaquette is 
composed of the above two interactions,
\begin{eqnarray}
H_{\text p} &=& - \sum_{\alpha\beta\gamma} 
                    \big[\mathbb{K}_1 ( c^{\dagger}_{1\alpha} c^{\phantom\dagger}_{6\alpha} 
                                 c^{\dagger}_{5\beta} c^{\phantom\dagger}_{4\beta}  
 c^{\dagger}_{3\gamma} c^{\phantom\dagger}_{2\gamma} + h.c.)
\nonumber \\
 &&       
+ \mathbb{K}_2 ( c^{\dagger}_{1\alpha} c^{\phantom\dagger}_{2\alpha} 
                                 c^{\dagger}_{3\beta} c^{\phantom\dagger}_{4\beta}  
                                  c^{\dagger}_{5\gamma} c^{\phantom\dagger}_{6\gamma}  
+ h.c. ) \big] + H^0_{\text{ex}}.
\end{eqnarray}
Based on the perturbative values $\mathbb{K}_1 = 6t_1^3/V_2^2,
\mathbb{K}_2= 6t_2^3/V_1^2$ and the fact that $t_1 > t_2$ and $V_1 > V_2$ in
LiZn$_2$Mo$_3$O$_8$, we think the relevant regime should be 
$ \mathbb{K}_1 \gg \mathbb{K}_2 $.
 
Because a strong PCO causes a strong modulation in the bond energy, 
the values of $\mathbb{K}_1$ and $\mathbb{K}_2$ for the resonating hexagon would be modified
from the effective Hamiltonian that is obtained from the perturbative analysis.
Likewise, the spin exchange in the resonating hexagon is enhanced from its perturbative value.
So we expect this treatment is a good approximation to understand the local spin physics. 

The Hilbert space of the Hamiltonian $H_{\text p}$ is spanned by the 
electron states that are labelled by the positions and the spins of the 
3 resonating electrons. Because the electrons are separated from each other
by one lattice site, the Hilbert space for the positions is quite limited. 
There are a total of 16 states labelled by
$\{ |\alpha\beta\gamma\rangle_{\text{A}} \equiv |2\alpha,4\beta,6\gamma \rangle,
|\alpha\beta\gamma \rangle_{\text{B}} \equiv | 1\alpha,3\beta,5\gamma \rangle  \}$
with $\alpha,\beta,\gamma = \uparrow, \downarrow$.  
Since the local Hamiltonian $H_{\text p}$ commutes with 
the total electron spin 
${\bf S}_{\text{tot}}$ and $S^z_{\text{tot}}$, 
we can use 
$\{ {\bf S}_{\text{tot}},  S^z_{\text{tot}}\}$ to label
 the states. From the spin composition rule for 3 spins ($\frac{1}{2}\otimes \frac{1}{2} 
\otimes \frac{1}{2} = \frac{1}{2} \oplus \frac{1}{2} \oplus \frac{3}{2} $), 
we have 2 pairs of ${\bf S}_{\text{tot}} =1/2$ states. 

The states with $S_{\text{tot}} = {3}/{2}$ are simply the (ferromagnetic) 
bonding and anti-bonding states
which are not favored by the AFM exchange. We find that when 
 $J > \frac{2}{3} (\mathbb K_1+ \mathbb K_2 -\sqrt{ \mathbb K_1^2 -\mathbb K_1
 \mathbb K_2 +\mathbb K_2^2  } ) 
=\mathbb K_2 - \frac{\mathbb K_2^2}{4 \mathbb K_1} + \mathcal{O} (\mathbb K_2^3)$,
the local ground states are 4 (bonding) states with $S_{\text{tot}} =1/2$. 
This local 4-fold degeneracy can be effectively 
characterized by 2 quantum numbers ($s^z,\tau^z$) with  
$s^z = \pm \frac{1}{2}$ and $\tau^z = \pm \frac{1}{2}$.
 $s^z$ refers to the total spin 
$s^z \equiv S^z_{\text{tot}} = \pm \frac{1}{2}$. 
We also introduce a pseudospin-${1}/{2}$ 
operator $\boldsymbol{\tau}$ whose physical meaning is explained below.
The wavefunctions of the four $| {\tau^z s^z} \rangle$ states are 
given by (to the order of $\mathcal{O} ({\mathbb{K}_2}/{\mathbb{K}_1})$),
\begin{eqnarray}
|{\uparrow \uparrow }\rangle &=& \frac{1}{2} \big[
 |{\uparrow\uparrow\downarrow} \rangle_{\text A} 
 - |{\uparrow\downarrow\uparrow} \rangle_{\text A} 
+ | {\downarrow \uparrow\uparrow}\rangle_{\text B}
- |{\uparrow\uparrow\downarrow} \rangle_{\text B} 
  \big]
  \label{equu}
\\
 |{\downarrow\uparrow} \rangle &=& \frac{\sqrt{3}}{6} \big[2 |{ \downarrow \uparrow\uparrow}
 \rangle_{\text A}
 - | { \uparrow\downarrow \uparrow} \rangle_{\text A} - |  {\uparrow \uparrow\downarrow}
  \rangle_{\text A}
  \nonumber \\
  && \quad\,  + 2|  {\uparrow\downarrow\uparrow} \rangle_{\text B} - | { \uparrow \uparrow\downarrow} \rangle_{\text B}
   - | {\downarrow \uparrow\uparrow} \rangle_{\text B} \big]
     \label{eqdu}
\end{eqnarray}
and $|{\uparrow \downarrow} \rangle$, $|{\downarrow \downarrow} \rangle$ 
are simply obtained by a time-reversal operation. 

%-----------------------------
\begin{figure}[ht]
{\includegraphics[width=8cm]{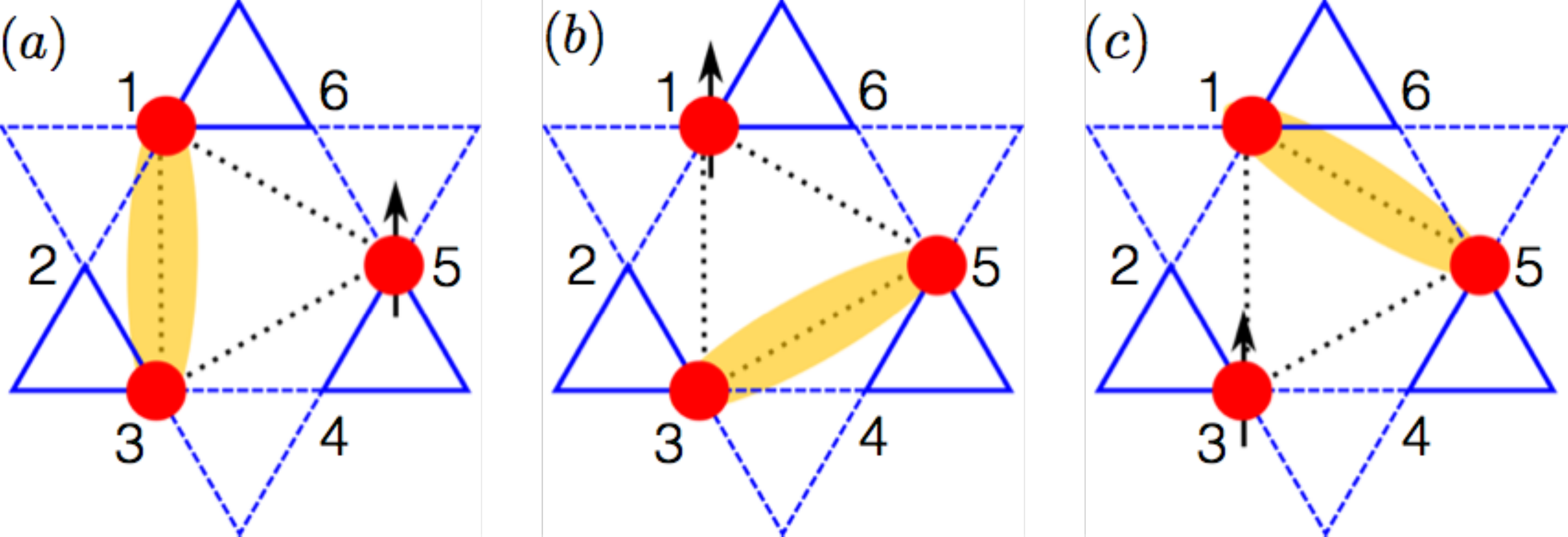}}
\caption{(Color online.) 
Three singlet positions that are related by the 3-fold rotation.  }
\label{fig13}
\end{figure}
%-----------------------------

What is the physical origin of this local 4-fold degeneracy? 
Clearly, the 2-fold degeneracy of $s^z =\pm 1/2$ arises from 
the time-reversal symmetry and the Kramers' theorem. 
The remaining 2-fold degeneracy comes from the point group symmetry of the resonating hexagon. 
This is easy to see if we 
freeze the positions of the 3 electrons. To be concrete, let us fix the 
electrons to the sites 1,3,5 in Fig.~\ref{fig13}. To optimize the
 exchange interaction, 2 electrons must form a spin singlet, 
 which leaves the remaining electron
as a dangling spin-1/2 moment. As shown in Fig.~\ref{fig13}, 
this singlet can be formed 
between any pair of the electrons and the different locations of the spin singlet 
are related by the 3-fold rotation. Even though there seems
to be 3 possible singlet positions, only 2 of them are linearly independent, which
 gives to the 2-fold $\tau^z$ degeneracy which survives even 
when the ring electron hopping is turned on. 
As a result, the pseudospin $\boldsymbol{\tau}$ is even under time-reversal 
and acts on the space of the singlet position or equivalently the dangling spin position. 
In fact, the two states in Eqs.~\eqref{equu} and \eqref{eqdu} comprise the E irreducible 
representation of the C$_{3v}$ point group.

\subsection{Kugel-Khomskii model}
\label{sec4B}

Now we consider the spin and pseudospin interaction between neighboring 
resonating hexagons. 
The neighboring resonating hexagons are connected by a ``bow-tie'' that 
is composed of one up and one down-triangle. 
The local moment interaction comes from the remaining exchange 
interaction between the 2 electron spins that reside on the four
outer vertices of the bow-tie. 
To be concrete, we consider the bow-tie that connects the 
two resonating hexagons at the ${\bf R}$ and ${\bf R} + {\bf a}_1$ (see Fig.~\ref{fig1}).
To derive the local moment interaction, one just needs to project the 
remaining electron spin exchange interaction onto the 4-fold ground state
manifold of each resonating hexagon. 
To this end, we first write down the inter-hexagon exchange interaction between the 
electrons at the bow-tie vertices,
\begin{eqnarray}
H'_{\text{ex}} &=& 
 - \frac{J\rq{}}{4} 
[  n_4 ( {\bf R} )  +   n_5 ({\bf R})  ]
[  n_1 (  {\bf R} + {\bf a}_1 )  +n_2 ({\bf R} + {\bf a}_1)]
\nonumber \\
&&+
J\rq{} [ {\bf S}_{4}( {\bf R}) n_{4} ({\bf R})  + {\bf S}_{5}({\bf R}) n_{5} ({\bf R})  ]
 \times 
[ {\bf S}_{1} ({\bf R}+{\bf a}_1) 
\nonumber \\
&& \times 
n_{1} ({\bf R}+{\bf a}_1)  
+ {\bf S}_{2} ({\bf R}+{\bf a}_1) n_{2} ( {\bf R}+{\bf a}_1 ) ],
\label{eq78}
\end{eqnarray}
where we have considered the exchange interactions for electrons at 
all 4 pairs of the sites. The exchange paths of these pairs
all go through the center vertex of the bow-tie and thus are of equal length.  As a result,
we only introduce one exchange coupling $J'$ for the four pairs in the above 
equation. Moreover, since $J\rq{}$ is the exchange coupling between the spins 
after the system develops the PCO, clearly $J'$ should be smaller than the 
intra-hexagon exchange coupling $J$ in Eq.~\eqref{eq74}. 

We project $H'_{\text{ex}}$ onto the local ground state manifold
at resonating hexagon sites ${\bf R}$ and ${\bf R}+{\bf a}_1$
and then express the resulting interaction in terms of the spin and 
pseudospin operators. The effective interaction on other bonds can be obtained similarly. 
The final local moment interaction reduces to a 
Kugel-Khomskii model\cite{Kugel82} that is defined on the ETL, 
which to the order of $\mathcal{O}( {\mathbb K_2}/{\mathbb K_1} )$ 
is given as
\begin{eqnarray}
H_{\text{KK}} &=& \frac{J'}{9} \sum_{{\bf R}} \sum_{\mu=x,y,z}  
\big[ {\bf s} ({\bf R}) \cdot {\bf s} ({{\bf R}+{\bf a}_{\mu} } ) \big] 
\nonumber \\
&& \times 
[  1 +4  \pi^{\mu} ({\bf R} ) ]
[  1 - 2 \pi^{\mu} ({{\bf R}+{\bf a}_{\mu}} ) ]
\label{eq79}
\end{eqnarray}
where the new set of pseudospin operators are defined as 
$\pi^{x,y} ({\bf R})  = -\frac{1}{2} \tau^z ({\bf R}) \mp  \frac{\sqrt{3}}{2} \tau^x ({\bf R} ), 
\pi^z ({\bf R}) = \tau^z ({\bf R})$, 
and ${\bf a}_{x} = {\bf a}_{1}, {\bf a}_y = {\bf a}_2$ and 
$ {\bf a}_z =- {\bf a}_1 -{\bf a}_2$. 
In Eq.~\eqref{eq79}, the exchange coupling is significantly reduced after the 
projection compared to the original exchange coupling in Eq.~\eqref{eq78}.

Since the pseudospin $\boldsymbol{\tau}$ does not directly 
couple to the external magnetic field, the low-temperature 
Curie-Weiss temperature 
($\Theta_{\text{CW}}^{\text L}$) 
and Curie constant ($\mathcal{C}^{\text L}$) are 
straightforward to compute from $H_{\text{KK}}$,
\begin{eqnarray}
\Theta_{\text{CW}}^{\text L} &=& - \frac{z_t s(s+1)}{3} \frac{J'}{9}, 
\,\, 
 \mathcal{C}^{\text L} = \frac{g^2 \mu_{\text B}^2 s(s+1) }{3k_{\text B}} \frac{N_\Delta}{3},
 \label{cwl} 
\end{eqnarray}
where $z_t=6$ is the coordination number for nearest neighbors of the 
triangular lattice. The above results are again consistent with the lower temperature 
1/3 Curie-constant of the spin susceptibility in LiZn$_2$Mo$_3$O$_8$. 

This Kugel-Khomskii model involves the spin-spin interaction, 
the pseudospin-pseudospin interaction and also the spin-pseudospin interaction, 
which make the model analytically intractable. 
In the absence of the spin-pseudospin interaction, 
the Heisenberg spin exchange model would favor the classical 
120-degree state. The presence of spin-pseudospin interaction, however, competes with 
the Heisenberg term, destabilizes the 120-degree state and may potentially favor 
a spin liquid state. Such a spin liquid, if exists, could be smoothly connected to
the U(1) QSL of the intermediate coupling regime in Sec.~\ref{sec3C}.
We leave this question for the future work. 

Despite its complicated form, the Kugel-Khomskii model becomes tractable in 
the presence of a strong external magnetic field.  
We apply a strong magnetic field to fully polarize the local spin moments
such that $s^z = 1/2$ in every ETL unit cell
but at the same time keep the field from polarizing all the electron spins
in the system. The remaining active local moments are the pseudospins $\boldsymbol{\tau}$,
and the interaction between them is a ferromagnetic compass model
on the ETL, 
\begin{equation}
H_{\text{r-KK}} = -\frac{2J'}{9}\sum_{\bf R} \sum_{\mu = x,y,z} 
\pi^\mu ({\bf R}) \pi^\mu ({\bf R} + {\bf a}_{\mu}).
\end{equation}

Using a standard Luttinger-Tisza method,\cite{PhysRev.70.954} 
we find that the classical ground state of $H_{\text{r-KK}}$ 
has a U(1) accidental degeneracy, {\it i.e.} any 
ferro- (${\bf q}=0$) state with the pseudospin orienting in $xz$ plane is 
a classical ground state. We parametrize the classical pseudospin ordering as 
\begin{equation}
\boldsymbol{\tau}_{\text{cl}} = \frac{1}{2} ( \cos \vartheta\, \hat{z} 
+ \sin \vartheta \, \hat{x})
\label{eq83}
\end{equation}
with $\vartheta \in [0,2\pi) $.
Quantum fluctuations lift
this classical U(1) degeneracy and lead to pseudospin orderings.  We find
$\vartheta =\pi/6 +n\pi/3  $ ($n\in$ integer) is selected by the quantum zero
point energy in a linear spin-wave analysis (see Appendix.~\ref{secapp3}). 
Although the total magnetization of each resonating hexagon is $\langle s^z\rangle=1/2$,
 the pseudospin ordering leads to a modulation of the spin 
 ordering inside the resonating hexagon which then breaks the 
 3-fold rotational symmetry around the center of the resonating hexagon.

\section{Discussion}

\subsection{Thermal transition and susceptibility crossover: applications to LiZn$_2$Mo$_3$O$_8$}
\label{sec5A} 

Since both the intermediate coupling and strong coupling regimes 
of the type-II CMI (PCO)
give a consistent explanation for the low-temperature ``1/3 anomaly'' in spin 
susceptibility of LiZn$_2$Mo$_3$O$_8$, 
we then propose that the Mo system in LiZn$_2$Mo$_3$O$_8$ 
is in the type-II CMI and also develops the PCO at low temperatures. 
We now discuss high-temperature properties of the system that would
be in the type-II CMI (PCO) phase at low temperatures.

The PCO at low temperatures breaks the lattice symmetry of the Kagome lattice, which
indicates a thermal phase transition at a finite temperature (see Fig.~\ref{fig2}). 
This phase transition is expected to occur at 
$T^{\ast} \sim \mathcal{O}( \mathbb{K}_1) = \mathcal{O} (t_1^3/V_2^2) $
(because $\mathbb{K}_1 \gg \mathbb{K}_2 $) when the 
local electron resonance in the elementary hexagons loses the 
quantum phase coherence. This thermal transition 
is found to be first order in a Landau theory analysis 
for a clean system (see Appendix.~\ref{secapp4}).
In reality, LiZn$_2$Mo$_3$O$_8$ is influenced by various disorders or 
impurities (e.g. the mixed Li/Zn sites 
and mobile Li ions)\cite{Sheckelton12}. For example, impurities 
would broaden the charge ordering transition\cite{McMillan75}. 
This may explain why a sharp transition is not observed 
in the experiments\cite{Sheckelton12}. 
Nevertheless, the experiments do observe a peak around 100K in 
heat capacity\cite{Sheckelton12} which might be related to the smeared-out 
phase transition.

\begin{figure}[ht]
{\includegraphics[width=6cm]{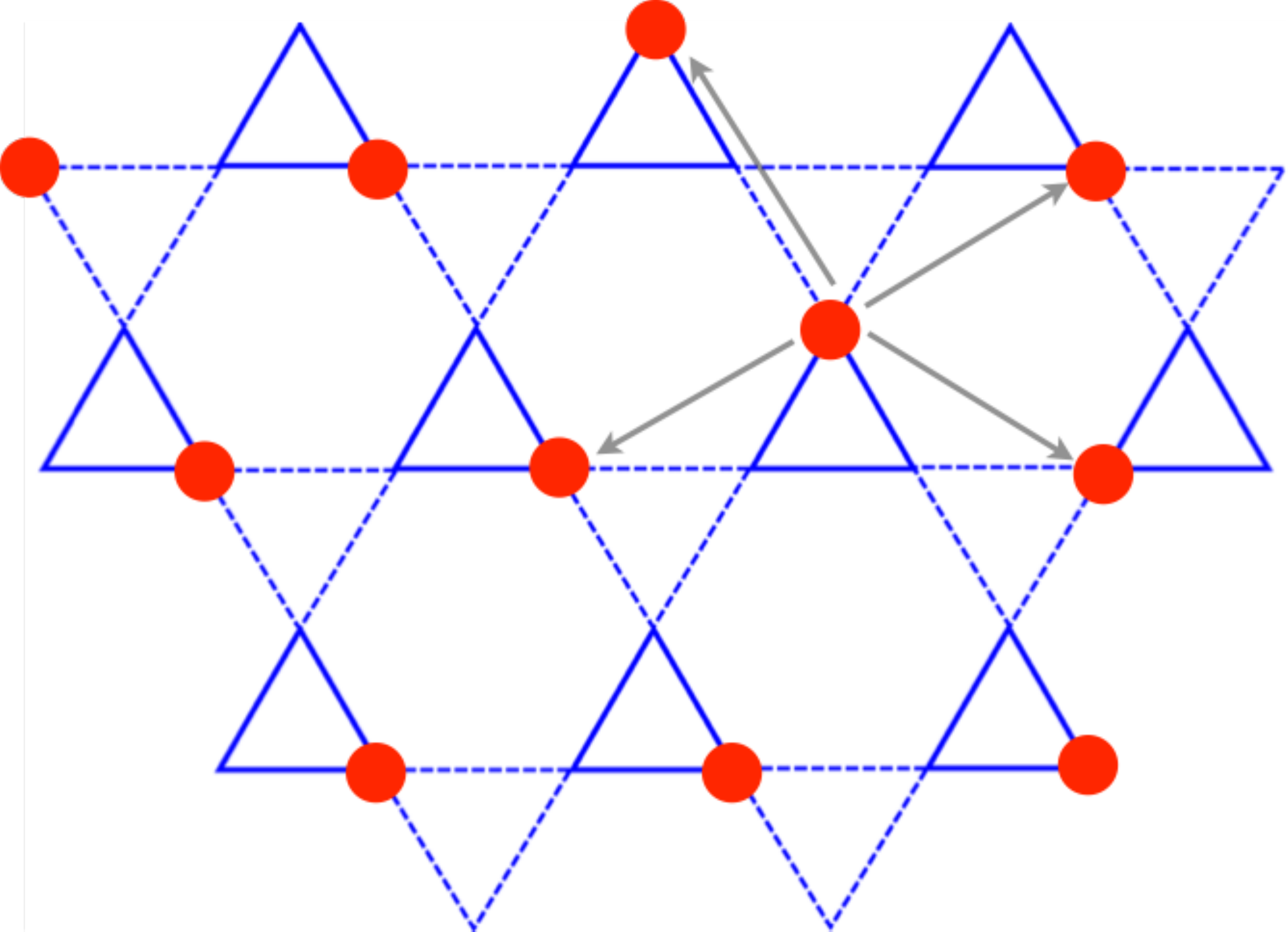}}
\caption{(Color online.) 
The snapshot of the electron occupation in the KCI. 
The arrows indicate the 4 neighboring occupied sites near one occupied site. }
\label{fig15}
\end{figure}

What happens if we increase the temperature above 
the thermal transition at $T^{\ast}$? As we explain in Sec.~\ref{sec1} and the plot
in Fig.~\ref{fig2}, the system can still be regarded as the type-II CMI 
if the thermal fluctuations do not significantly violate the electron 
localization in each triangle. 
Because of the extensive degeneracy of the electron occupation configuration,
such an intermediate temperature regime ($T^{\ast} < T < T^{\ast\ast}= \mathcal{O}(V_2)$)
is named as Kagome charge ice (KCI) regime in Sec.~\ref{sec1}
and the ``charge ice rule'' corresponds to the ``one electron per triangle'' 
localization condition of the type-II CMI. 
In the KCI regime, all the electron spins are active and contribute 
to the Curie-Weiss law that is governed by 
\begin{equation}
\Theta_{\text{CW}}^{\text M} = - \frac{{z}_{\text{bt}} s(s+1)  }{3} J_{\text M},
\,\,
\mathcal{C}^{\text M} = \frac{g^2 \mu_{\text B}^2 s(s+1) }{3k_{\text B}}
 {N_{\vartriangle}},
\label{eq114}
\end{equation}
where ${z}_{\text{bt}} =4$ is the number of neighboring bow-tie structures of 
one occupied Kagome lattice site (see Fig.~\ref{fig15}). 
The Curie constant in the KCI regime is 3 times the low-temperature Curie constant in Eq.~\eqref{cwl}.
In Eq.~\eqref{eq114}, we have assumed the spin exchange interaction in every 
bow-tie structure takes the form like Eq.~\eqref{eq78} with an exchange coupling $J_{\text M}$. 
If we set $J_{\text M} = J'$, we have 
$\Theta_{\text{CW}}^{\text M}=6\Theta_{\text{CW}}^{\text L}$ (see Eq.~\eqref{cwl}). 

It is not clear whether this KCI regime does exist from the current
experiments in LiZn$_2$Mo$_3$O$_8$. 
It is likely that the crossover temperature $T^{\ast\ast}$ is close to 
the transition temperature $T^{\ast}$ such that the KCI regime is 
too narrow to be clearly observed. 
In addition, the charge ice rule would be spoiled by the local charge
impurities which can also destabilizes the KCI.

If the temperature is above $T^{\ast\ast}$ but 
much lower than the inter-site repulsion $V_1$ in the up-triangles, 
this system can be thought as the high temperature limit of the type-I CMI. 
The electrons form localized $S=\frac{1}{2}$ moments on the up-triangles. 
Like the KCI, all the spins are active and contribute to
the Curie-Weiss law. So the high-temperature Curie-Weiss law is governed by
\begin{equation}
 \Theta_{\text{CW}}^{\text H} = - \frac{z_t s(s+1)}{3} {J_{\text H}}, 
\,\,
 \mathcal{C}^{\text H}= \frac{g^2 \mu_{\text B}^2 s(s+1) }{3k_{\text B}}
{N_{\vartriangle}},
\label{CWh}
\end{equation}
where $z_t=6$ is the coordination number for nearest neighbors in the triangular lattice. Here 
the triangular lattice is formed by the up-triangles. 
So ${\mathcal C}^{\text H} = {\mathcal C}^{\text M} = 3 {\mathcal C}^{\text L}$. 
In Eq.~\eqref{CWh}, we assume a Heisenberg exchange interaction 
between the electron spins in the nearest-neighbor up-triangles 
with an exchange coupling $J_{\text H}$. 
If we set $J_{\text H} = J'$, we have $\Theta_{\text{CW}}^{\text H} = 
9 \Theta_{\text{CW}}^{\text L}$(see 
Eq.~\eqref{cwl}). 

We now comment on the $S=1/2$ cluster spin picture
that is assumed in the Refs.~\onlinecite{Sheckelton12,Flint13}. 
In this picture, each Mo$_3$O$_{13}$ cluster
contains one unpaired electron 
which is delocalized over the three Mo atoms in the cluster and 
this delocalized electron gives rise to the $S=1/2$ cluster spin in 
each Mo$_3$O$_{13}$ cluster.
In our theory for LiZn$_2$Mo$_3$O$_8$, however,
this cluster spin picture is only valid
if the system is in the type-I CMI.
Although the $S=1/2$ cluster spin is probably not a valid 
description of the magnetic properties of the low-temperature type-II CMI,
it can become valid in the high-temperature type-I CMI 
above the crossover temperature $T^{\ast\ast}$.

The inter-site repulsion $V_1$ is expected to be a large energy scale, so we 
do not discuss any further finite-temperature crossover when the temperature
is further increased.

\subsection{Further experimental predictions for LiZn$_2$Mo$_3$O$_8$}
\label{sec5B}

In the case of a U(1) QSL with spinon Fermi surfaces (as well as the PCO)
of the type-II CMI, we expect the usual behaviors of a 2D U(1) QSL with 
spinon Fermi surfaces would show up. That is, the specific heat $C_v \sim T^{2/3}$,
and a Pauli-like spin susceptibility in the low temperature limit\cite{Lee05}. 
The crossover in the behavior of the spin susceptibility 
from the local moment Curie-Weiss regime to the 
Pauli-like behavior is expected to happen at the 
temperature set by the bandwidth of active spinon bands (see Sec.~\ref{sec3C}), 
or equivalently, by the low-temperature Curie-Weiss temperature 
$|\Theta_{\text{CW}}^{\text L}|$. This crossover temperature
should be very small because of the suppressed Curie-Weiss temperature at low temperatures.
As a result, the Pauli-like spin susceptibility may be smeared out by 
various extrinsic factors like local magnetic impurities at very low temperatures. 
Likewise, even though the $C_v/T$ experiences a upturn 
below 10K in the absence of external magnetic fields, 
it is likely that the nuclear Schottky anomalies may 
complicate the specific heat data. 

On the other hand, the apparently gapless spectrum of the spin excitations in the inelastic 
neutron scattering measurement\cite{Mourigal14} is consistent with the gapless spinon
Fermi surface of our U(1) QSL. Moreover,
the measurements of relaxation rate from both NMR ($1/(T_1 T)$) and 
 $\mu$SR ($\lambda T^{-1}$) also indicate gapless spin-spin correlations.\cite{Sheckelton14}
In our U(1) QSL, the reduction of the spinon bandwidth due to the PCO increases 
the density of the low-energy magnetic excitations. This would lead to a 
 low-temperature upturn of the spin-lattice relaxation, which is in fact observed in 
NMR and $\mu$SR experiments\cite{Sheckelton14}. 

In future experiments, it might be interesting to apply a 
pressure to the material and to drive the system from the type-II CMI to 
the type-I CMIs and/or to the FL-metal. This can not only confirm our 
phase diagram but also provide an opportunity to explore the interesting 
phase transitions and the critical Fermi surfaces that may occur in the system. 
Since the type-I CMI is also a U(1) QSL, the large exchange energy scale in
the type-I CMI may provide a wide temperature window
to study the intrinsic properties of the QSL at low temperatures. 
A direct measurement of the PCO at low temperatures is crucial for our theory. 
To this end, a high resolution X-ray scattering measurement can be helpful. 
Moreover, the presence of local quantum entanglement within the resonant
hexagon may be probed optically by measuring the local exciton excitations. 
Furthermore, if the system is in a U(1) QSL with a spinon Fermi surface, 
the low-temperature thermal conductivity can be an indirect probe of 
the low-energy spinon excitation, and a direct measurement of the 
correlation of the emergent U(1) gauge field might be possible because 
the strong spin-orbit coupling of the Mo atoms can enhance the coupling
between the spin moment and the spin texture\cite{Lee13}.

\subsection{Other Mo based cluster magnets}
\label{sec5C}

%-----------------------------
\begin{table}[t]
\centering
\begin{tabular}{lccccc}
\hline\hline 
& [Mo-Mo]$_{\text u}$ & [Mo-Mo]$_{\text d}$ & $\lambda$ & e$^-$/Mo$_3$ & Ref
\\
\hline\hline
LiZn$_2$Mo$_3$O$_8$ & 2.6\AA & 3.2\AA & 1.23 & 7 & [\onlinecite{Sheckelton12}]
\\
Li$_2$InMo$_3$O$_8$ & 2.54\AA & 3.25\AA & 1.28 & 7 & [\onlinecite{Gall13}]
\\
ScZnMo$_3$O$_8$ & 2.58\AA & 3.28\AA  & 1.27 & 7 & [\onlinecite{TORARDI85}]
\\
\hline\hline
\end{tabular}
\caption{Mo-Mo bond lengths, 
 anisotropic parameters ($\lambda$) 
and number of electrons per Mo$_3$O$_{13}$ cluster 
 for three different cluster magnets. 
The electron number is counted from stoichiometry.}
\label{tab3}
\end{table}

The compounds that incorporate the Mo$_3$O$_{13}$ cluster unit
represent a new class of magnetic materials called ``cluster magnets''.  
Several families of materials, such as M$_2$Mo$_3$O$_8$ (M=Mg,Mn,Fe,Co,Ni,Zn,Cd),
 LiRMo$_3$O$_8$ (R=Sc,Y,In,Sm,Gd,Tb,Dy,Ho,Er,Yb) and other related variants\cite{McCarroll57,McCarroll77,Gall13,TORARDI85}, 
fall into this class. The magnetic properties of most materials 
have not been carefully studied so far. 
In Tab.~\ref{tab3}, we list three cluster magnets with odd 
number of electrons in the Mo$_3$O$_{13}$ cluster unit. 
We introduce a phenomenological parameter $\lambda$ to characterize
the anisotropy of the Mo Kagome lattice, which is defined as 
the ratio between inter-cluster (or down-triangle) and intra-cluster (or up-triangle) 
Mo-Mo bond lengths,
\begin{equation}
\lambda = \frac{\text{[Mo-Mo]}_{\text d}}{\text{[Mo-Mo]}_{\text u}}. 
\end{equation}
According to our theory, more anisotropic systems tend to favor the QSL of the type-I CMI. 
As shown in Tab.~\ref{tab3}, Li$_2$InMo$_3$O$_8$ has a larger anisotropic
parameter than LiZn$_2$Mo$_3$O$_8$. 
Unlike LiZn$_2$Mo$_3$O$_8$, the spin susceptibility of 
Li$_2$InMo$_3$O$_8$ does not show the ``1/3 anomaly''
but is instead characterized by one Curie-Weiss temperature $\Theta_{\text{CW}} = -207$K
down to 25K.\cite{Gall13}
Moreover, the Curie constant is consistent with one unpaired spin-1/2 moment 
per Mo$_3$O$_{13}$ cluster in the type-I CMI. 
Below 25K, the spin susceptibility of Li$_2$InMo$_3$O$_8$ 
saturates to a constant, which is consistent with the Pauli-like spin susceptibility 
for a spinon Fermi surface U(1) QSL. Besides the structural and spin susceptibility data,
very little is known about Li$_2$InMo$_3$O$_8$.  Thus,
more experiments are needed to confirm the absence of magnetic ordering in 
Li$_2$InMo$_3$O$_8$ and also to explore the magnetic properties of ScZnMo$_3$O$_8$
and other cluster magnets.

\acknowledgements 
We thank P. A. Lee, A. Essin, A. Burkov, L. Balents and M. Hermele for helpful discussion,
and especially T. McQueen for email correspondence and conversation 
that clarify the experimental results.  
This work was supported by the NSERC, CIFAR, and 
Centre for Quantum Materials at the University of Toronto. 
We also acknowledge NSF grant no.~PHY11-25915 for supporting the visitor program 
at the Kavli institute for theoretical physics during the the workshop ``Frustrated Magnetism and Quantum Spin 
Liquids'', where the current work was initiated.  

\bibliography{ref}

\begin{thebibliography}{44}
\bibitem [{not()}]{note2}%
  \BibitemOpen
  \href@noop {} {}\bibinfo {note} {\textnormal{The intra-group interaction of
  different groups should be of the same order }}\BibitemShut {NoStop}%
\end{thebibliography}

\appendix

\section{Theory of type-I CMI at $V_2 =0$} 
\label{secapp1} 

Here we explore a limiting case of the model in which the inter-site 
repulsive interaction $V_2$ is vanishing and study the type-I CMI 
driven by the remaining interaction $V_1$ in the framework of 
the slave-rotor mean-field theory rather than the slave-particle construction
of Sec.~\ref{sec3}. 

The kinetic part of the Hubbard model can be easily diagonalized and the 
electrons form 3 bands with the following dispersions,
 \begin{eqnarray}
\epsilon^{1,2}_ {\bf k} & = & 
- \frac{1}{2} \big(t_1 +t_2  \pm [9t_1^2 -6t_1 t_2 + 9t_2^2 
+8t_1 t_2 \big( \cos {\bf k}\cdot {\bf b}_1
\nonumber \\
&&
  + \cos {\bf k} \cdot {\bf b}_2 + \cos ({\bf k}\cdot {\bf b}_1 
  + {\bf k}\cdot {\bf b}_2) \big) ]^{1/2}  \big), 
  \label{eqapp1}
  \\
\epsilon^3_{\bf k} & = & 
t_1 + t_2,
\label{eqapp2}
\end{eqnarray}
where ${\bf b}_1$ and ${\bf b}_2$ are the 2 elementary lattice vectors
of the underlying triangular Bravais lattice (see Fig.~\ref{fig1}). 
These 3 electron bands are well-separated from each other 
and only touch at certain discrete momentum points. 
In particular, the lowest two bands have Dirac-point band
touchings at the Brillouin zone corners when $t_1 = t_2$. 
With the 1/6 electron filling, the electrons only fill a half the lowest band 
$\epsilon^1_{\bf k}$ and the ground state is a FL-metal
when the interaction $V_1$ is weak.

Now we consider the effects of the interactions. A strong on-site Hubbard-$U$ 
interaction only suppresses double electron occupation on a single site and cannot 
cause electron localization due to the 1/6 electron filling. The strong inter-site 
repulsion $V_1$ penalizes the double occupancy in the up-triangles and drives a 
Mott transition to the type-I CMI. Because the number of the up-triangles is equal 
to the total electron number, there is exactly one electron in each up-triangle in the type-I CMI.

Using the same slave-rotor formulation that we used in the beginning of Sec.~\ref{sec2}, 
we obtain the same Hamiltonians for the spin and charge sectors as in Eqs.~\eqref{eq5} 
and \eqref{eq6}
except here we have $V_2 =0$. 
We adopt a mean-field approximation and rewrite the 
charge sector Hamiltonian as
\begin{eqnarray}
H_{\text{ch}} &=& -2  {J}_1^{\text{eff}} 
\sum_{\langle ij \rangle \in \text{u} } 
\cos( \theta_i - \theta_j) - 2{J}_2^{\text{eff}} 
\sum_{ \langle ij \rangle \in \text{d}} \cos( \theta_i - \theta_j)
\nonumber \\
&+&   \frac{V_1}{2} 
\sum_{{\bf r}\in \text{u}}\mathbb{L}_{\bf r}^2  
+ (h + \frac{5V_1}{2}) \sum_{ {\bf r} \in \text{u}  }  \mathbb{L}_{\bf r}
+ \frac{U-V_1}{2} \sum_i  L_i^2 
\nonumber \\
&+&  \text{constant},
\label{eqapp3}
\end{eqnarray}
where 
${\bf r}$ labels the position of the center of each up-triangle or the Kagome unit cell. 
As we explain in Sec.~\ref{sec2}, 
a lattice site can be either labelled by the index $i$ or the combination of the 
unit cell position ${\bf r}$ and the sublattice index $\mu = \text{A,B,C}$. 
In Eq.~\eqref{eqapp3}, we have introduced an angular momentum 
operator $\mathbb{L}_{\bf r}$ that is defined as 
 \begin{eqnarray}
 \mathbb{L}_{\bf r}& =& L_{{\bf r}\text{A}} + L_{{\bf r}\text{B}} + L_{{\bf r}\text{C}} 
 + \frac{1}{2}
 \\
 &\equiv& \sum_{\mu,\sigma} 
 f^{\dagger}_{{\bf r}\mu\sigma}f^{\phantom\dagger}_{{\bf r}\mu\sigma} - 1,
 \end{eqnarray}
 where ${\bf r} \in \text{u}$. 
 $\mathbb{L}_{\bf r}$ measures the total electron occupation on the up-triangle at ${\bf r}$. 
 Moreover, from the allowed values of $L_{{\bf r}\mu}$, we know that $\mathbb{L}_{\bf r}$ 
 can take $-1,0,1,2$. 
 Finally, because $L_i = \pm 1/2$, the last term in the second line of Eq.~\eqref{eqapp3} 
 also reduces to a constant and thus can be dropped from the Hamiltonian.

It is convenient to define the conjugate variable for the new angular variable 
$\mathbb{L}_{\bf r}$. 
We introduce a super-rotor operator $e^{\pm i \Theta_{\bf r}}$ whose physical meaning is to 
create and annihilate an electron charge in the up-triangle at ${\bf r}$, respectively.  
Clearly, we should have 
\begin{equation}
\Theta_{\bf r} \equiv \frac{1}{3} 
 ( \theta_{{\bf r} \text{A}}  + \theta_{{\bf r}\text{B}} + \theta_{{\bf r}\text{C}}),
\end{equation}
and
\begin{equation}
[\Theta_{\bf r}, \mathbb{L}_{{\bf r}' }   ] = i \delta_{{\bf r}{\bf r}'}. 
\end{equation}
  
Returning to the charge sector Hamiltonian in Eq.~\eqref{eqapp3},
we note that 
$[\theta_{{\bf r}'\mu} - \theta_{ {\bf r}'\nu},  \mathbb{L}_{\bf r}] = 0 $
for any choice of $\mu$ and $\nu$. 
Hence, the hopping terms on the bonds of the up-triangles 
commute with the $V_1$ interaction term and we can then set 
\begin{equation}
 \theta_{{\bf r}\text A} = \theta_{{\bf r}\text B} = \theta_{{\bf r}\text C} 
 \equiv \Theta_{\bf r}
 \label{eq17}
\end{equation}
for each up-triangle. 
Nevertheless, the hopping terms on the bonds of the down-triangles 
do not commute with the $V_1$ interaction. 
Increasing the inter-site repulsion $V_1$ penalizes the kinetic 
energy gain through hoppings on the down-triangle bonds 
(or equivalently through hoppings between different up-triangles)
and causes the electron localization in the up-triangles.
On the other hand, the electrons remain mobile inside each up-triangle and 
can still gain kinetic energy through internal hoppings within the up-triangle. 
Therefore, the system is locally ``metallic'' within each up-triangle and 
remains so even in the type-I CMI when the interaction $V_1$ becomes dominant. 
 
%-----------------------------
\begin{figure}[t]
{\includegraphics[width=6cm]{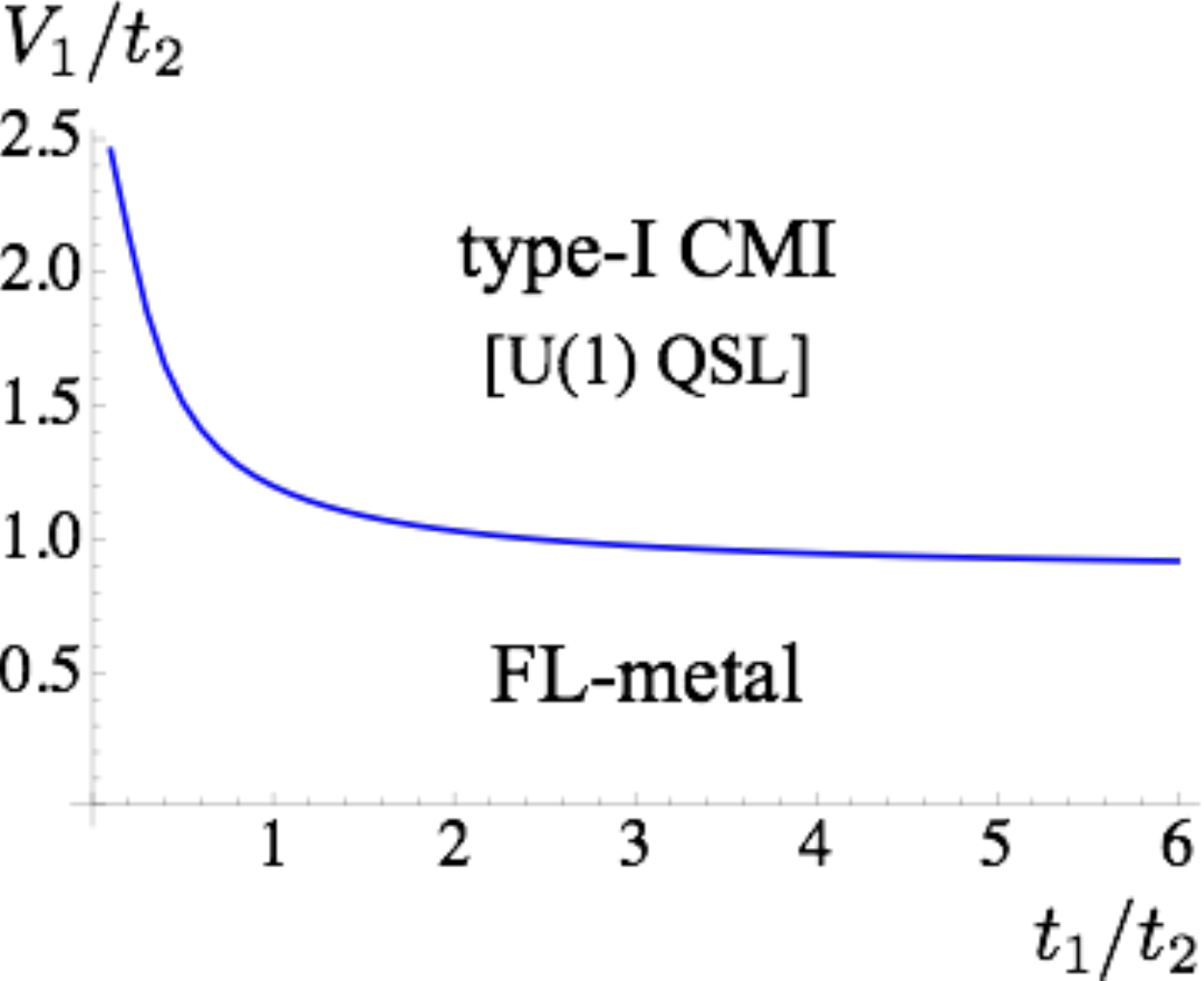}}
\caption{(Color online.) The slave-rotor mean-field phase diagram at $V_2=0$.
We exclude the 120-degree state in the strong coupling limit ($V_1 \gg t_2$). 
}
\label{fig3}
\end{figure}
%-----------------------------
 
Using the local metallic condition in Eq.~\eqref{eq17} to optimize the 
intra-up-triangle hopping, we obtain a reduced rotor Hamiltonian that is 
defined on the triangular lattice formed by the centers of the up-triangles, 
\begin{eqnarray}
\tilde{H}_{\text{ch}} &=& 
- 2 {J}_2^{\text{eff}} \sum_{ \langle {\bf r} {\bf r}\rq{} \rangle } 
\cos (\Theta_{\bf r} - \Theta_{{\bf r}\rq{}} )
+ \sum_{{\bf r} \in \text{u}} \frac{V_1}{2} \mathbb{L}_{\bf r}^2 \nonumber\\
&& + h^{\text{eff}} \sum_{{\bf r} \in \text{u}} \mathbb{L}_{\bf r},
\end{eqnarray}
where $\langle {\bf r} {\bf r}' \rangle$ labels two neighboring up-triangles and 
$h^{\text{eff}} = h + 5V_1/2$. 

Now it is clear that, the relevant degrees of freedom for the Mott transition is 
the super-rotor mode $e^{i\Theta_{\bf r}}$. When this super-rotor mode is condensed 
and $\langle e^{i\Theta_{\bf r}} \rangle \neq 0$, we obtain the FL-metal phase. 
When this super-rotor mode is gapped and 
$\langle e^{i\Theta_{\bf r}} \rangle = 0$,
the type-I CMI with the electrons localized on the up-triangles is obtained. 

In the type-I CMI, there exists charge coherence within the up-triangle clusters (or it is 
``locally metallic"). Thus the gauge field fluctuations within the up-triangles clusters become
massive due to the Higgs mechanism. However, the gauge fluctuations along the links between
two nearby up-triangles remain gapless and we represent it by $a_{{\bf r}{\bf r}\rq{}  }$ for 
two up-triangles at ${\bf r}$ and ${\bf r}\rq{}$.
The reduced rotor Hamiltonian $\tilde{H}_{\text{ch}}$ and the spinon 
Hamiltonian $H_{\text{sp}}$ are then invariant under the following U(1) gauge transformation
\begin{eqnarray}
\left\{
\begin{array}{l}
f^\dagger_{{\bf r}\mu\sigma} \rightarrow f^\dagger_{ {\bf r}\mu\sigma} e^{- i\chi_{\bf r}},
\vspace{1.5mm}
\\
\Theta_{\bf r} \rightarrow \Theta_{\bf r} + \chi_{\bf r},
\vspace{1.4mm}
\\
a_{{\bf r}{\bf r}\rq{}} \rightarrow a_{{\bf r}{\bf r}\rq{}} + \chi_{\bf r} - \chi_{{\bf r}\rq{}},
\end{array}
\right.
\end{eqnarray}
for ${\bf r}, {\bf r}' \in \text{u} $.

In the type-I CMI, the spinon mean-field Hamiltonian $H_{\text{sp}}$ 
describes the hopping of the spinons between the Kagome lattice sites.
As a result, the spinon band structure is identical to the electron band structure in 
Eqs.~\eqref{eqapp2} and \eqref{eqapp3} except for the modified hopping parameters $t_1^{\text{eff}}$
and $t_2^{\text{eff}}$. Thus, the spinons fill a half of the lowest spinon band, leading to a 
spinon Fermi surface. The resulting state in the spin sector is a U(1) QSL with a spinon Fermi surface
and the low-energy physics is described by the spinon Fermi surface coupled to a fluctuating 
U(1) gauge field. It is generally believed that, the U(1) QSL is in the deconfined phase due to the 
presence of the spinon Fermi surface. When the super-rotor mode is condensed, the U(1) gauge 
field picks up a mass via the Higgs' mechanism and the charge rotor and fermionic spinons are then
combined back to the original electron. 

Now we solve the reduced charge sector Hamiltonian $\tilde{H}_{\text{ch}}$
and the spin sector Hamiltonian $H_{\text{sp}}$ self-consistently for the phase diagram 
and the Mott transition. We follow the standard procedure and implement the coherence
state path integral for the super-rotor variable. We integrate out the field $\mathbb{L}_{\bf r}$
and obtain the partition function, 
\begin{equation}
\mathbb{Z} = \int \mathcal{D} {\Phi}^\dagger \mathcal{D} {\Phi} \mathcal{D} \lambda 
e^{- \mathcal{S} -  \sum_{{\bf r} \in \text{u} } \int d\tau \lambda_{\bf r} ( | {\Phi}_{\bf r} |^2 -1 )  },
\label{eq19}
\end{equation}
with the effective action $\mathcal{S}$ 
\begin{equation}
\mathcal{S} = \int d\tau  \sum_{{\bf r}\in \text{u}} \frac{1}{2 V_1} |\partial_{\tau} 
{\Phi}_{\bf r}  |^2  - 
J_2^{\text{eff}} \sum_{\langle {\bf r} {\bf r}\rq{}  \rangle } 
( {\Phi}^\dagger_{\bf r} {\Phi}^{\phantom\dagger}_{{\bf r}\rq{} }  + h.c. )
\end{equation}
and $ {\Phi}^{\dagger}_{\bf r} \equiv e^{i\Theta_{\bf r}}$. 
Because $ \sum_{{\bf r} \in \text{u}} \langle \mathbb{L}_{\bf r} \rangle = 0$, 
the parameter $h^{\text{eff}}$ 
is required to vanish so we drop this term in the action. In Eq.~\eqref{eq19}, we 
introduce the Lagrange multiplier $\lambda_{\bf r}$ to enforce 
the unimodular constraint $| {\Phi}_{\bf r} | =1$ for each up-triangle. 
We make a uniform saddle point approximation by setting $\lambda_{\bf r} = \lambda$. 
Upon integrating out the ${\Phi}$ fields, we obtain the following 
saddle point equation in the Mott insulating phase, 
\begin{equation}
\frac{1}{S_{\text{BZ}}}  \int d^2 {\bf k} \frac{V_1}{ \omega_{\bf k}} =1,
\label{saddleEq}
\end{equation}
where $S_{\text{BZ}}$ is the area of the first Brioullin zone of the triangular lattice
and $\omega_{\bf k}$ is the dispersion of the super-rotor mode with
\begin{equation}
\omega_{\bf k} = \big[
2V_1 
\big(
\lambda - 2J_2^{\text{eff}} ( \cos {\bf k}\cdot {\bf b}_1
+ \cos {\bf k}\cdot {\bf b}_2 + \cos {\bf k}\cdot ({\bf b}_1 + {\bf b}_2)
\big) 
\big]^{\frac{1}{2}}.
\end{equation}
The super-rotor mode is condensed when the dispersion $\omega_{\bf k}$ becomes 
gapless. This occurs when $\lambda = 6 J_2^{\text{eff}}$. 
Combining this with the super-rotor saddle point equation Eq.~\eqref{saddleEq}
and the spinon-sector mean-field theory, we solve for the 
mean-field phase diagram that is depicted in Fig.~\ref{fig3}.  
Here we do not consider the 
possibility of magnetic ordering in the strong Mott regime. 
For a small (large) $V_1/t_2$, we obtain a FL-metal
(a U(1) QSL with a spinon Fermi surface).
The Mott transition is continuous and of the quantum
XY type in the mean-field theory, which is expected to be 
so even after including the U(1) gauge fluctuations.\cite{Senthil08July1} 
The phase boundary of the Mott transition is understood as follows. 
For smaller (larger) $t_1/t_2$, the electrons gain more (less)
kinetic energy from the $t_2$ hopping or the inter-up-triangle hopping, 
and thus, a larger (smaller) critical $V_1/t_2$ is needed to localize
the electrons in the up-triangles. In particular, 
in the limit of $t_1/t_2 \rightarrow \infty$, the extended 
Hubbard model with $V_2 = 0$ and 1/6 electron filling is
equivalent to a triangular lattice Hubbard model at half-filling
where the triangular lattice is formed by the up-triangles. 
Therefore, the U(1) QSL with a Fermi surface 
in the type-I CMI is smoothly connected to the U(1) QSL 
with a Fermi surface in the triangular lattice Hubbard model 
at the half-filling\cite{Lee05}.

\section{Mean-field theory for the type-II CMI} 
\label{secapp2} 

\subsection{Slave-particle mean-field theory} 
\label{secapp2A} 

In this section, we explain how to solve the slave-particle mean-field theory in Sec.~\ref{sec3B}. 
We combine the string sector with the spinon and charged-boson sectors, and solve the mean-field
equations self-consistently. 
%The charge-boson mean-field Hamiltonian is 
%an interacting model. 
To solve the bosonic mean-field Hamiltonian, we first introduce a rotor variable $\phi_{\bf r}$ 
that is conjugate to the U(1)$_{\text{ch}}$ charge operator $Q_{\bf r}$ with
\begin{equation}
[\phi_{\bf r}, Q_{\bf r}] = i,
\end{equation}
and hence
\begin{eqnarray}
&& \bar\Phi_{\bf r} = e^{-i \phi_{\bf r}},\\
&& \bar\Phi^\dagger_{\bf r} \bar\Phi^{\phantom\dagger}_{\bf r} =1. 
\end{eqnarray}
We then carry out the same procedure as in Appendix.~\ref{secapp1}. 
We implement the coherent
state path integral for the $\Phi_{\bf r}$ fields and integrate out the $Q_{\bf r}$ field. 
The resulting partition function for the up-triangle subsystem is given by
\begin{equation}
\mathbb{Z}_{\text u} = \int {\mathcal D} \bar\Phi^\dagger {\mathcal D} 
\bar\Phi {\mathcal D} \lambda
e^{- {\mathcal S}_{\text u} - \sum_{{\bf r} \in \text{u}} \int d\tau
\lambda_{\bf r} ( | \bar\Phi_{\bf r}|^2 -1 )  }, 
\end{equation}
where 
the Lagrange multiplier $\lambda_{\bf r}$ is used to implement the unimodular constraint for the 
$\Phi$ field at each ${\bf r}$ site. 
The effective action $\mathcal{S}_{\text u}$ for the up-triangle subsystem is 
\begin{equation}
\mathcal{S}_{\text u} = \int d\tau \sum_{{\bf r} \in \text{u}} 
\frac{1}{2V_1} | \partial_{\tau} \bar\Phi_{\bf r}  |^2
- \bar{J}_1 \sum_{\langle {\bf r} {\bf r}' \rangle \in \text{u}} 
(
\bar\Phi_{\bf r}^\dagger \bar\Phi_{\bf r}^{\phantom\dagger} 
+ h.c.
),
\end{equation}
where $\langle {\bf r}{\bf r}'\rangle$ refers to the nearest-neighbor sites 
on the up-triangle subsystem. 
The rest of the treatment is identical to what we did to the super-rotor mode in Appendix.~\ref{secapp1}
and we can then find the critical $V_1/\bar{J}_1$ at which the boson 
is condensed. 
Similar effective action can be straight-forwardly written down for the down-triangle subsystem and
is given by
\begin{equation}
\mathcal{S}_{\text d} = \int d\tau \sum_{{\bf r} \in \text{u}} 
\frac{1}{2V_2} | \partial_{\tau} \bar\Phi_{\bf r}  |^2
- \bar{J}_2 \sum_{\langle {\bf r} {\bf r}' \rangle \in \text{d}} 
(
\bar\Phi_{\bf r}^\dagger \bar\Phi_{\bf r}^{\phantom\dagger} 
+ h.c.
),
\end{equation}
where $\langle {\bf r}{\bf r}'\rangle$ refers to the nearest-neighbor sites 
on the down-triangle subsystem. 

Notice that the above actions for up- and down-triangle subsystems
look identical to the action of the super-rotor mode in Appendix~\ref{secapp1} 
except for different labelling of couplings and operators. This indicates the close
connection between this slave-particle approach used here and the slave-rotor formulation used
in Appendix~\ref{secapp1}. 
In the region of the phase diagram where the two approaches overlap, they 
should give qualitatively the same results. 
The difference is that, the slave-rotor approach in Appendix~\ref{secapp1} does not take into 
account of the reduction of the spinon or electron bandwidth due to the 
on-site Hubbard interaction, while the current approach takes care of that 
through the string parameters.  
As a result, we expect that the slave-rotor approach probably overestimates 
the stability of the FL-metal phase.

\subsection{String mean-field theory} 
\label{secapp2B} 

Here we explain the string mean-field theory in Sec.~\ref{sec3C} in details. 
To solve the combined 
Hamiltonian of $\tilde{H}_{\text{sp}}$ and $\tilde{H}_{\text{ring}}$ self-consistently, 
we obtain the effective spinon hoppings by evaluating 
the boson or rotor hopping amplitudes with respect to the variational ground state $\Psi(\{z_i \})$,
\begin{eqnarray}
\langle L^+_{\mu} ( {\bf R}) L^-_{\nu} ({\bf R}) \rangle & = &
\langle L^+_{\mu} ( {\bf R}) \rangle \langle L^-_{\nu} ({\bf R}) \rangle
\end{eqnarray}
where $\mu,\nu$ label the sublattices.
We also evaluate the parameter $M_{ijklmn}$ against
the spinon hopping Hamiltonian. Using the Wick theorem, we have
\begin{eqnarray}
M_{ijklmn} &=& \quad
\sum_{\alpha\beta\gamma} 
\langle f^\dagger_{i\alpha} f^{\phantom\dagger}_{j\alpha} \rangle
\langle f^\dagger_{k\beta} f^{\phantom\dagger}_{l\beta} \rangle
\langle f^\dagger_{m\gamma} f^{\phantom\dagger}_{n\gamma} \rangle
\nonumber \\
&& +  \sum_{\alpha} \langle f^\dagger_{i\alpha} f^{\phantom\dagger}_{n\alpha} \rangle
\langle f^\dagger_{k\alpha} f^{\phantom\dagger}_{j\alpha} \rangle
\langle f^\dagger_{m\alpha} f^{\phantom\dagger}_{l\alpha} \rangle
\nonumber \\
&& +  \sum_{\alpha} \langle f^\dagger_{i\alpha} f^{\phantom\dagger}_{l\alpha} \rangle
\langle f^\dagger_{m\alpha} f^{\phantom\dagger}_{j\alpha} \rangle
\langle f^\dagger_{k\alpha} f^{\phantom\dagger}_{n\alpha} \rangle
\nonumber \\
&&- \sum_{\alpha\beta}  \langle f^\dagger_{i\alpha} f^{\phantom\dagger}_{j\alpha} \rangle
\langle f^\dagger_{k\beta} f^{\phantom\dagger}_{n\beta} \rangle
\langle f^\dagger_{l\beta} f^{\phantom\dagger}_{m\beta} \rangle
\nonumber \\
&&- \sum_{\alpha\beta}  \langle f^\dagger_{i\alpha} f^{\phantom\dagger}_{l\alpha} \rangle
\langle f^\dagger_{k\beta} f^{\phantom\dagger}_{j\beta} \rangle
\langle f^\dagger_{m\beta} f^{\phantom\dagger}_{n\beta} \rangle
\nonumber \\
&&- \sum_{\alpha\beta}  \langle f^\dagger_{i\alpha} f^{\phantom\dagger}_{n\alpha} \rangle
\langle f^\dagger_{m\beta} f^{\phantom\dagger}_{j\beta} \rangle
\langle f^\dagger_{k\beta} f^{\phantom\dagger}_{l\beta} \rangle
\label{eq73}
 \\
&=& \chi_{ij}^3 + \frac{\chi_{ik}^3}{4} + \frac{\chi_{il }^3}{4}
       -\frac{3}{2} \chi_{ij} \chi_{ik} \chi_{il}, 
\label{eq74a}
\end{eqnarray}
where we have defined the $\chi$ variable as 
\begin{equation}
\chi_{ij} = \sum_{\alpha} \langle f^{ \dagger}_{i \alpha} f^{\phantom\dagger}_{j \alpha}  \rangle,
\end{equation}
and we have also used the three-fold rotational symmetry as well as the reflection symmetry of the 
hexagon in Eq.\eqref{eq74a} so that
\begin{eqnarray}
&&\chi_{ij} = \chi_{kl} = \chi_{mn}
\\
&&\chi_{ik} = \chi_{km} = \chi_{mi} = \chi_{jl} = \chi_{ln} = \chi_{nj} 
\\
&& \chi_{ik} = \chi_{jm} = \chi_{kn} .
\end{eqnarray}

Strictly speaking, the above self-consistent mean-field theory can only be applied to the 
string condensed phases in the terminology of Levin and Wen in Ref.~\onlinecite{Levin04}. 
The string condensed phase in our problem corresponds to the deconfined phase of the 
U(1)$_{\text{ch}}$ gauge theory in the charge sector. For a string condensed phase or the 
deconfined phase of the U(1)$_{\text{ch}}$ gauge theory, we can apply the string mean-field 
theory or gauge mean-field theory by setting 
$\langle L^{\pm}_i \rangle \equiv \langle l^{\pm}_i \rangle \neq 0  $.  
As we discussed previously, the U(1)$_{\text{ch}}$ gauge field can be in 
a deconfined phase because of the gapless spinon Fermi surface, and if it happens, 
the PCO phase may be more properly labelled as the PCO$^{\ast}$ 
phase (with fractionalized excitations). Even if the U(1)$_{\text{ch}}$ gauge field is in the confining phase, 
since our goal is to understand the reconstruction of the spinon band structure 
due to the PCO, the above mean-field theory does qualitatively captures this part of the physics. 
So for our purpose, we essentially put the U(1)$_{\text{ch}}$ gauge field of the charge 
sector into the deconfined phase and introduce the PCO on top of that in the above mean-field theory. 
Moreover, since the spinon Fermi surface always exists even when the PCO appears, 
the U(1)$_{\text{sp}}$ gauge field is expected to be in the deconfined phase and the spinon 
degrees of freedom remains to be a valid description of the low-energy sector of the system 
even when the charge sector U(1)$_{\text{ch}}$ gauge field is confining.

\section{Linear spin-wave theory for the Kugel-Khomskii model in a strong magnetic field}
\label{secapp3} 

The classical U(1) ground state degeneracy of the reduced Kugel-Khomskii model in Sec.~\ref{sec4B} 
is lifted when the quantum fluctuation is turned on. 
We study this quantum order by disorder phenomenon using the linear spin wave theory. 
We use the Holstein-Primakoff (HP) bosons to represent the pseudospin operator,
\begin{eqnarray}
\boldsymbol{\tau}({\bf R}) \cdot \hat{\boldsymbol{\tau}}_{\text{cl}} 
&=& \frac{1}{2} - a^\dagger ({\bf R}) a({\bf R})
\\
\boldsymbol{\tau}({\bf R})\cdot ( \hat{y} \times  \hat{\boldsymbol{\tau}}_{\text{cl}} )
&=& \frac{1}{2} [ a({\bf R}) + a^\dagger ({\bf R}) ],
\end{eqnarray}
where $\hat{\boldsymbol{\tau}}_{\text{cl}} =
 {\boldsymbol{\tau}}_{\text{cl}}/|{\boldsymbol{\tau}}_{\text{cl}} |  $
 and $ {\boldsymbol{\tau}}_{\text{cl}}$
is given in Eq.~\eqref{eq83}.
 We keep the quadratic term in $a$ and $a^\dagger$, and write
the reduced Kugel-Khomskii model as
\begin{eqnarray}
H_{\text{r-KK}} &=& E_{\text{cl}} + \sum_{{\bf k} \in \text{BZ2}} \big[
2 A_{\bf k} a_{\bf k}^\dagger a_{\bf k}^{\phantom\dagger} 
\nonumber 
\\
&&
+ B_{\bf k} (a_{\bf k} a_{-{\bf k}} + a^\dagger_{\bf k} a^\dagger_{-{\bf k}})\big],
\end{eqnarray}
where ``BZ2'' is the Brioullin zone of the ETL (see Fig.~\ref{fig12}d) and 
\begin{widetext}
\begin{eqnarray}
E_{\text{cl}} &=& -\frac{J'}{12} \frac{N_{\Delta}}{3} ,
\\
A_{\bf k} &=&  \frac{2J'}{9}\big[ - \frac{\sin^2 ( \vartheta - \pi/3)}{4} 
\cos ({\bf k}\cdot {\bf a}_x)
 -  \frac{\sin^2 ( \vartheta +\pi/3)}{4} 
\cos ({\bf k}\cdot {\bf a}_y)
 - \frac{\sin^2 \vartheta}{4} \cos ({\bf k} \cdot {\bf a}_z) +  \frac{3}{4} \big],
\\
B_{\bf k} &=&  \frac{2J'}{9} \big[ - \frac{\sin^2 ( \vartheta - \pi/3)}{4}  
\cos ({\bf k} \cdot {\bf a}_x)
 - \frac{\sin^2 ( \vartheta+ \pi/3)}{4}  \cos ({\bf k} \cdot {\bf a}_y)-
 \frac{\sin^2 \vartheta}{4} \cos ({\bf k} \cdot {\bf a}_z)\big]  .
\end{eqnarray}
\end{widetext}

\begin{figure}[ht]
{\includegraphics[width=6.5cm]{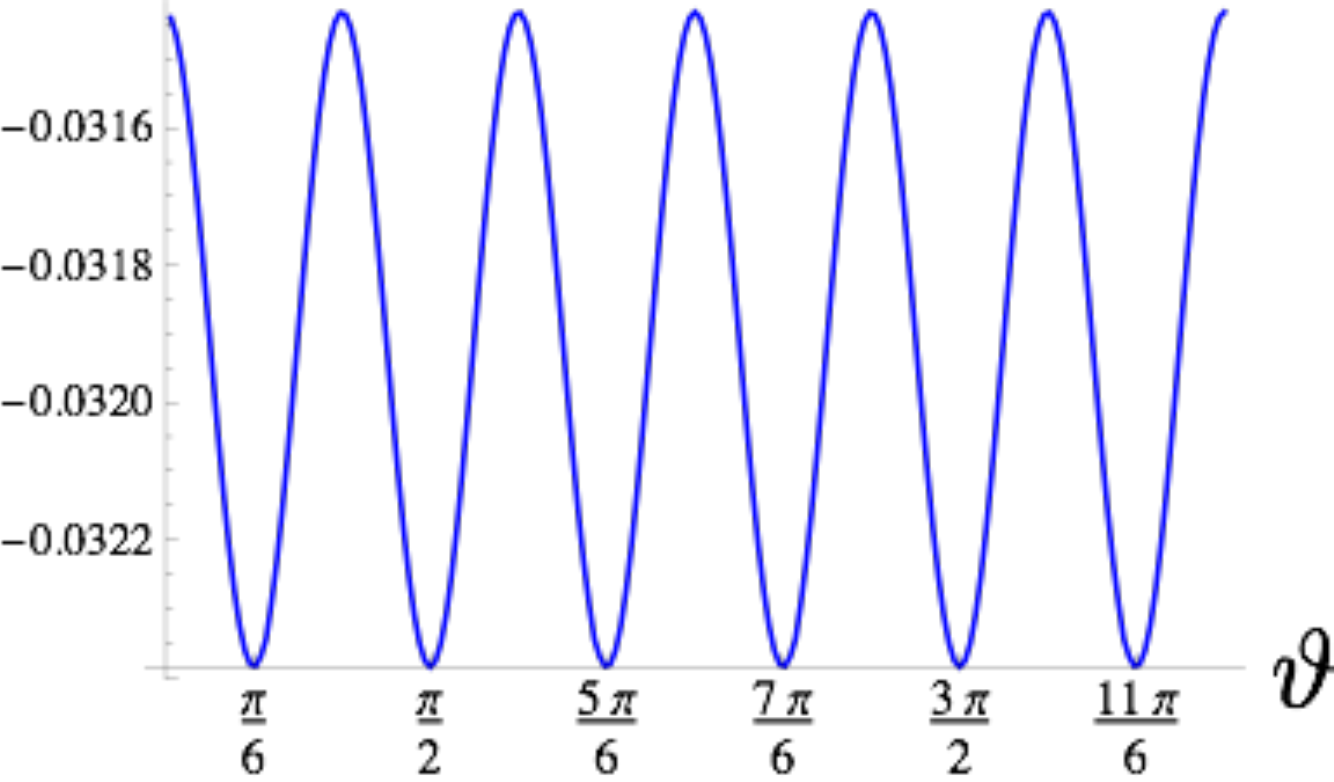}}
\caption{(Color online.) 
The quantum correction to the classical ground state energy per unit cell of the ETL. The energy unit is set to $2J'/9$.
}
\label{fig14}
\end{figure}

The pseudospin wave Hamiltonian is then diagonalized by a Bogoliubov 
transformation and is given by 
\begin{eqnarray}
H_{\text{r-KK}} &=& E_{\text{cl}} +\sum_{{\bf k} \in \text{BZ2}}
[\frac{\omega_{\bf k}}{2}  -  A_{\bf k} ]+ \sum_{{\bf k} \in \text{BZ2}}
\omega_{\bf k} \alpha^\dagger_{\bf k} 
\alpha_{\bf k}^{\phantom\dagger} 
\label{eq90}
\end{eqnarray}
where the pseudospin wave mode reads 
\begin{equation}
\omega_{\bf k} = 2 \sqrt{A_{\bf k}^2 - B_{\bf k}^2},
\end{equation}
which is gapless at the $\Gamma$ point due to the accidental U(1) degeneracy. 
This pseudo-Goldstone mode is expected to be gapped if the interaction between 
HP bosons is included. 
From Eq.~\eqref{eq90}, the quantum correction to the ground state energy is 
\begin{equation}
\Delta E = \sum_{{\bf k} \in \text{BZ2}}[\frac{\omega_{\bf k}}{2} - A_{\bf k} ]. 
\end{equation}
In Fig.~\ref{fig14}, we plot the quantum correction as a function of the parameter $\vartheta$. 
The minima occur at $\vartheta = \pi/6+ n\pi/3$ ($n\in \text{integer}$). 

%-----------------------------
%-----------------------------

What is the physical consequence of the pseudospin ordering? 
To address this question, we consider the following product 
wavefunction which is appropriate for the ${\bf q} =0$ 
state for the pseudospin ordering,
\begin{equation}
|\Psi\rangle = \prod_{\bf R} \big[ \cos \frac{\vartheta}{2} 
| {\uparrow \uparrow} \rangle_{\bf R} 
+ \sin \frac{\vartheta}{2} | {\downarrow \uparrow} \rangle_{\bf R} \big],
\end{equation}
This variational wavefunction gives the classical pseudospin ordering 
in Eq.~\eqref{eq83}. From this wavefunction, we find that the electron 
density is uniform at every site within each resonating hexagon
and thus preserves the rotation and reflection symmetries. 
We then compute the local magnetization for each site within the 
resonating hexagon, 
\begin{eqnarray}
\langle n_{1\uparrow} - n_{1\downarrow}  \rangle &=& 
\langle n_{6\uparrow} - n_{6\downarrow}  \rangle 
 = \frac{1}{6}+ \frac{\sin (\vartheta-\frac{\pi}{6})}{3} ,
 \\
\langle n_{2\uparrow} - n_{2\downarrow}  \rangle &=& 
\langle n_{3\uparrow} - n_{3\downarrow}  \rangle
=\frac{1}{6} + \frac{\sin \vartheta}{3},
 \\
\langle n_{4\uparrow} - n_{4\downarrow}  \rangle &=& 
\langle n_{5\uparrow} - n_{5\downarrow}  \rangle 
=\frac{1}{6} - \frac{\sin( \vartheta + \frac{\pi}{6})}{3}.
\end{eqnarray}
Although the total local magnetization of each resonating hexagon is 
$\langle s^z \rangle = 1/2$,
the pseudospin ordering leads to the modulation of the spin ordering inside 
each resonating hexagon. The 3-fold rotational symmetry about the 
center of the resonating hexagon is explicitly broken
by the pseudospin ordering.

\section{Landau theory for the thermal transition at $T^{\ast}$} 
\label{secapp4} 

We work out the symmetry allowed Landau free energy for the PCO. 
The charge order parameter is introduced as 
\begin{equation}
{ n}_c ({\bf r})  = \bar{n} + 
\sum_{i=1,2,3} \text{Re} ( \Phi_i e^{i {\bf q}_i \cdot {\bf r}}  )
\end{equation}  
where $n_c ( {\bf r})$ is a coarse-grained electron density at 
the position ${\bf r}$, $\bar{n}$ is the uniform electron 
charge density, and $\{ {\bf q}_1, {\bf q}_2, {\bf q}_3 \}$
are 3 wavevectors corresponding to the charge modulation in the PCO. 
As shown in Fig.~\ref{fig12}d, 
${\bf q}_1, {\bf q}_2, {\bf q}_3$ are also the basis vectors of the Brioullin zone of the ETL.
 
The material LiZn$_2$Mo$_3$O$_8$ 
is described by the R$\bar{3}$m space group, which is taken as the symmetry 
of the high temperature normal phase. The R$\bar{3}$m space group is generated by five operations, 
including two translations,
\begin{eqnarray}
T_1 &:& {\bf r} \rightarrow {\bf r} + {\bf b}_1, 
\\
T_2 &:& {\bf r} \rightarrow {\bf r} + {\bf b}_2.
\end{eqnarray}
Next we consider the 2-fold and 3-fold rotations,
\begin{eqnarray}
\hat{R}_2 &:& (x,y) \rightarrow (-\frac{x}{2} + \frac{ \sqrt{3} y}{2}, 
\frac{\sqrt{3} x}{2} + \frac{y}{2}),
\\
\hat{R}_3 &:& (x,y) \rightarrow (-\frac{x}{2}-\frac{\sqrt{3} y}{2} , 
\frac{\sqrt{3} x }{2} -\frac{y}{2} ). 
\end{eqnarray}
and finally an inversion,
\begin{equation}
I : {\bf r} \rightarrow -{\bf r}. 
\end{equation}

From the symmetry operation on the electron density
\begin{equation}
{\mathcal O} : n_c({\bf r}) \rightarrow n_c( \mathcal{O}^{-1} {\bf r} ),
\end{equation}
we obtain the symmetry transformation for the PCO order parameter, 
\begin{eqnarray}
T_1 &:& \Phi_i \rightarrow \Phi_i e^{-i {\bf q}_i \cdot {\bf b}_1}, \\
T_2 &:& \Phi_i \rightarrow \Phi_i e^{-i {\bf q}_i \cdot {\bf b}_2}, \\
\hat{R}_2&:& \left\{
\begin{array}{l}
\Phi_1 \rightarrow \Phi_2 \\
\Phi_2 \rightarrow \Phi_1 \\
\Phi_3 \rightarrow \Phi_3
\end{array}
\right. ,\\
\hat{R}_3&:& \left\{ 
\begin{array}{l}
\Phi_1 \rightarrow \Phi_3  \\
\Phi_2 \rightarrow \Phi_1 \\
\Phi_3 \rightarrow \Phi_2
\end{array}
\right. , \\
I &:& \Phi_i \rightarrow \Phi^{\ast}_i.
\end{eqnarray}

From the above symmetry operation, we then determine 
the symmetry allowed Landau free energy
in the vincinity of the phase transition (up to third order), 
\begin{eqnarray}
F &=& c_{2,1} \sum_{i } |\Phi_i|^2 
+ c_{2,2} \sum_{i\neq j} \Phi_i^{\ast} \Phi_j 
+ c_{3,1} \sum_{i}  \text{Re} (\Phi_i^3 )
\nonumber \\ 
&+& c_{3,2} \sum_{i\neq j}\text{Re} (\Phi_i^2 \Phi_j + \Phi_i \Phi_j^2) 
  + c_{3,3} \text{Re} (\Phi_1 \Phi_2 \Phi_3)
  \nonumber \\
  &+& \mathcal{O} ( \Phi_i^4)
\end{eqnarray}
Due to the commensurate charge order, the cubic order umklapp term is allowed in the free energy.
For the PCO, we have $|\Phi_1| =| \Phi_2 | = |\Phi_3|$. From the presence of a cubic term in the free energy, we expect a first order phase transition for the clean system.

\end{document}